\documentclass[aps,prd,preprint, nofootinbib,superscriptaddress]{revtex4-1}
\usepackage{amsmath,amssymb,amsbsy}
\usepackage{color}
\usepackage{graphicx}
\usepackage{ulem}
\usepackage{cancel}

\newcommand{\beqa}{\begin{eqnarray}}
\newcommand{\eeqa}{\end{eqnarray}}

\newcommand{\kcotd}{k \cot \delta_0(k)}

\newcommand{\EREp}{ERE$_{k^2>0,BE}$\ }
\newcommand{\EREn}{ERE$_{k^2<0}$\ }
\begin{document}
\preprint{YITP-17-24}
\preprint{UTCCS-P-102}
\preprint{RIKEN-QHP-301}

\title{
Are two nucleons bound in lattice QCD for heavy quark masses?\\
-- Consistency check with L\"uscher's finite volume formula --\\
}

\newcommand{\Tsukuba}{
Center for Computational Sciences, University of Tsukuba, Tsukuba 305-8577, Japan
}

\newcommand{\Kyoto}{
Center for Gravitational Physics, 
Yukawa Institute for Theoretical Physics, Kyoto University, Kitashirakawa Oiwakecho, Sakyo-ku, 
Kyoto 606-8502, Japan
}

\newcommand{\Riken}{
Theoretical Research Division, Nishina Center, RIKEN, Wako 351-0198, Japan
}

\newcommand{\RikenB}{
iTHEMS Program and iTHES Research Group, RIKEN, Wako 351-0198, Japan
}

\newcommand{\Nihon}{
Nihon University, College of Bioresource Sciences, Kanagawa 252-0880, Japan
}

\newcommand{\RCNP}{
Research Center for Nuclear Physics (RCNP), Osaka University, Osaka 567-0047, Japan
}

\author{Takumi Iritani}
\affiliation{\Riken}

\author{Sinya~Aoki}
\affiliation{\Kyoto}
\affiliation{\Tsukuba}

\author{Takumi Doi}
\affiliation{\Riken}
\affiliation{\RikenB}

\author{Tetsuo Hatsuda}
\affiliation{\Riken}
\affiliation{\RikenB}

\author{Yoichi~Ikeda}
\affiliation{\RCNP}

\author{Takashi~Inoue}
\affiliation{\Nihon}
 
\author{Noriyoshi~Ishii}
\affiliation{\RCNP}

\author{Hidekatsu Nemura}
\affiliation{\Tsukuba}

\author{Kenji Sasaki}
\affiliation{\Kyoto}

\collaboration{HAL QCD Collaboration}

\begin{abstract}
  On the basis of the 
  L\"uscher's finite volume formula,  a simple test (consistency check or sanity check) is introduced and applied to inspect the 
  recent claims of the existence of  the nucleon-nucleon ($NN$) bound state(s) for heavy quark masses in lattice QCD. 
We show that the consistency between the scattering  phase shifts  at $k^2 > 0$ and/or  $k^2 < 0$
  obtained from the lattice data 
 and the behavior of phase shifts from the effective range  expansion (ERE) around $k^2=0$ 
  exposes the validity of the original lattice data,   otherwise such information is hidden 
  in the energy shift $\Delta E$ of the two nucleons on the lattice.
 We carry out this sanity check for all the lattice results in the literature claiming the existence of the 
$NN$  bound state(s) for heavy quark masses, and find that 
(i)  some of the $NN$ data  show clear inconsistency between the behavior of ERE at $k^2 > 0$ and that at $k^2 < 0$,
(ii)   some of the $NN$ data exhibit singular  behavior of the low energy parameter (such as the divergent effective range) at $k^2<0$,
(iii) some of the $NN$ data have the unphysical residue for the bound state pole in S-matrix,
and (iv) the rest of the $NN$ data are inconsistent among themselves.
 Furthermore, we raise a caution of using the ERE in the case of the multiple bound states. 
  Our finding, together with the fake plateau problem  previously pointed out by the present authors,
  brings a serious doubt on the existence of 
  the $NN$ bound  states for  pion masses heavier than 300 MeV in the previous studies.
\end{abstract}

\maketitle

\section{Introduction}
\label{sec:introduction}
 In recent years, hadron-hadron interactions in lattice QCD have been investigated by two
approaches. The first approach is the direct method where the ground state energy is extracted
from the temporal correlation function on a finite lattice volume.  If the
interaction is attractive at low energies, the energy shift $\Delta E$ 
in the center of mass system defined by the ground state
energy of two-hadron relative to the sum of hadron masses is always negative in the finite volume: 
  For  bound states (scattering states), $\Delta E$ remains  negative (approaches to zero) 
     in the infinite volume limit.
If the interaction is repulsive, $\Delta E$ is positive in the finite volume, and 
 the scattering phase shift at the corresponding energy can be determined
via L\"uscher's finite volume formula~\cite{Luscher:1990ux}.  The second approach is the HAL QCD
method~\cite{Ishii:2006ec,Aoki:2008hh,Aoki:2009ji}, 
where the energy independent non-local potential between
hadrons is defined and extracted from the spacetime dependence of the
 Nambu-Bethe-Salpeter (NBS) wave function: Observables such as the
binding energies and the scattering phase shifts are obtained by solving
the Schr\"odinger-type equation with the potential.
The HAL QCD method has been extensively applied to 
various two-hadron systems~\cite{Aoki:2011ep,Aoki:2012tk, HALQCD:2012aa, Inoue:2010es,
Inoue:2011ai, Nemura:2008sp, 
Murano:2011nz,Murano:2013xxa,Aoki:2011gt, Aoki:2012bb, Sasaki:2015ifa,
Ikeda:2013vwa,Etminan:2014tya,Yamada:2015cra,Ikeda:2016zwx}
as well as three-hadron systems~\cite{Doi:2011gq} using the derivative expansion
with respect to the non-locality of potentials.
 
For the volume larger than the range of the interactions,
the asymptotic behavior of the NBS wave function encodes the phase shift of the S-matrix.
This phase shift can be extracted from the two-particle energy via
L\"uscher's finite volume formula~\cite{Luscher:1990ux} 
or from the potential through the Schr\"odinger equation~\cite{Aoki:2009ji, Aoki:2012tk}.
  As two methods utilize the property of the same NBS
  wave function \cite{Aoki:2009ji, Aoki:2012tk, Aoki:2010ry}, 
  they in principle give the same results, and
they indeed
agree quantitatively well with each other  in the case of
 the $I=2$ $\pi\pi$ scattering~\cite{Kurth:2013tua}, results for the
two-nucleon ($NN$)  for heavy quark masses show disagreement  (for example,
see Fig.~8 in Ref.~\cite{Doi:2012ab}):  All studies with the direct
method~\cite{Yamazaki:2011nd,Yamazaki:2012hi,Yamazaki:2015asa,Beane:2011iw,
Beane:2012vq,Beane:2013br,Orginos:2015aya,Berkowitz:2015eaa}
indicate that 
bound states appear in both  ${}^1S_0$
(dineutron) and ${}^3S_1$ (deuteron) channels.  On the other hand, the HAL QCD method 
 shows no bound states   in both
channels for heavy quarks \cite{Ishii:2006ec,Aoki:2008hh,Aoki:2009ji,Aoki:2011ep,Aoki:2012tk,
HALQCD:2012aa,Inoue:2011ai}.

\begin{figure}[t]
\centering
  \includegraphics[width=0.48\textwidth]{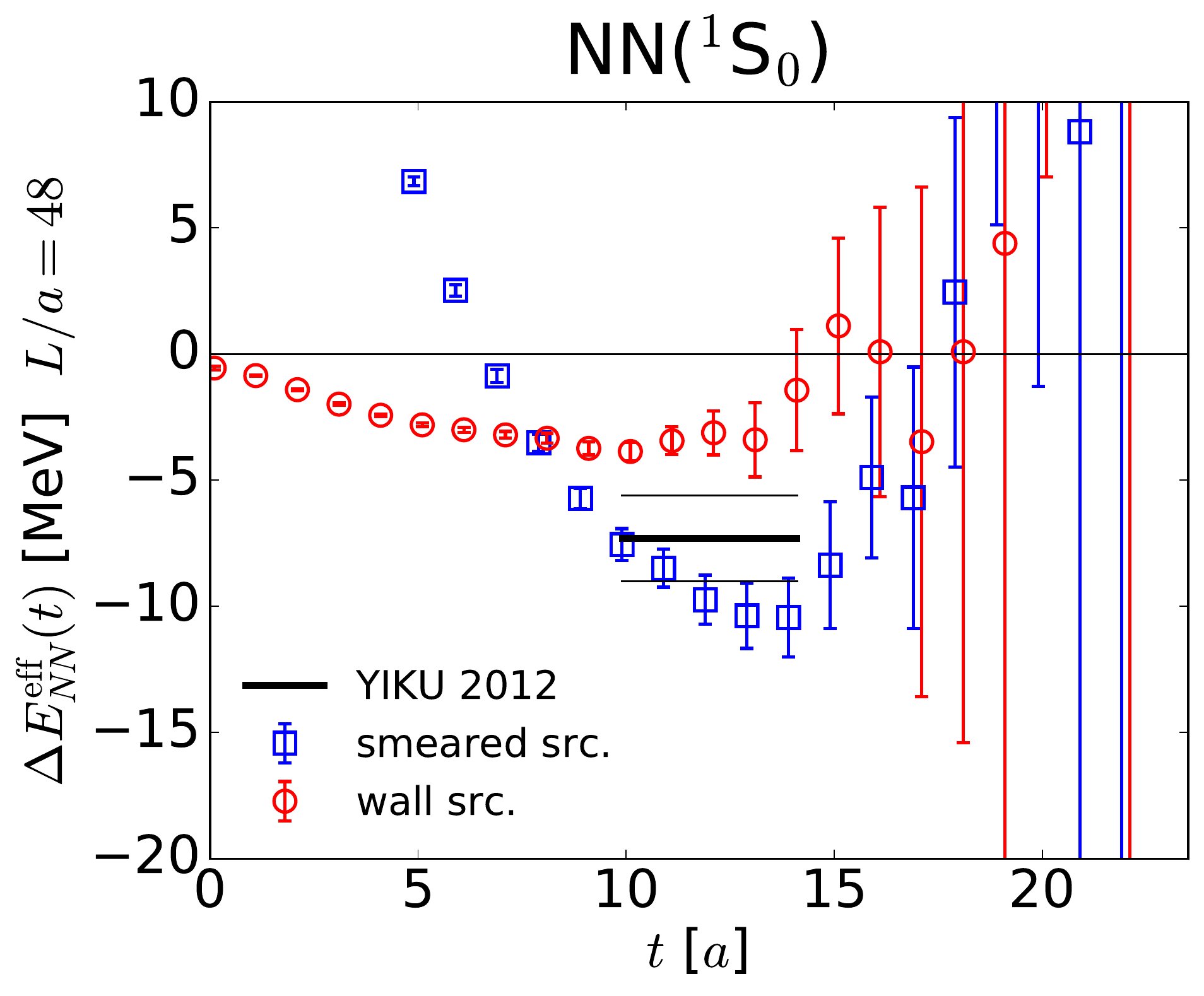}
  \includegraphics[width=0.48\textwidth]{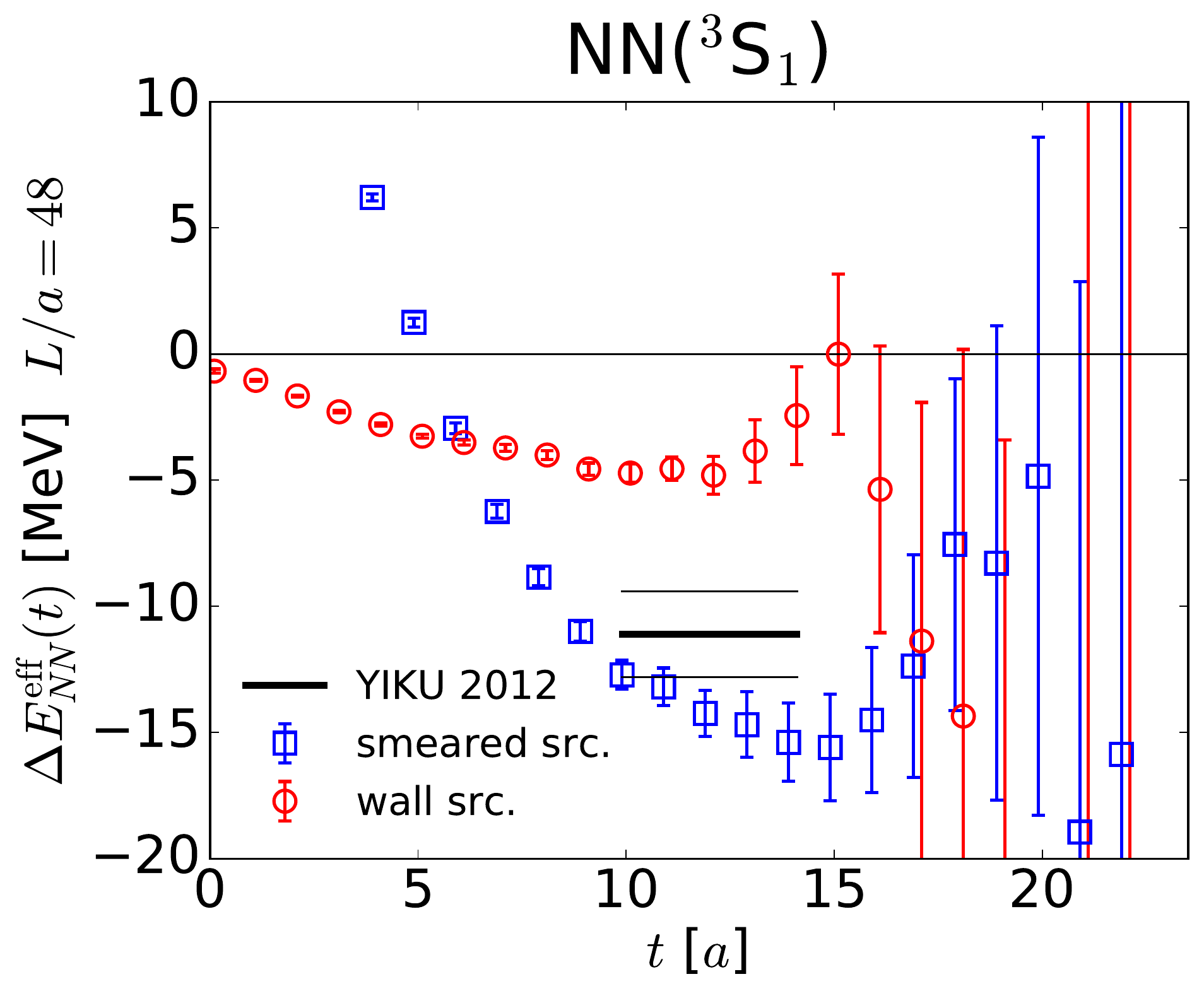} 
 \caption{ Effective energy shift, $\Delta E_{NN}^{\rm eff}(t) = E_{NN}^{\rm
   eff}(t) -2 m_N^{\rm eff}(t)$, in the $NN$ (${}^1S_0$) channel (Left) and the
   $NN$(${}^3S_1$) channel (Right)  at $m_\pi = 0.51$ GeV, $L=4.3$ fm and $a\simeq 0.09$ fm,
   from the smeared source (blue squares) and the wall source (red circles)
   with the non-relativistic operator. Here $E_{NN}^{\rm  eff}(t)$ and
   $m_N^{\rm eff}(t)$ are the effective energy of $NN$ and the effective mass
   of $N$, respectively.  The black solid line represents the fit to the
   plateau of data in Ref.~\cite{Yamazaki:2012hi}, in which $\Delta E_{NN}^{\rm
   eff}(t)$ was calculated from the same smeared source on the same gauge
   configurations but with smaller statistics.  These figures are adapted from
   Ref.~\cite{Iritani:2016jie}.
}
 \label{fig:E_eff}
\end{figure}

In our previous papers~\cite{Iritani:2015dhu,Iritani:2016jie,Iritani:2016xmx}, 
we have studied the origin of this discrepancy.
The direct method is  based on the plateau fitting of the effective energy shift $\Delta E_{\rm eff}(t)$
as a function of the imaginary time $t$.
In principle, one can make a reliable calculation by taking sufficiently large $t$
compared to the inverse of the excitation energy,
while relatively small time regions $t \simeq 1-2$ fm were used in all previous studies.
We pointed out that the plateau identification in the direct method
 for such small imaginary time regions
suffers a serious systematic bias from the excited-state contaminations.
Such a bias is inevitable,  since the multi-baryon on the lattice
 has elastic scattering states whose excitation energies approach zero as the lattice volume increases.
  We have demonstrated this situation, by using mock data,  that even  the 10\%  
contamination of the excited state can easily produce
fake plateaux  (which we called ``mirage''  in \cite{Iritani:2016jie}) at small $t$.  
 Moreover, we have shown that such fake plateaux are indeed observed in lattice data 
  \cite{Iritani:2016jie}. An example  with real data is recapitulated in Fig.~\ref{fig:E_eff},
where plateaux for $\Delta E_{\rm eff}(t)$ are found to be inconsistent 
between smeared and wall quark sources for
 $NN$ source operators.\footnote{A strong sink operator dependence is also observed with the smeared quark source. See appendix A in~\cite{Iritani:2016jie}.}

\begin{figure}[h]
\centering
  \includegraphics[width=0.48\textwidth]{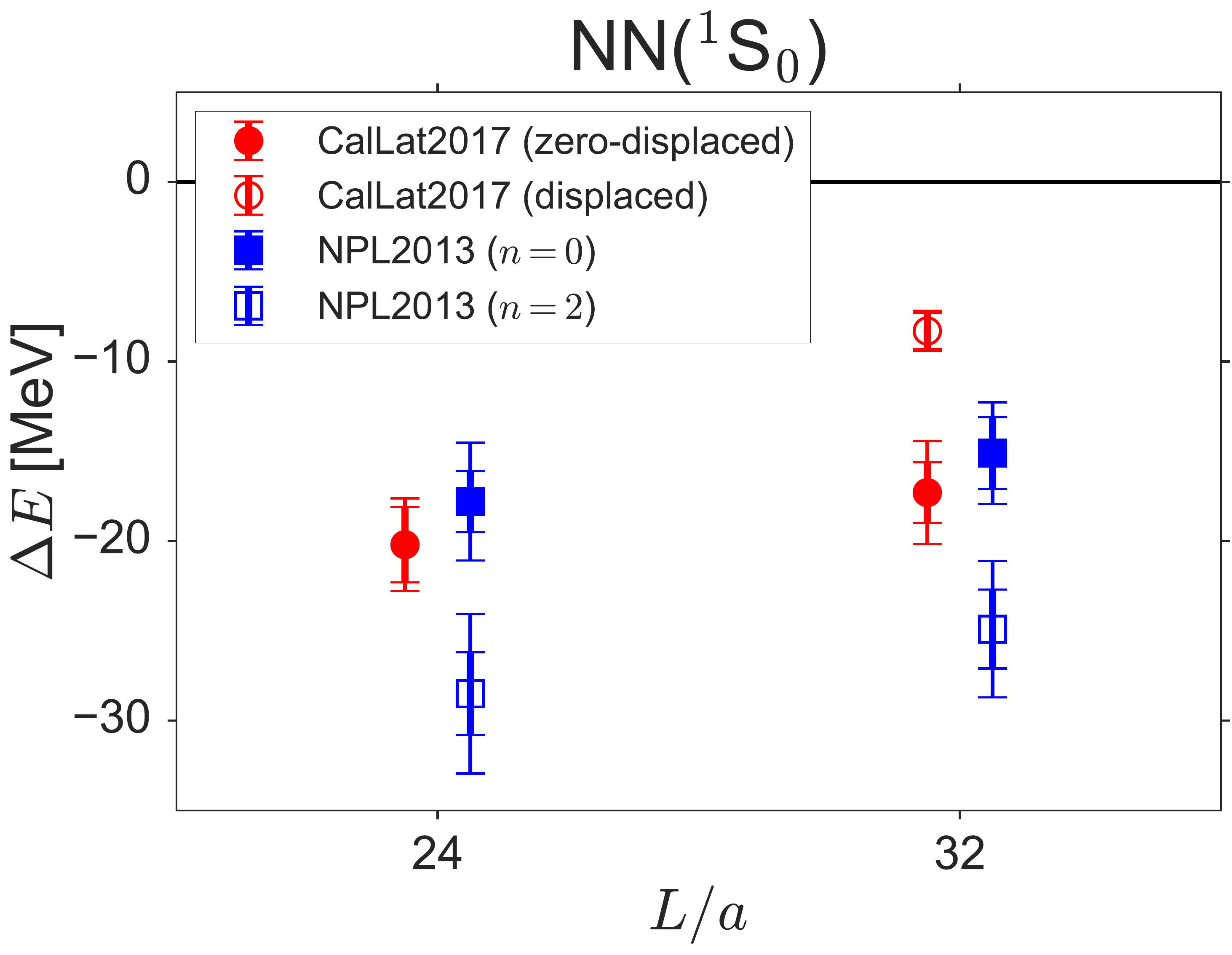}
  \includegraphics[width=0.48\textwidth]{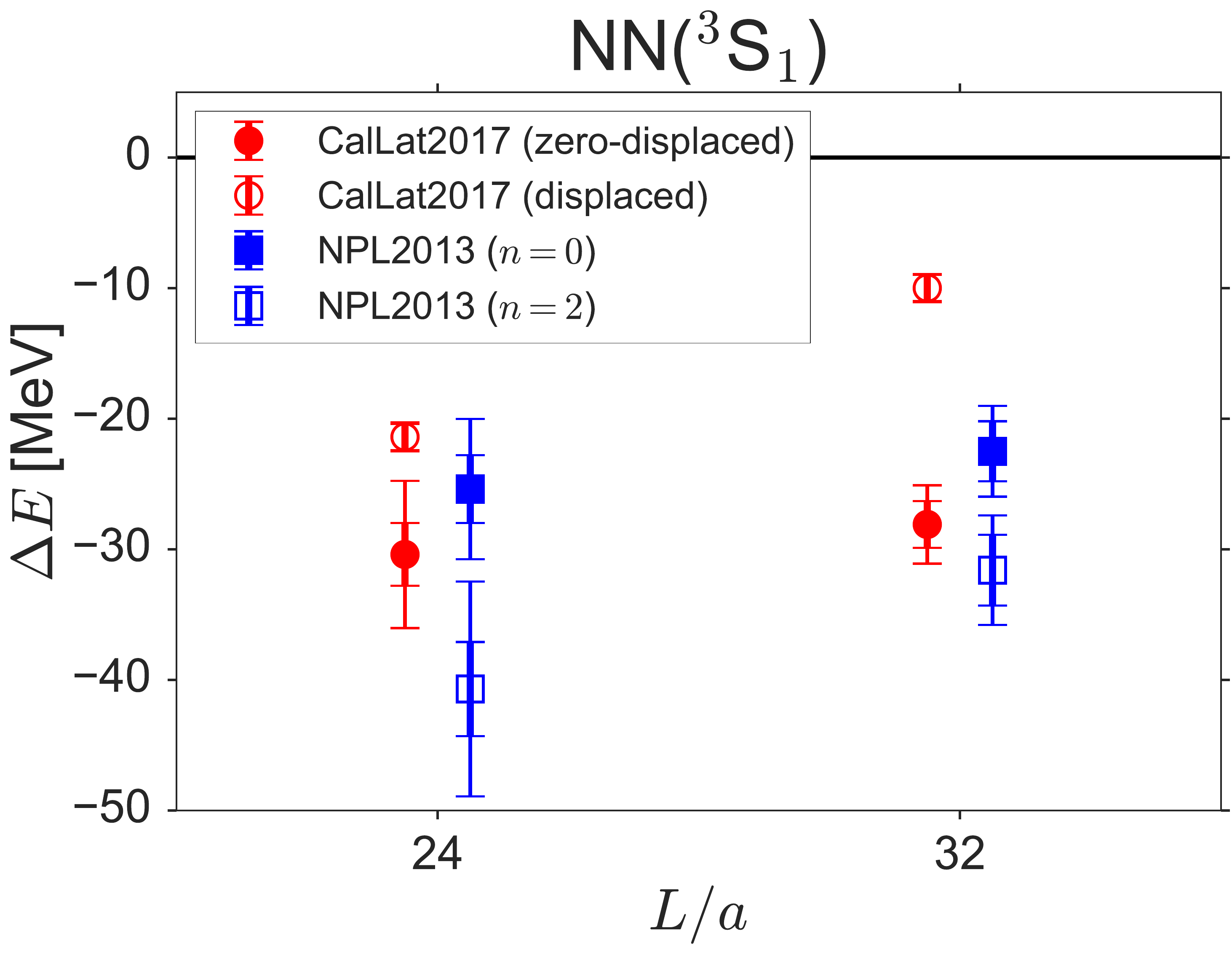}
 \caption{ 
The energy shifts  $\Delta E$
 on the $ L/a=24$ and $32$ lattices,
   in the $NN$ (${}^1S_0$) channel (Left) and the
   $NN$(${}^3S_1$) channel (Right),
   from CalLat2017 (red circles)
   and from NPL2013 (blue squares).
   Inner and outer error bars represent the statistical errors
   and statistical and systematic errors added in quadrature, respectively.
Red filled (open) circles for CalLat2017
   are obtained from zero (non-zero) displaced two-nucleon source operator in the center of mass system.
   Blue filled (open) squares for NPL2013 are obtained from zero displaced one
   in the center of mass system (the $n=2$ boosted system).
}
 \label{fig:Delta_E:CalLat_vs_NPL2013}
\end{figure}

Inconsistent plateaux in the direct method
are also observed in other  studies 
 claiming the existence of the $NN$ bound states.
In CalLat2017~\cite{Berkowitz:2015eaa}
and NPL2013~\cite{Beane:2013br,Beane:2012vq} papers,
$NN(^1S_0)$ and $NN(^3S_1)$ were studied 
with the 3-flavor degenerate quark masses
at $m_\pi=0.81$ GeV 
and $a \simeq 0.145$ fm.
The same gauge configurations with the spatial extension 
$L/a=24$ and $32$ are used among these studies.
They exclusively employ the smeared quark source 
\footnote{
  While both of CalLat2017 and NPL2013 employed the gaussian smearing,
  the detailed implementations are slightly different from each other:
  CalLat2017 employed the Coulomb gauge fixing while NPL2013 employed the gauge covariant smearing.
  Parameters for the gaussian are also different.
  We thank Dr.~A.~Walker-Loud for the information.
}
 to construct the single-nucleon operator.
It is then used to construct several types of two-nucleon source operators:
CalLat2017 studied both zero and non-zero displacements between two nucleons.
 NPL2013 used only  zero displacement between two nucleons, while 
 the center of mass is boosted with the momentum, $\vec{P} = (2\pi/L)\cdot \vec{n}$.

For the energy shift $\Delta E$ at each $L$,
 the results of CalLat2017 and NPL2013 must agree with each other within errors
 no matter what kind of displacement is taken or what kind of boost is given as long as the boost is not too large.  
The latter is due to the  fact that  the data at $n \equiv |\vec{n}| =0$  and $n =2$
are almost identical on these volumes according to
the finite volume formula~\cite{Luscher:1990ux,Rummukainen:1995vs}.\footnote{Their difference is less than 1.0\% (0.2\%) at $\Delta E \le -15$ MeV for $L/a=24$ $(32)$. }
  The actual lattice results, however, exhibit significant inconsistency as shown in 
 Fig.~\ref{fig:Delta_E:CalLat_vs_NPL2013}.
\footnote{In $NN(^1S_0)$ channel on $L/a=24$, 
datum corresponding to the non-zero displacement  was not given in CalLat2017.} 
This is another manifestation of the fake plateau (mirage) problem described in \cite{Iritani:2016jie}.

Note here that 
CalLat2017 interpreted two values of $\Delta E$ in their data
as the indication of the existence of two states with $\Delta E < 0$ by
speculating  that the source with zero (non-zero) displacement  couples dominantly to 
the deeper (shallower) bound state. 
However, such interpretation can be 
justified only after a sophisticated variational analysis ~\cite{Luscher:1990ck} is performed.
\footnote{Also, 
 the ERE used  by CalLat2017  for two states with $\Delta E < 0$ cannot be theoretically justified
 as will be discussed in the next section and 
 appendix~\ref{app:NPL2013-CalLat2017}.
}

The above observations cast
strong doubt on the existence of the $NN$ bound states claimed by using 
the direct method. Note that the method  has been
  abused in the previous literature without careful analysis of  a large systematic bias due to
   the excited state contamination as discussed in Ref.~\cite{Iritani:2016jie}.
For further inspection of  the results obtained by the direct method,
 we  introduce an alternative and simpler test (consistency check or sanity check) in this paper on the basis of 
 the L\"uscher's finite volume formula.
  The basic idea is to investigate the behaviors of the scattering 
  phase shifts in the region  of negative squared momentum $k^2 < 0$: 
   Consistency between the lattice data as a function of $k^2$ 
 and the  effective range  expansion (ERE) around $k^2=0$ 
 exposes the reliability or unreliability of the lattice data, the information  otherwise hidden 
  in the energy shift $\Delta E$.

In Sec.~\ref{sec:FVF}, we discuss  the theoretical basis behind our sanity check.
In Sec.~\ref{sec:data}, we summarize all the $NN$ data sets to be analyzed in this paper,
 together with tables of numerical data  in appendix~\ref{app:data-table}.
They are taken from the previous literature claiming the $NN$ bound states for heavy quarks.
 In Sec.~\ref{sec:diagnostic}, sanity checks of these $NN$ data are presented in detail.
 Sec.~\ref{sec:conclusion} is devoted to  conclusion and discussions.
In appendix.~\ref{app:demo-well}, we demonstrate typical behaviors of the phase shift  using analytic solutions for the square well potential. The phase shifts of NPL2013 and CalLat2017 will not be considered in the main text but given in appendix~\ref{app:NPL2013-CalLat2017}, as the mirage problems are already observed. 
Typical examples of the phase shifts with hyperons are presented in appendix~\ref{app:other-BB}.
 Data used in the paper are collected in appendix~\ref{app:data-table}.
 We note that a preliminary account of this study was given in Ref.~\cite{Aoki:2016dmo}.

\section{Finite volume formula}
\label{sec:FVF}

The L\"uscher's finite volume formula~\cite{Luscher:1990ux} 
(and the extensions thereof, e.g., for boosted systems~\cite{Rummukainen:1995vs}
and arbitrary spin/partial waves~\cite{Briceno:2013lba, Briceno:2014oea})
provides a  relation between the scattering phase
shifts and the energies on a  finite box.
If we focus on the elastic S-wave scattering of two baryons with identical mass $m$
in the center of mass system,
the scattering phase shift $\delta_0(k)$ is given by
\begin{equation}
  k \cot \delta_0 (k) = \frac{1}{\pi L} \sum_{\vec n\in \mathbf{Z}^3}\frac{1}{\vec n^2 -q^2},
  \qquad q=\frac{k L}{2\pi}, 
  \label{eq:kcot_delta}
\end{equation}
where $k$ is defined through $\Delta E = E_{BB}- 2m_B \equiv 2\sqrt{k^2+m_B^2} - 2m_B$   with
$E_{BB}$ being the energy of the two-baryon  state
measured in lattice QCD on a finite box with the spatial extension $L$. 
Only the discrete sets of points $(k^2, \kcotd)$ which satisfy 
the L\"uscher's finite volume formula are realized on  a given volume.
Vice versa, by measuring the energy of the two-particle system on a box,
the scattering phase shift at the corresponding energy can be extracted from lattice QCD.
If the interaction between two hadrons is attractive, we have $\Delta E < 0$ ($k^2 < 0$), so that
 Eq.~(\ref{eq:kcot_delta}) provides 
a way to make  analytic continuation of
$k\cot\delta_0(k)$ to the negative $k^2$ region.

\begin{figure}[h]
\centering
  \includegraphics[width=0.45\textwidth]{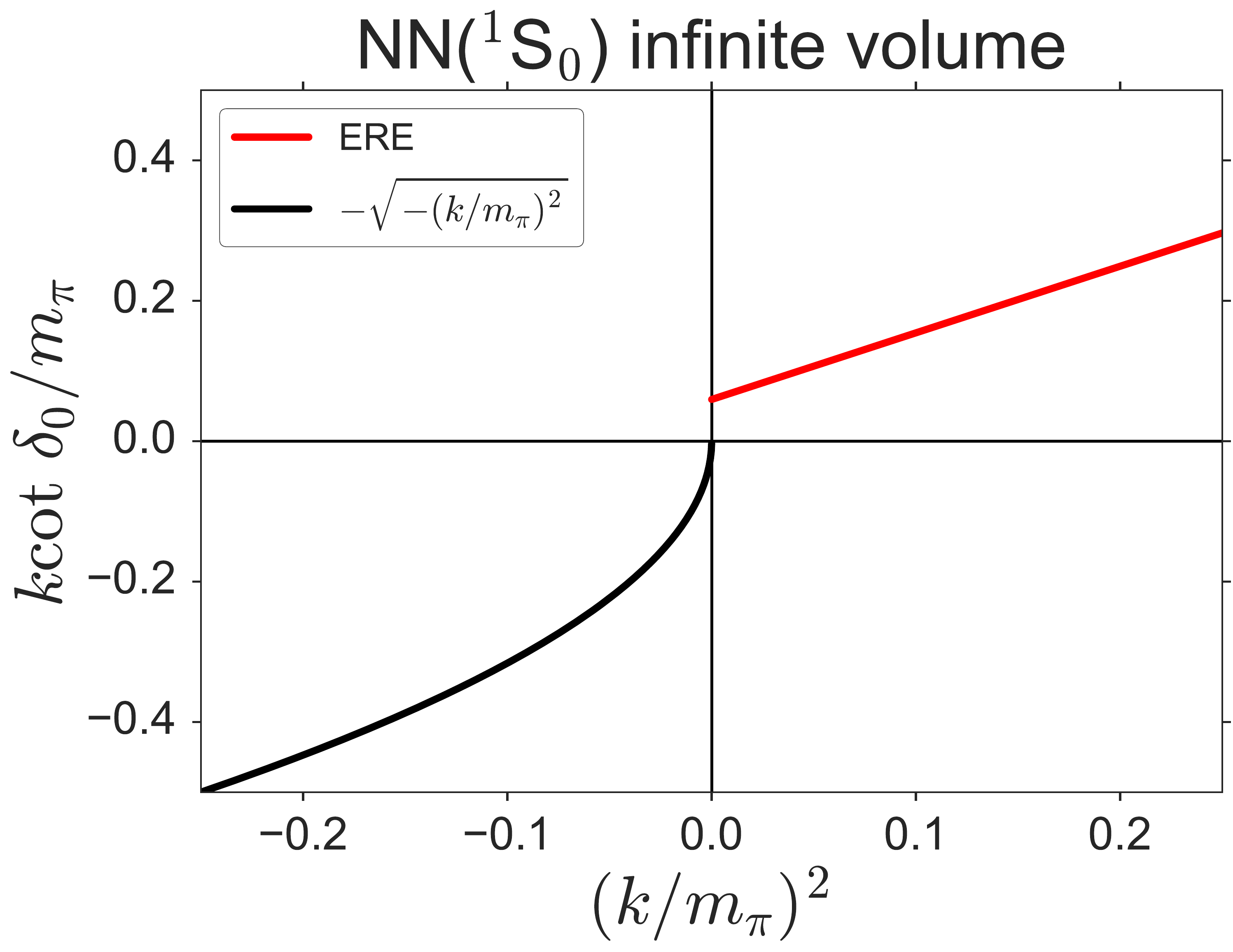}
  \includegraphics[width=0.45\textwidth]{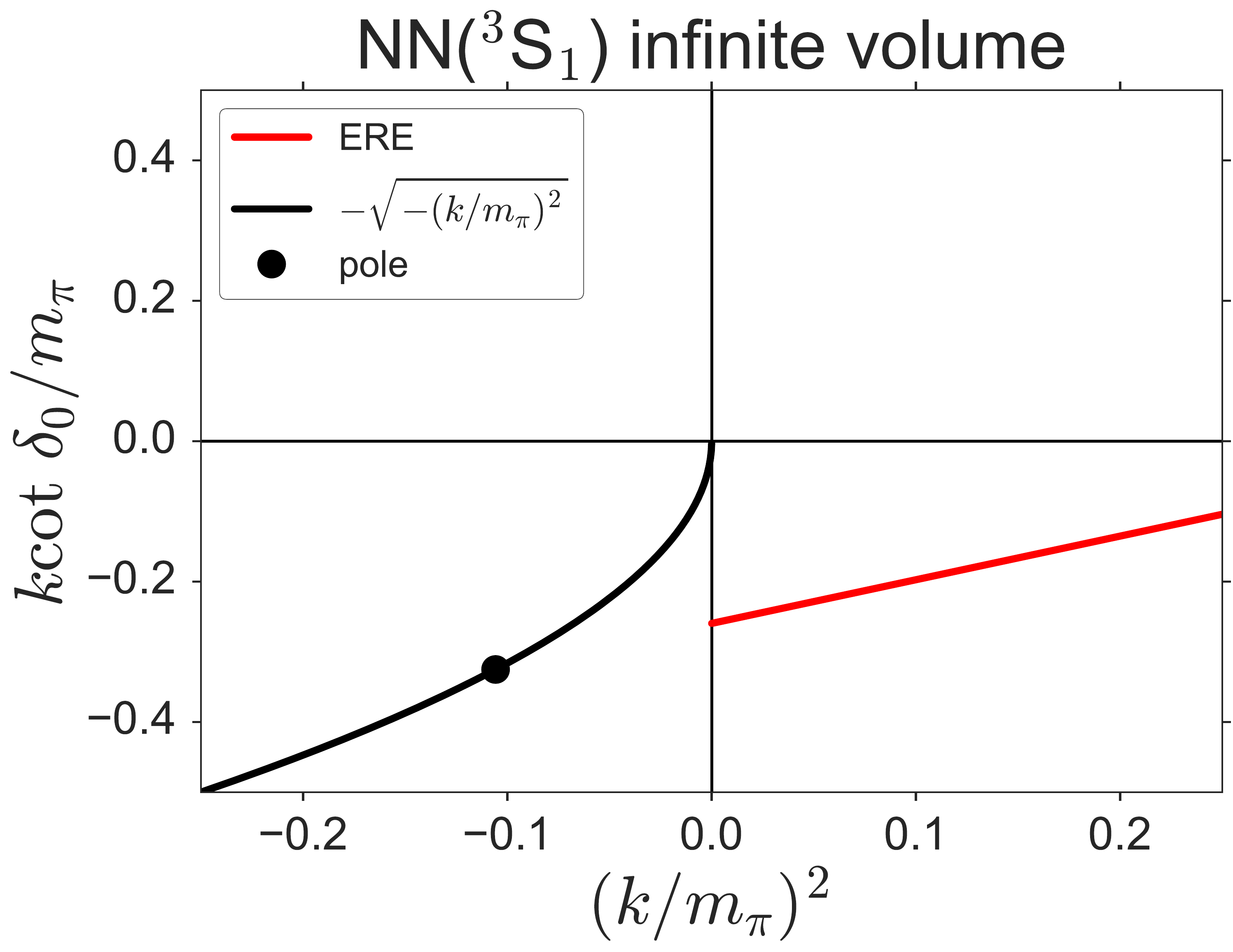}
  \includegraphics[width=0.45\textwidth]{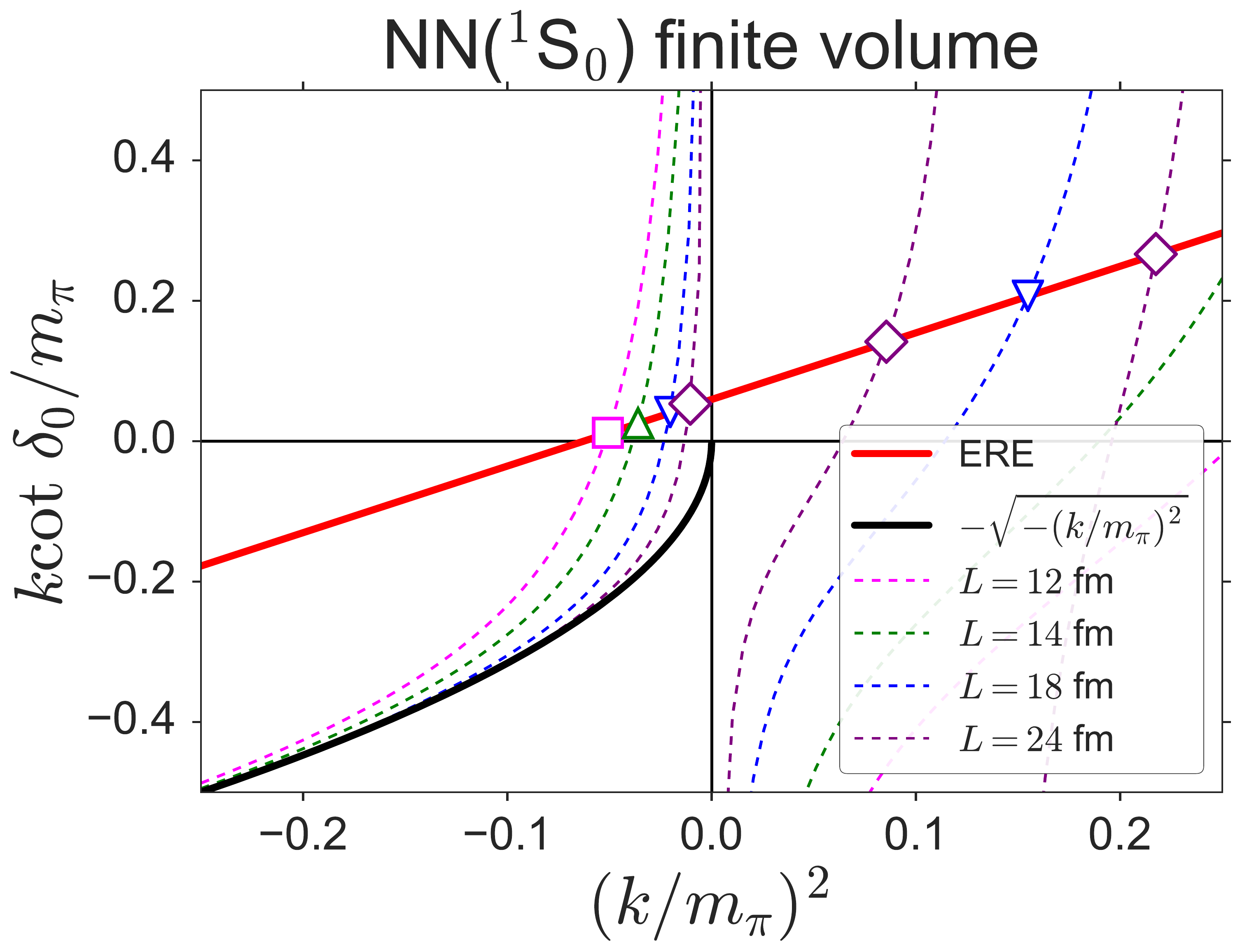}
  \includegraphics[width=0.45\textwidth]{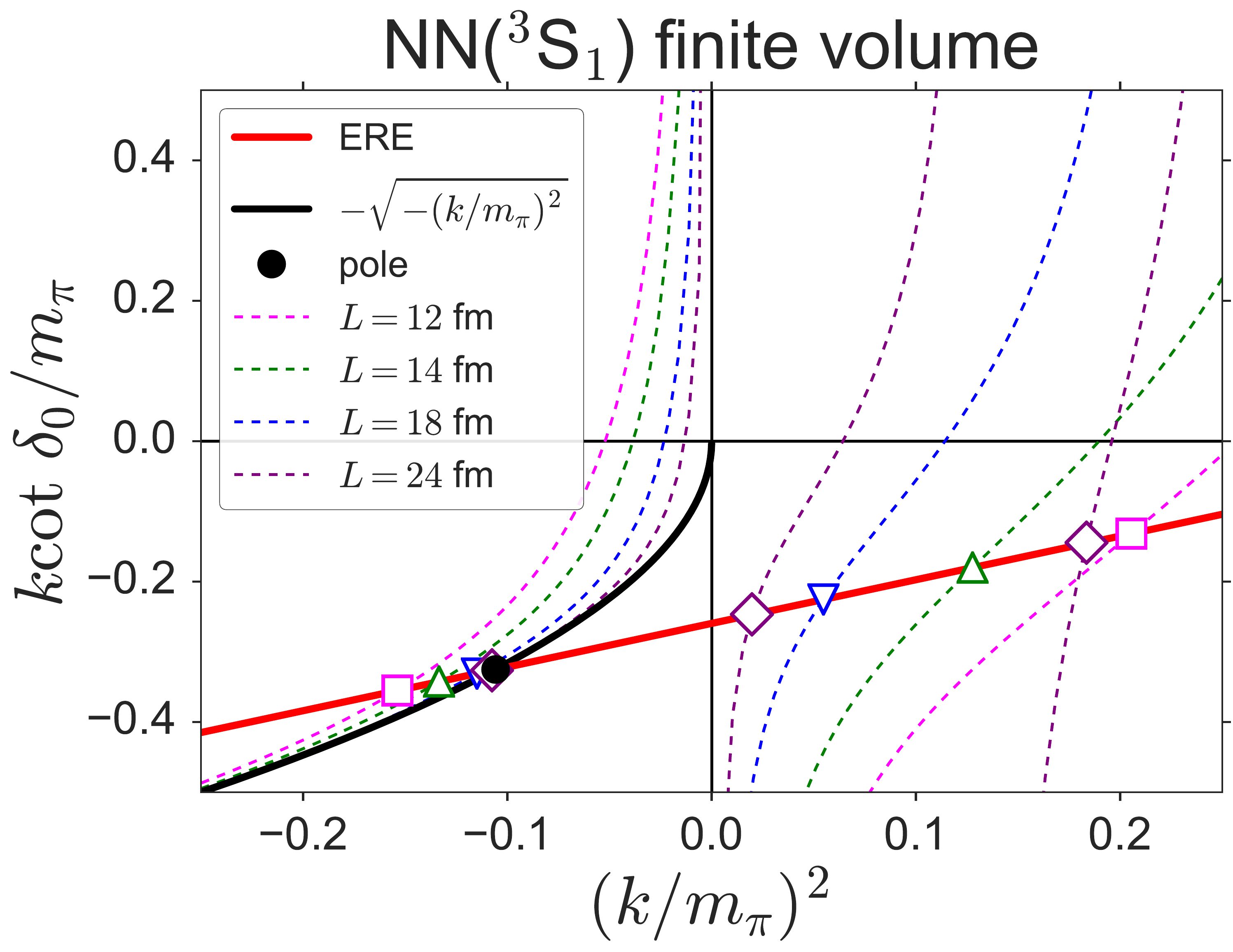}
  \caption{
    The relation between $\kcotd/m_\pi$ and $(k/m_\pi)^2$ in the infinite volume
    for $NN(^1S_0)$ (Upper Left) and $NN(^3S_1)$ (Upper Right) with $m_\pi = 0.14$ GeV.
    The red solid lines denote empirical ERE relations
    and the black solid lines are the condition for the bound state pole.
    In the upper right figure, the bound state is identified as the filled black point.
    The lower panels show
    the relation between $\kcotd/m_\pi$ and $(k/m_\pi)^2$ on finite volumes.
    The colored dashed lines represent the L\"uscher's formula
    for each finite volume $L$.
    Realized on each volume are the discrete points
      which satisfy both the L\"uscher's formula and the ERE relation,
      as shown by open squares, up/down triangles and diamonds
      for $L = 12, 14, 18, 24$ fm, respectively.
  }
 \label{fig:kcot}
\end{figure}

The relation between $k^2$ and $k\cot\delta_0(k)$ characterizes the
underlying baryon-baryon interaction at low energies,  which can be 
best seen  through the effective range expansion (ERE) around $k^2=0$;
\begin{equation}
  k\cot\delta_0 (k) = \frac{1}{a_0} + \frac{r_0}{2} k^2+ \sum_{n=2}^\infty P_0^{(n)} k^{2n} ,
  \label{eq:ERE}
\end{equation}
where $a_0$, $r_0$ and $P^{(n)}_0$ are the scattering parameters representing 
the scattering length, the effective range
and shape parameters, respectively.

In the upper panels of Fig.~\ref{fig:kcot},
we illustrate  the ERE  up to next-to-leading order (NLO)  by the red lines
 in which the empirical $NN$ scattering lengths ($a_0$) and effective ranges ($r_0$) are used.
Fig.~\ref{fig:kcot} (Upper Left)
corresponds to the ${NN}(^1S_0)$ case with $a_0 m_\pi = 16.8$ and $r_0 m_\pi = 1.9$
with no bound state in the infinite volume $(L=\infty$).
In Fig.~\ref{fig:kcot} (Upper Right), we show the ERE line 
corresponding  to the ${NN}(^3S_1)$ case with $a_0 m_\pi = -3.8$ and $r_0 m_\pi = 1.3$.
The bound state pole (deuteron) can be identified as  
the point where  $k \cot\delta_0(k) /m_{\pi} = -\sqrt{-(k/m_{\pi})^2}$ is satisfied (the 
filled black circle).

For finite volumes $(L < \infty )$, two-particle spectra are quantized, so that
only the  discrete values  satisfying the L\"uscher's formula   Eq.~(\ref{eq:kcot_delta})
are realized on the ERE line. 
They  are indicated by the open square, up/down triangle and diamond symbols 
in Fig.~\ref{fig:kcot} (Lower Left) and (Lower Right), where 
Eq.~(\ref{eq:kcot_delta})  is drawn by the dashed lines for 
different values of the lattice volume $L=$ 12, 14, 18, 24 fm. 
As the volume becomes larger, the state density increases for $k^2  \ge 0$ 
to form the continuous ERE line. On the other hand, for $k^2 <0$,   the discrete points 
constitute a sequence which  leads to an accumulation point corresponding to either
the $k^2=0$ scattering state at the threshold energy (Lower Left) or the bound state pole (Lower Right).

\begin{figure}[h]
\centering
  \includegraphics[width=0.45\textwidth]{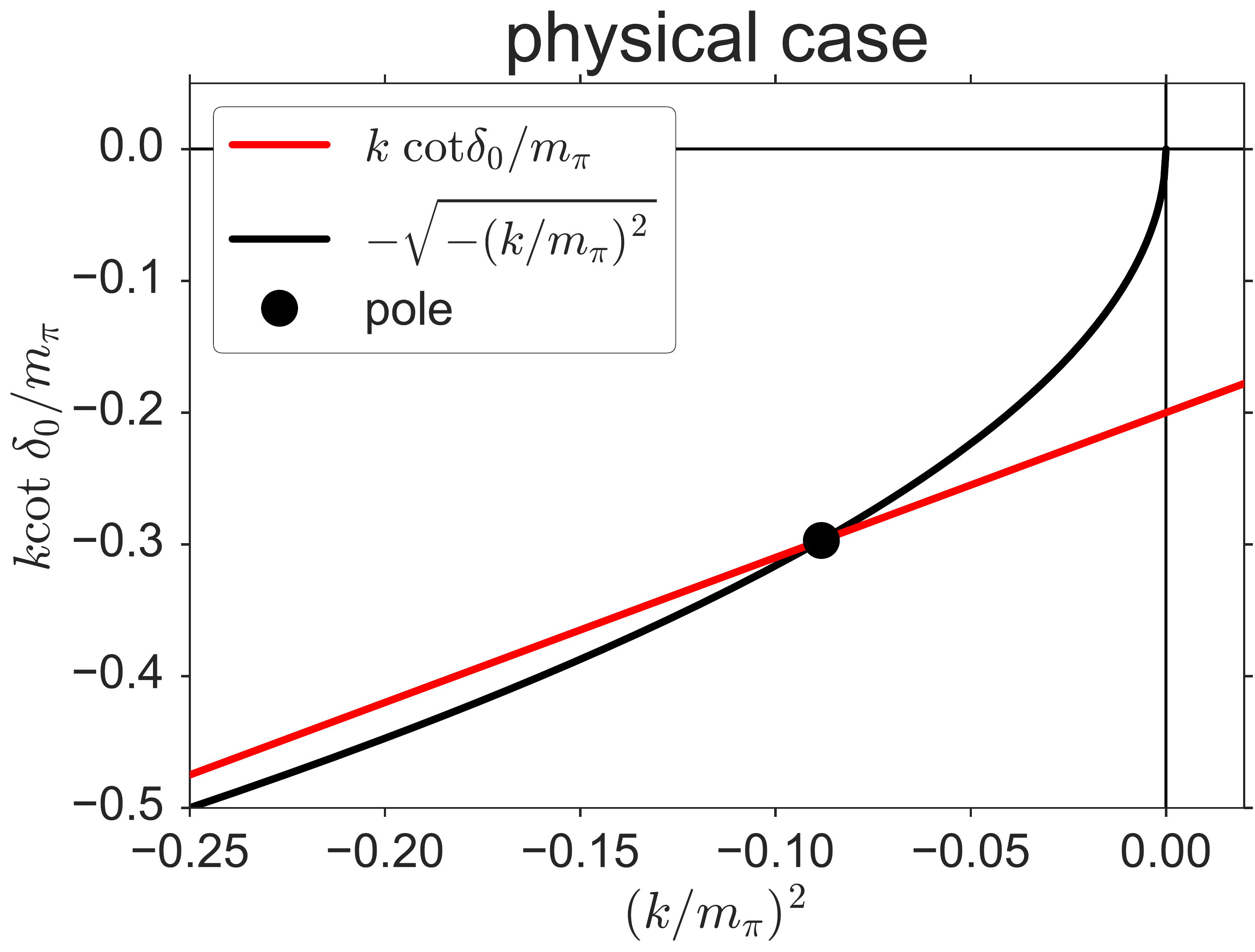}
  \includegraphics[width=0.45\textwidth]{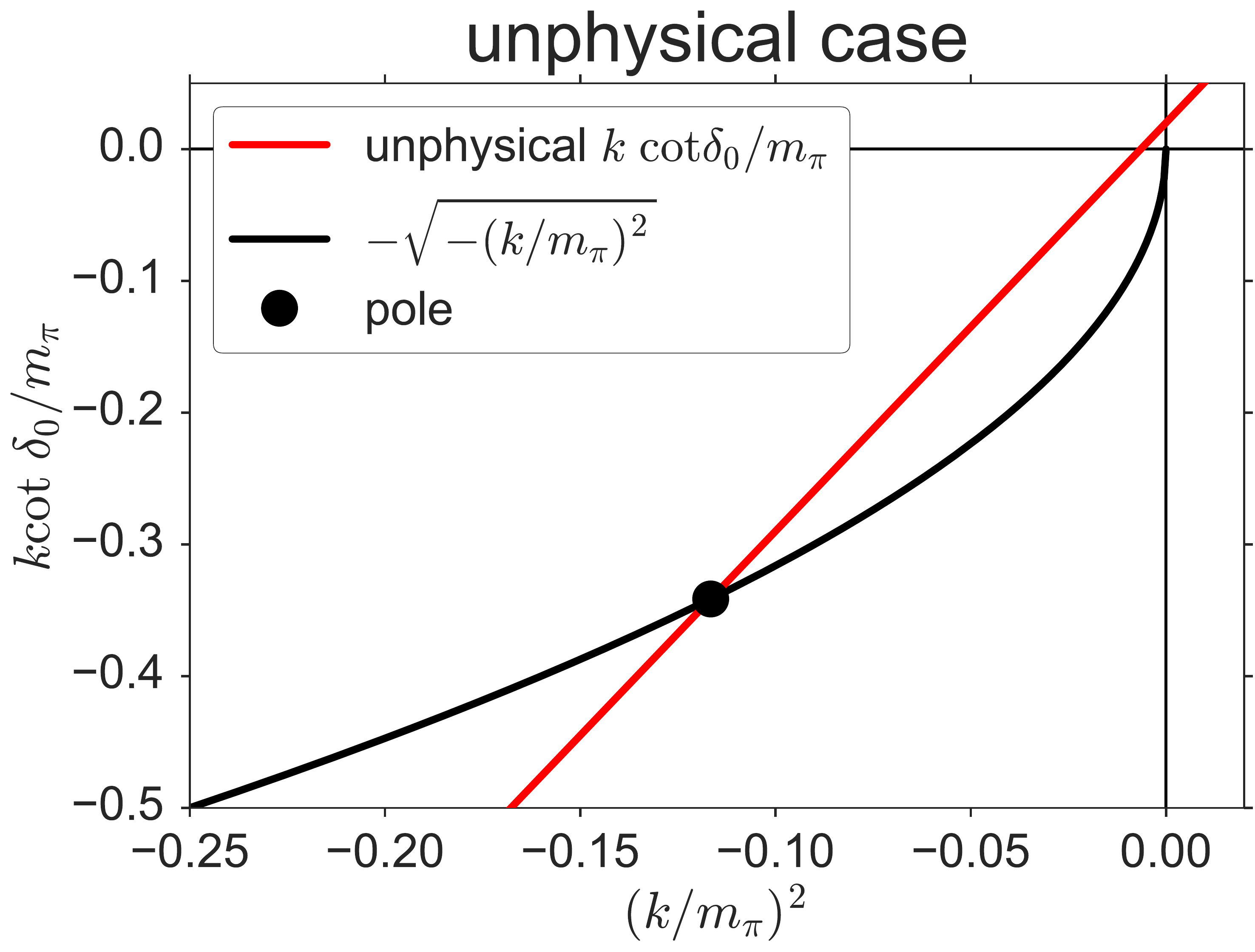}
 \caption{
Schematic illustration for the system with the pole satisfying
the physical condition (Left) or the pole having
the unphysical residue (Right).
The red solid lines denote $\kcotd$ obtained by fitting the lattice QCD data at finite volumes.
The bound state is identified as the crossing point of $\kcotd$ and 
 the bound state condition (black solid line), as 
indicated by the black solid circle.
In the left (right) figure, 
the behavior of $\kcotd$  near the bound state pole
is consistent (inconsistent) with the condition, Eq.~(\ref{eq:kcotd:bound}),
and thus the bound state pole has a physical (unphysical) residue in the S-matrix.}
 \label{fig:kcot_demo:1pole}
\end{figure}

It is in order here to discuss general properties 
of $(k^2, \kcotd)$ obtained from the analytic properties of the S-matrix for systems with bound state(s).  
Suppose we have a bound state at momentum, $k = i\kappa_b$ with $\kappa_b > 0$. Then 
the S-matrix, $S(k)=e^{2i\delta_0(k)}$, has the corresponding simple pole and simple zero 
at $k = i\kappa_b$ and $k = -i\kappa_b$, respectively.  
By using  the identity, 
\begin{eqnarray}
\kcotd = ik\cdot \frac{S(k) + 1}{S(k) - 1} ,
\label{eq:kcotd_vs_S}
\end{eqnarray}
one obtains the bound state condition, $\kcotd = -\sqrt{-k^2}$ at $k^2 = -\kappa_b^2$,
as mentioned above.
In addition,  the S-matrix near the  pole corresponding to the bound state is known to obey the formula~\cite{Sitenko},
\begin{eqnarray}
S(k \sim i\kappa_b) \simeq \frac{-i \beta_b^2}{k-i\kappa_b}, 
\label{eq:S-matrix:bound}
\end{eqnarray}
where $\beta_b^2$ is real and  positive for physical poles.
Consequently, the
S-matrix with a pure imaginary momentum near the bound state pole diverges as
\begin{eqnarray}
\left.  \lim_{\epsilon\rightarrow 0} S(k)\right\vert_{k= i(\kappa_b \pm \epsilon)} = \mp \lim_{\epsilon\rightarrow 0}\frac{\beta_b^2}{\epsilon} \rightarrow \mp \infty .
\label{eq:S-matrix:bound:lim}
\end{eqnarray}
Also, we have
\begin{eqnarray}
\frac{d}{dk^2} \left[ \kcotd - (-\sqrt{-k^2})\right]\Bigg|_{k^2= - \kappa_b^2} =  - \frac{1}{\beta_b^2} < 0,
\label{eq:kcotd:bound}
\end{eqnarray}
which implies that the slope of $k\cot\delta_0(k)$  as a function of $k^2$  must be smaller than that of $-\sqrt{-k^2}$ at the bound state pole.
We note here that the conditions~(\ref{eq:S-matrix:bound}), (\ref{eq:S-matrix:bound:lim}) and (\ref{eq:kcotd:bound}) 
hold as long as $\kappa_b^2$ is smaller than the possible lowest-lying left-hand singularity,
\footnote{These conditions may not be valid beyond the left-hand singularity \cite{Sitenko,Ma:1947zz,Goldberger-Watson,Bargmann}.}
while the ERE,  Eq.~(\ref{eq:ERE}),  is valid only for small $k^2$.

In Fig.~\ref{fig:kcot_demo:1pole} (Left),
we show an example 
for a system with one bound state 
which satisfies the condition~(\ref{eq:kcotd:bound}).  
(Here, for simplicity,  we assume that the binding energy is sufficiently small, so that the NLO ERE  is valid.)  This corresponds to the situation of the deuteron pole shown in Fig.~\ref{fig:kcot} (Right panels)
 except for the small S/D mixing.
In Fig.~\ref{fig:kcot_demo:1pole} (Right), we  show an unphysical case 
which does not satisfy the condition~(\ref{eq:kcotd:bound}). If the fit of the lattice data indicates such behavior,
it is a clear evidence that the data are not reliable. 

Let us now consider the case where there exist multiple bound states.  Then 
the conditions~(\ref{eq:S-matrix:bound:lim}) and (\ref{eq:kcotd:bound}) 
must be satisfied for each bound state.
This poses a further constraint on the behavior of $(k^2, \kcotd)$.
 To illustrate this, consider the system with two bound states 
at $k=i\kappa_{b_1}$ and $i\kappa_{b_2}$ with $\kappa_{b_1} > \kappa_{b_2} > 0$.
Then we have $S(k)\Big|_{k = i(\kappa_{b_1} - \epsilon)} = +\infty $ and 
$S(k)\Big|_{k = i(\kappa_{b_2}+\epsilon)} = -\infty$.  Since 
 $S(k)$ is real for pure imaginary $k$, (for it is defined 
 by the ratio of the Jost functions \cite{Sitenko,Goldberger-Watson}),
 there exists at least one $\kappa_c$ between $\kappa_{b_1}$ and $\kappa_{b_2}$
 which satisfies $S(k)\Big|_{k = i(\kappa_c \pm \epsilon)} = 1 \pm \epsilon $. 
 Combining this with the identity~(\ref{eq:kcotd_vs_S}),  we obtain,
\begin{eqnarray}
\kcotd\Big|_{k^2 = -(\kappa_c\pm \epsilon)^2} = \mp \infty ,
\end{eqnarray}
 i.e.
the  $k\cot\delta_0(k)$ must diverge at least once between two bound state poles. 
 The generalization of this to the case with more than two bound states is straightforward.

\begin{figure}[h]
\centering
  \includegraphics[width=0.45\textwidth]{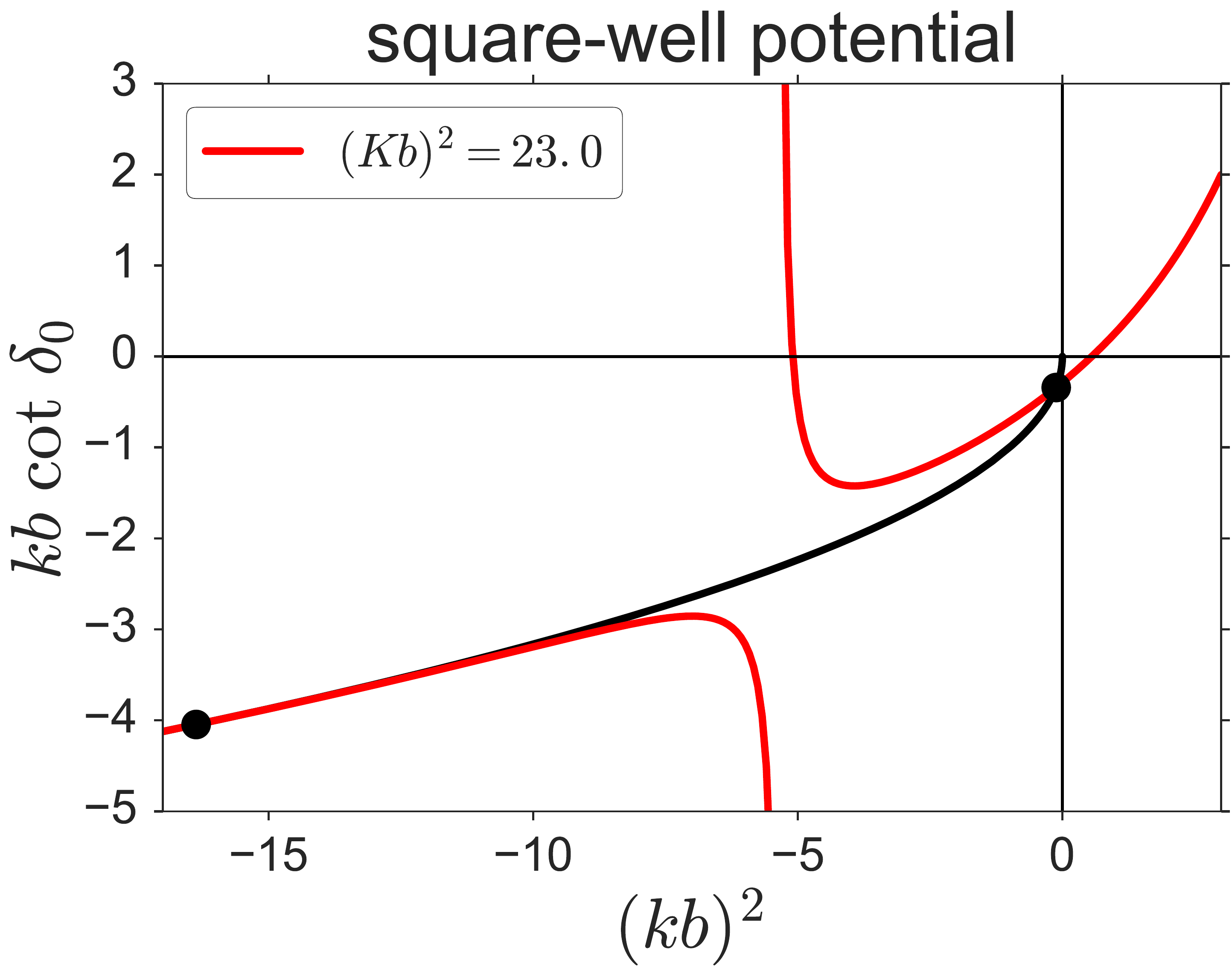}
  \includegraphics[width=0.47\textwidth]{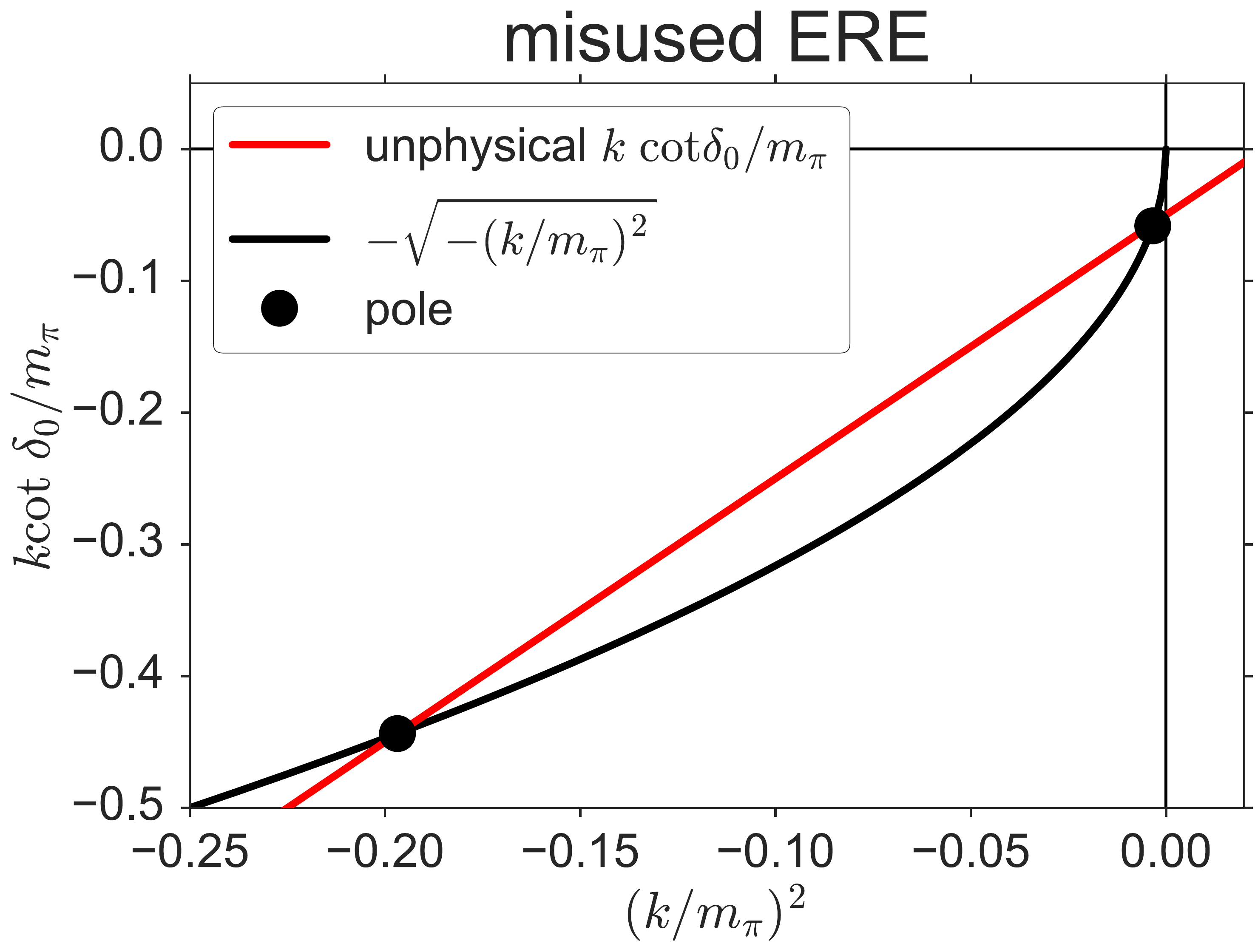}
 \caption{
(Left) The $\kcotd$  (the red line) for a 3-dimensional square-well potential.
The radius of the potential is denoted by $b$ and the potential depth is chosen so that
there exist two bound states. 
The black solid line is the condition for the bound state poles, which are denoted by
the black solid circles.   Note that $\kcotd$ diverges between two bound states.
(Right) Illustration of a misuse of  ERE beyond its convergence radius, where the left crossing point between
the red line and black line violates the condition~(\ref{eq:kcotd:bound}).
}
 \label{fig:kcot_demo:2pole}
\end{figure}

Shown in Fig.~\ref{fig:kcot_demo:2pole} (Left) is  a case with two bound states, taken from exactly solvable 
 3-dimensional square-well potential with the radius $b$ 
 \footnote{See appendix ~\ref{app:demo-well} for notations and analysis.}.
The deeply bound state at  $(k b)^2 \simeq -16.4$ and  the shallow bound state 
 at $ (k b)^2 =-0.1$ are denoted by the black solid circles, while the 
$\kcotd$ is plotted by the red solid  line.  One finds  the condition~(\ref{eq:kcotd:bound}) is 
satisfied for both bound states, so that they are indeed physical.
Note here that $\kcotd$ diverges between two bound states at $(k b)^2 \simeq -5.4$, so that ERE of $\kcotd$ 
around $k^2=0$ has clearly finite convergence radius.

Fig.~\ref{fig:kcot_demo:2pole} (Right) illustrates a case where ERE is erroneously  applied 
beyond the convergence radius. The unphysical crossing violating the condition~(\ref{eq:kcotd:bound}) at the 
 deeper pole indicates that the use of ERE is incorrect. 
 In the real lattice data, we do not know the black solid circles from the beginning.
 They are rather obtained as a result of the  fitting of the lattice QCD data which are all located above the black solid
 line for $k^2 < 0$.  If one finds that the naive ERE fitting of the lattice data shows the situation such as  Fig.~\ref{fig:kcot_demo:2pole} (Right) ,
 i.e. the unphysical crossing of the red line and the black line, one needs to try the proper fitting of $\kcotd$ without using ERE or 
  to doubt the original lattice data.

Having now established the general properties of  $\kcotd$ at $k^2<0$,
we present its novel applications  in lattice QCD assuming that there is at most 
one bound state whose binding energy is small enough within the convergence radius of 
 ERE around $k^2=0$, Eq.~(\ref{eq:ERE}).  Then one may 
extract the scattering parameters at $k^2=0$ such as the scattering length $a_0$ and the effective range $r_0$
 through the ERE fitting of the lattice data either at $k^2>0$ or  at $k^2<0$ (or both). 
Such an analysis for the data at $k^2 < 0$
with   the exact  L\"uscher's formula 
 has never been conducted in previous lattice studies  for two-baryon systems 
in the direct method~\cite{Yamazaki:2011nd,Yamazaki:2012hi,Yamazaki:2015asa,
Beane:2011iw,Beane:2012vq,Beane:2013br,Orginos:2015aya},
except for the one in Ref.~\cite{Berkowitz:2015eaa}.
(See also \cite{Dudek:2016wcf,Briceno:2016qmn,Mohler:2017ibi} for reviews with meson(s).)

Furthermore, the method  can be used to test the reliability of 
 lattice data, which we  call   a  ``sanity check'':  Self-inconsistent 
 and/or singular behaviors of 
  ERE lines around $k^2=0$ 
  and/or the unphysical behaviors as shown in
Figs.~\ref{fig:kcot_demo:1pole} (Right) and \ref{fig:kcot_demo:2pole} (Right)
indicate that the systematic errors of  the original $\Delta E$ on the lattice are substantially underestimated.
   A main source of the systematic errors is likely to be the 
 excited state contaminations, which generate fake plateaux in the  direct method, 
 as pointed out in~\cite{Iritani:2016jie} and  recapitulated in Sec.~\ref{sec:introduction}.
 A potential danger of  this fake plateaux applies to
 $NN$ data in Refs.~\cite{Yamazaki:2011nd,Yamazaki:2012hi,Yamazaki:2015asa,
Beane:2011iw,Beane:2012vq,Beane:2013br,Orginos:2015aya,Berkowitz:2015eaa}.
 In addition,  the general properties of $\kcotd$ at $k^2 < 0$ region tells us
the proper use of the ERE for claiming more than one bound state as we discussed above.
This applies to the data of Ref.~\cite{Berkowitz:2015eaa}.

In the next sections, we apply this sanity check to existing lattice data
which claim existences of bound states for two-baryon systems at heavy pion masses.

\section{Data sets}
\label{sec:data}

\begin{table}[tbh]
\centering
\begin{tabular}{|cr|c|c|c|c|c|c|c|}
\hline
Name & Ref. & $N_f$ & $a$ [fm] & $L$ [fm] & $m_\pi$ [GeV] & $m_N$ [GeV] & $m_\Lambda$ [GeV] & $m_\Xi$ [GeV] \\
\hline
 YKU2011 &\cite{Yamazaki:2011nd}  & 0 & 0.128   & 3.1, 4.1, 6.1, 12.3 & 0.80 & 1.62  & --- & --- \\
 YIKU2012&\cite{Yamazaki:2012hi}  & 2+1 & 0.090 & 2.9, 3.6, 4.3, 5.8 & 0.51 & 1.32 & --- & --- \\
 YIKU2015&\cite{Yamazaki:2015asa} & 2+1 & 0.090 & 4.3, 5.8 & 0.30 & 1.05 & --- & --- \\
 NPL2012 & \cite{Beane:2011iw}    & 2+1    & 0.123 (aniso.) & 2.9, 3.9 & 0.39 & 1.17 & 1.23 & 1.34 \\
 NPL2013 &\cite{Beane:2013br,Beane:2012vq}  & 3 & 0.145 & 3.5$^{(*)}$, 4.6$^{(*)}$, 7.0$^{(*)}$  & 0.81 &  1.64 & 1.64 & 1.64 \\ %
 NPL2015 &\cite{Orginos:2015aya} & 2+1 & 0.117 & 2.8, 3.7, 5.6 & 0.45 & 1.23 & 1.31 & 1.42  \\
 CalLat2017 & \cite{Berkowitz:2015eaa} & 3 & 0.145 & 3.5, 4.6 & 0.81 & 1.64 & 1.64 & 1.64 \\ 
 \hline
\end{tabular}
\caption{Summary of references for lattice data used in this paper. 
NPL2013 and CalLat2017  employed the same set of lattice configurations.
NPL2012 employed the anisotropic lattice with $a_s/a_t \simeq 3.5$ where $a_s
(\equiv a)$ and $a_t$ are spatial and temporal lattice spacings, respectively. \\
$^{(*)}$ We use the lattice spacing $a = 0.1453(16)$~fm given in NPL2013
   for $L$ in the present table.   
   }
\label{tab:references}
\end{table}

 Lattice data to be  checked 
are summarized in Table~\ref{tab:references}.  
Numerical results of $(k/m_\pi)^2$ and
$k\cot\delta_0(k)/m_\pi$  together with  $\Delta E$ are recapitulated in the tables in
appendix~\ref{app:data-table}:
Table~\ref{tab:data1} for
YKU2011~\cite{Yamazaki:2011nd},
Table~\ref{tab:data1b} for YIKU2012~\cite{Yamazaki:2012hi} and YIKU2015~\cite{Yamazaki:2015asa}, 
 Table~\ref{tab:data3} for NPL2012~\cite{Beane:2011iw}, 
 Tables~\ref{tab:data3a} and~\ref{tab:data4} for
NPL2013~\cite{Beane:2013br,Beane:2012vq}, 
 Table~\ref{tab:data5} for NPL2015~\cite{Orginos:2015aya},
 Table~\ref{tab:data6} for CalLat2017~\cite{Berkowitz:2015eaa}.
For YKU2011, NPL2013, NPL2015 and CalLat2017,  
data for excited  states are also given. 
(We tabulated only the data below the possible lowest-lying left-hand singularity,  $ |(k/m_\pi)^2| < 0.25$.)
Two nucleon source operators  with zero displacement under quark-source smearing
are  employed in all these literature. CalLat2017 used non-zero displacement additionally as mentioned in Sec.~\ref{sec:introduction}. 

Strictly speaking,
$^3S_1$ channel mixes with $l=2$ partial wave ($^3D_1$ channel) due to the presence of the tensor interaction.
In addition, each of $^1S_0$ and $^3S_1$ channels mixes with $l=4,6,\cdots$ partial waves due to the breaking of the rotational symmetry on a cubic  box. 
In the above references, however, binding energies of $NN(^1S_0)$ and $NN(^3S_1)$ are extracted
without explicitly taking into account these higher partial waves.
Correspondingly, 
 if the numerical values of  $\Delta E$, $(k/m_\pi)^2$ and  $k\cot\delta_0(k)/m_\pi$ are not explicitly given 
in the references in Table~\ref{tab:references}, we calculate them by using the 
L\"uscher's formula for S-wave, Eq.~(\ref{eq:kcot_delta}).%
\footnote{For this conversion,
  the statistical/systematic errors for
  pion and baryon masses are neglected since they are much smaller compared to other errors.}
  Both statistical and systematic errors evaluated in the original references  are taken into account in the tables in appendix~\ref{app:data-table}.
 The  systematic errors originating from the scale setting given
in NPL2012, NPL2013, NPL2015 are not considered, since we  analyze
only the dimensionless quantities normalized by  $m_\pi$ in this paper.

Although we  focus on  the $NN$ states in this paper,
 we also tabulate
$\Lambda\Lambda (^1S_0)$ and $\Xi\Xi (^1S_0)$ states (NPL2012),
and two octet-baryon states in 
 {\bf 1},  {\bf 8$_\mathrm{A}$} and {\bf 10} irreducible representations of flavor SU(3)  (NPL2013)
 in appendix~\ref{app:data-table}.

\section{Sanity check for each lattice data}
\label{sec:diagnostic}

\subsection{NPL2015}
\label{subsec:NPL2015}

\begin{figure}[h!]
  \centering
  \includegraphics[width=0.49\textwidth]{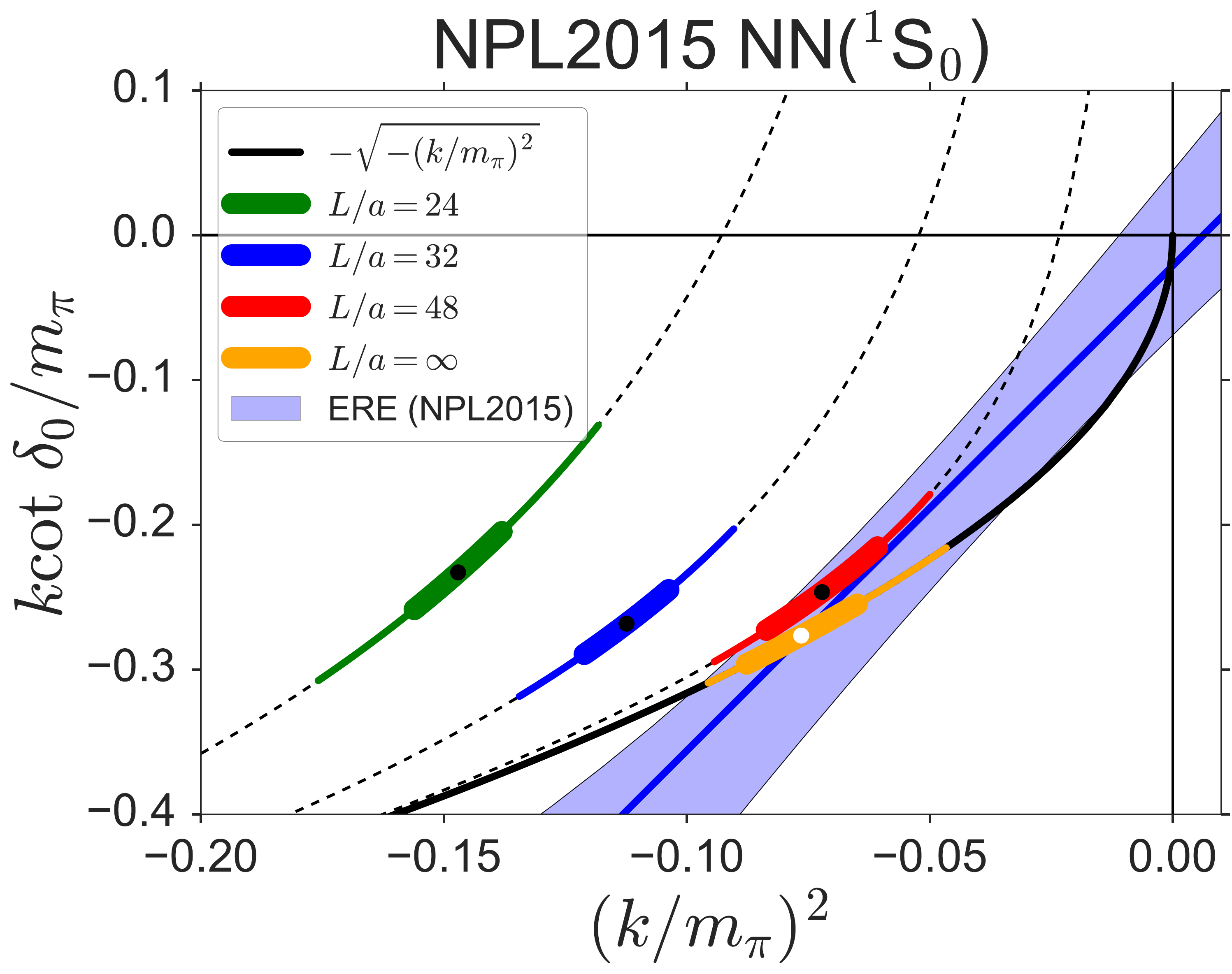}
  \includegraphics[width=0.49\textwidth]{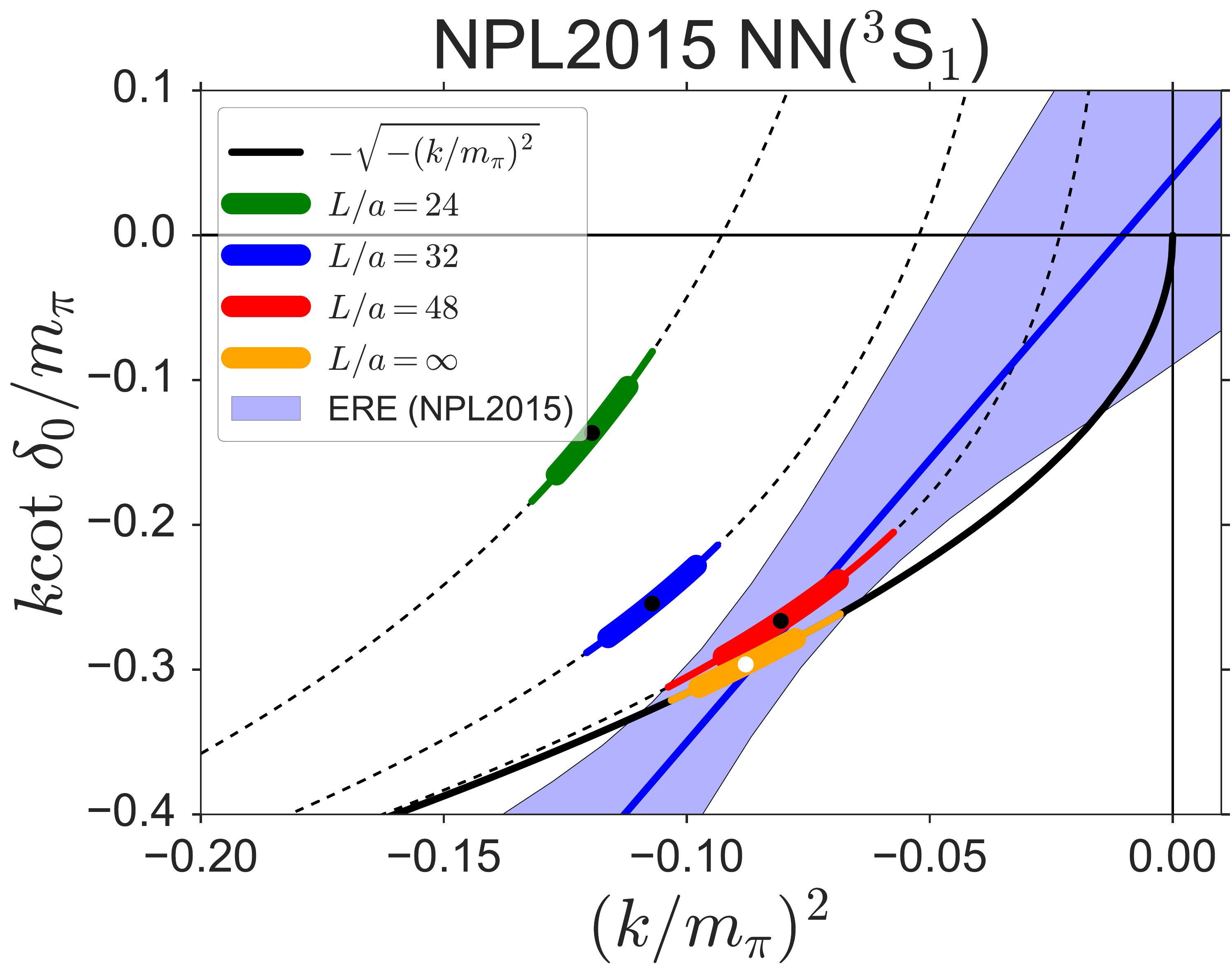}
  \includegraphics[width=0.49\textwidth]{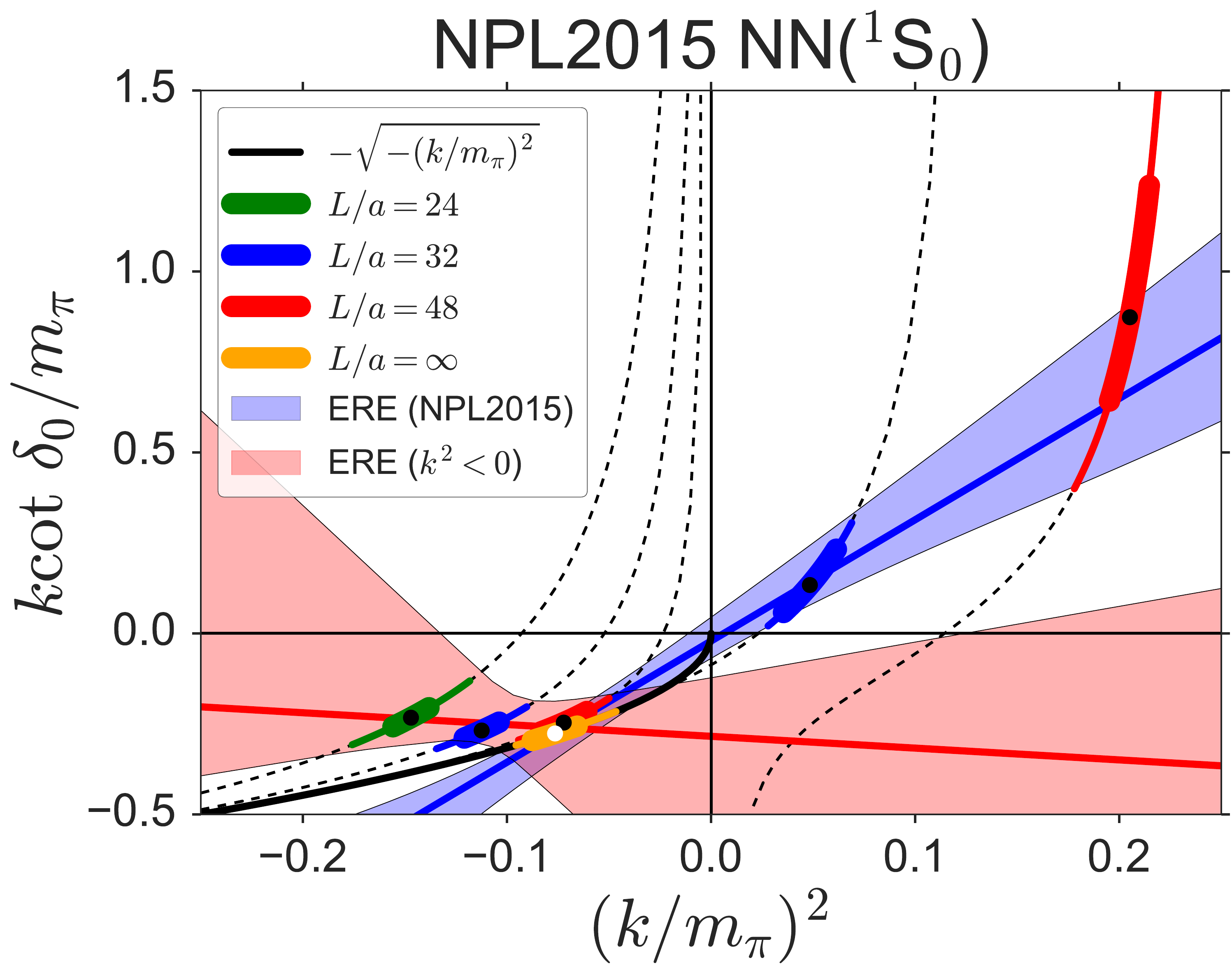}
  \includegraphics[width=0.49\textwidth]{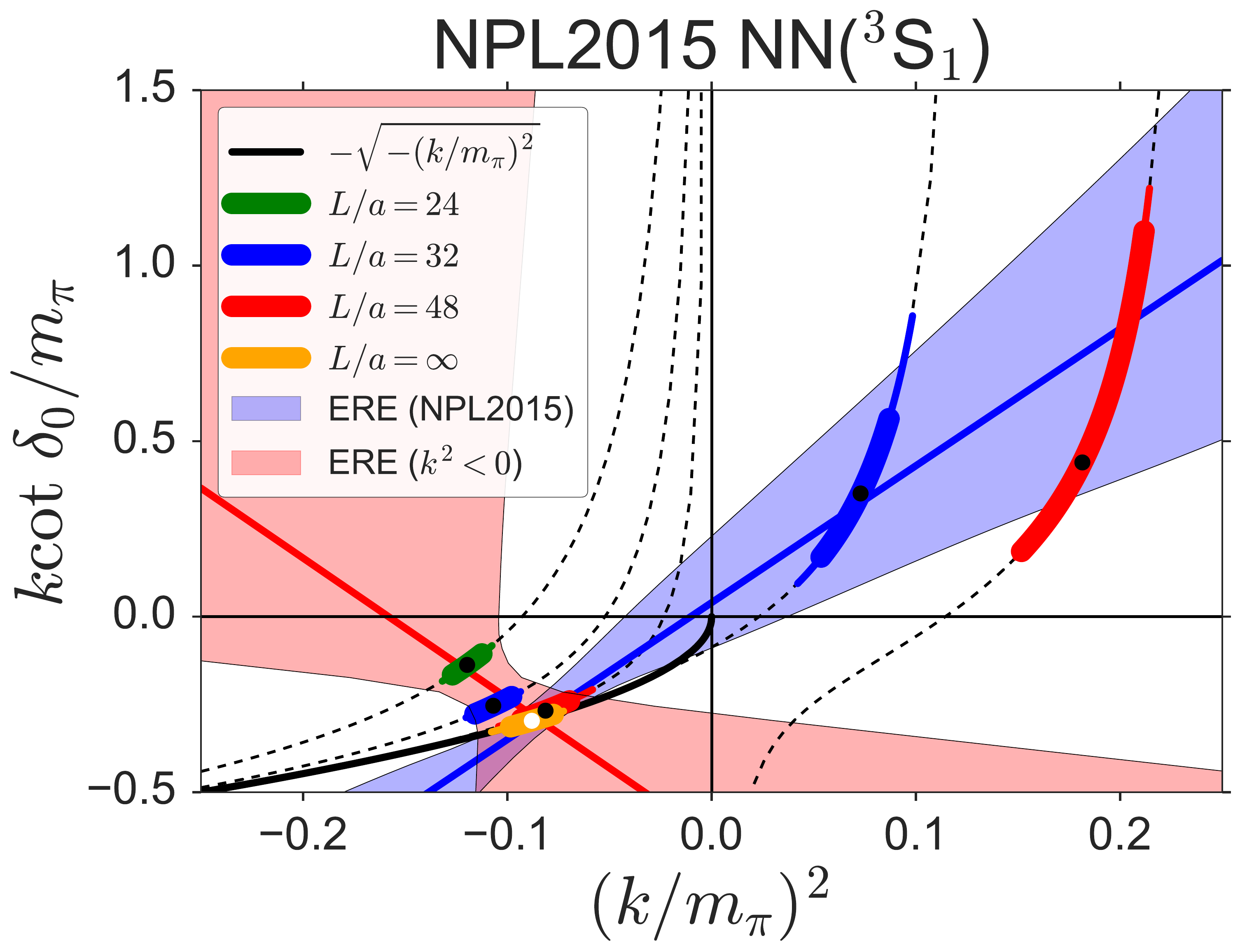}
  \caption{$k\cot\delta_0(k)/m_\pi$ as a function of $(k/m_\pi)^2$ for $NN$($^1$S$_0$) (Left) and
    $NN$($^3$S$_1$) (Right) of NPL2015.
    Black dashed lines correspond to L\"uscher's formula for each finite volume, while the black solid line
    represents the bound state condition that $ -\sqrt{-(k/m_\pi)^2}$.
    Upper panels show the data at $(k/m_\pi)^2 < 0$, while lower ones include the data at $(k/m_\pi)^2 > 0$.
    Light blue bands correspond to ERE with statistical and systematic errors added in quadrature
    given in NPL2015, obtained from data $k^2 >0$ and the binding energy 
    in the infinite volume; this is called \EREp in the text.
   Light red bands in lower panels correspond to the ERE 
 obtained by  using data at $k^2 < 0$ on finite volumes; this is called \EREn in the text.
   The red (blue) lines in the middle of the red (blue) bands correspond to the best-fits.
 }
  \label{fig:kcot_NPL2015}
\end{figure}

We first consider the data from NPL2015,
in which $NN(^1S_0)$ and $NN(^3S_1)$ were studied
in (2+1)-flavor QCD at $m_\pi=0.45$ GeV.
The data contain not only the ground states ($k^2 <0$) but also 
excited states ($k^2 >0$), 
and thus are particularly useful data set for the full sanity check.

Fig.~\ref{fig:kcot_NPL2015} shows $k\cot\delta_0(k)/m_\pi$ as a function
of $(k/m_\pi)^2$ for $NN$($^1$S$_0$) (Left) and $NN$($^3$S$_1$) (Right).
Upper panels focus on the data at $k^2 < 0$, while 
lower panels include data at $k^2 > 0$.
Black dashed lines in the
figures represent the behavior of Eq.~(\ref{eq:kcot_delta}) for each volume,
 and the black solid lines represent  $- \sqrt{-(k/m_\pi)^2}$. 
 The lattice data $k\cot\delta_0(k)/m_\pi$ on finite volumes are shown by the 
 solid circles together with statistical (systematic) errors denoted by the
  thick (thin) line segments.
  
NPL2015 claims  the existence of bound states in both channels indicated by the open circles,
where the binding energies
were obtained by the infinite volume extrapolation using the data at $k^2 < 0$
with the asymptotic expansion~\cite{Beane:2003da,Sasaki:2006jn} of the
L\"uscher's formula.
In NPL2015, ERE parameters up to NLO
were also determined using the data on the finite volume at $k^2 > 0$
 below the lowest-lying left-hand singularity
together with  the binding energy in the infinite volume (open circles).
We call this fit as \EREp. Corresponding EREs with 
statistical and systematic errors added in quadrature are shown by the 
light blue bands  in the figures.
 As clearly seen in upper panels of Fig.~\ref{fig:kcot_NPL2015},
for both channels, 
the \EREp determined in NPL2015 has  wrong intersection with
the bound state condition  in a same way as Fig.~\ref{fig:kcot_demo:1pole} (Right).
 
To further check the reliability of the data, we perform
 the ERE fit using the data only at $k^2 < 0$ on the finite volumes $(L/a=24, 32, 48)$,
which we refer to \EREn. The results are shown by the light red bands in lower panels of Fig.~\ref{fig:kcot_NPL2015}.
The two ERE bands (light red and light blue) in the figures   are clearly inconsistent with each other 
for both  channels.
Indeed, the scattering parameters obtained by \EREp and  \EREn do not agree with each other 
  in magnitude and/or sign
as summarized in Tab.~\ref{tab:ERE-NPL2015},
despite that
\EREp and \EREn should be consistent with each other
as shown in Fig.~\ref{fig:kcot} (Lower Right).
This observation casts a serious doubt on the reliability of the lattice data of NPL2015.

\begin{table}[t]
\centering
\begin{ruledtabular}
\begin{tabular}{cc|cc|cc} 
Name & Ref.                                            & \multicolumn{2}{c|}{$NN(^1S_0)$}                                            & \multicolumn{2}{c}{$NN(^3S_1)$}               \\ 
      &                                                & $(a_0 m_\pi)^{-1}$                   & $r_0 m_\pi$                           & $(a_0 m_\pi)^{-1}$                        & $r_0 m_\pi$   \\ \hline
\EREp & \cite{Orginos:2015aya}                         & $-0.021(^{+0.036}_{-0.028})(^{+0.063}_{-0.032})$  &  $6.7(^{+1.0}_{-0.8})(^{+2.0}_{-1.3})$   & $0.04(^{+0.10}_{-0.07})(^{+0.17}_{-0.08})$   & $7.8(^{+2.2}_{-1.5})(^{+3.5}_{-1.7})$   \\ 
\EREn & this work                                      & $-0.28(^{+0.06}_{-0.07})(^{+0.10}_{-0.23})$      & $-0.65(^{+1.05}_{-1.18})(^{+1.82}_{-4.71})$  & $-0.63(^{+0.18}_{-0.49})(^{+0.19}_{-2.02})$  & $-8.0(^{+3.4}_{-9.1})(^{+3.7}_{-37.5})$    \\ 
\end{tabular}
\end{ruledtabular}
\caption{
Summary of the scattering parameters obtained from NPL2015 data~\cite{Orginos:2015aya}.
\EREp is the ERE fit using   data at $k^2 > 0$ and the binding energy in the infinite volume.
\EREn is the ERE fit using  data at $k^2 < 0$ on finite volumes.
}
\label{tab:ERE-NPL2015}
\end{table}

What causes these inconsistencies?
The first possibility is that the volume is
too small for the finite volume formula (\ref{eq:kcot_delta})  applicable.
This is, however, unlikely by the fact that  $m_\pi L\ge 6.4$ in NPL2015.
The second possibility is that
the ERE up to NLO has large truncation errors.
However, this is also unlikely since  
the data under consideration are well 
 below the lowest-lying left-hand singularity at $\vert(k/m_\pi)^2\vert=0.25$.
The third and most plausible possibility is that the energy shifts
$\Delta E$ in NPL2015 are incorrect 
due to contaminations from excited states nearby.
Indeed, $\Delta E$ in NPL2015  are extracted from the data at $t \simeq 0.6-1.5$ fm,
 while fake plateaux due to contamination from the excited states  can 
 easily  appear at $t \simeq 1-2$ fm
as demonstrated in Ref.~\cite{Iritani:2016jie} and recapitulated in Sec.~\ref{sec:introduction}.

To summarize, the unphysical behavior of 
 \EREp as well as  the  inconsistency between \EREp and \EREn
  exposed by our sanity check indicate that 
  $\Delta E$ in NPL2015 is not reliable enough to
  claim the existence of $NN$ bound states at $m_{\pi}=0.45$ GeV.

\subsection{YKU2011}
\label{subsec:YKU2011}

Next we consider YKU2011,
in which $NN(^1S_0)$ and $NN(^3S_1)$ were studied
in quenched QCD at $m_\pi=0.80$ GeV.
As in the case of NPL2015, the data in YKU2011 contain both the ground states ($k^2 <0$) and excited states ($k^2 >0$)
and serve as the useful data set for the sanity check.

\begin{figure}[t]
  \centering
  \includegraphics[width=0.46\textwidth]{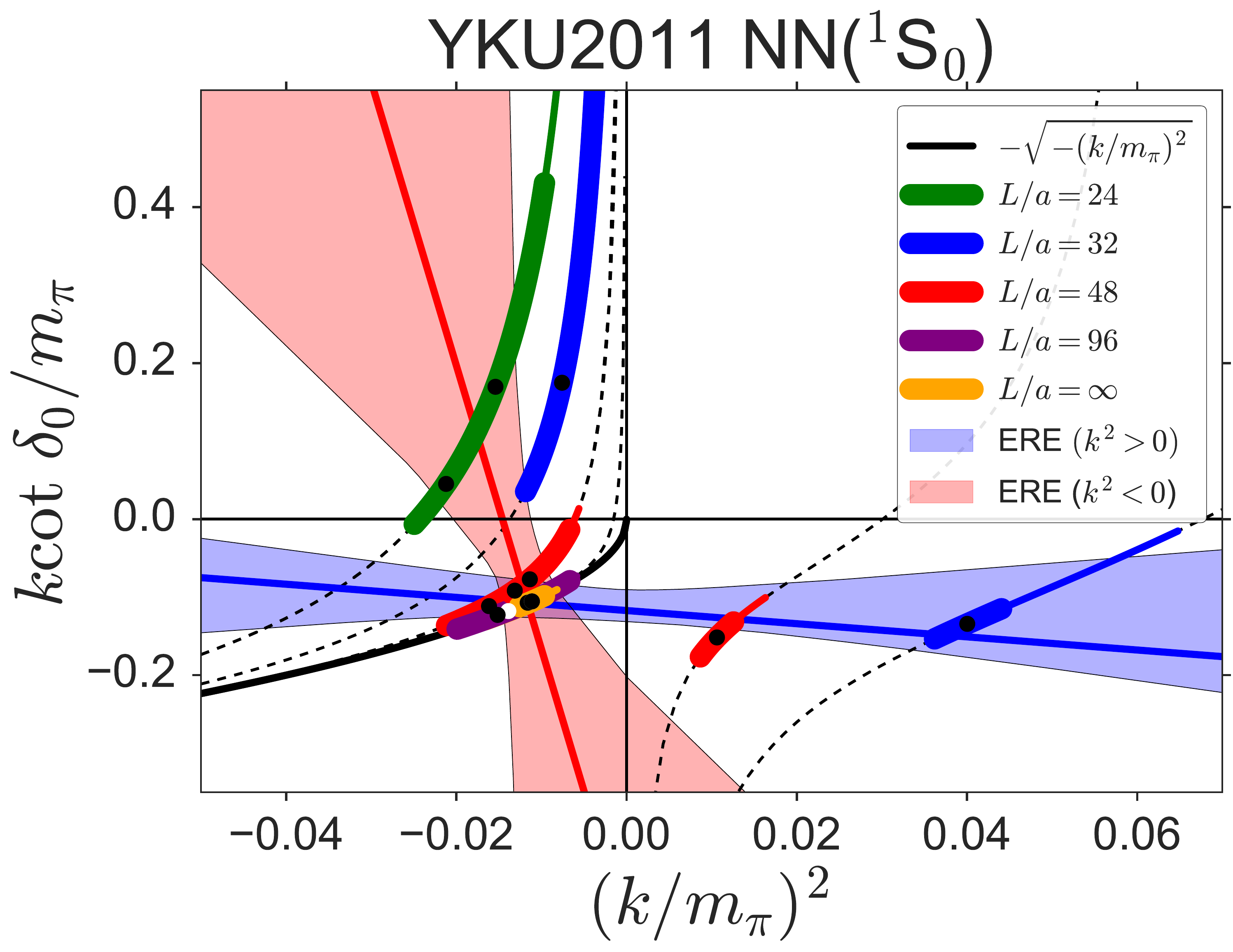}
  \includegraphics[width=0.46\textwidth]{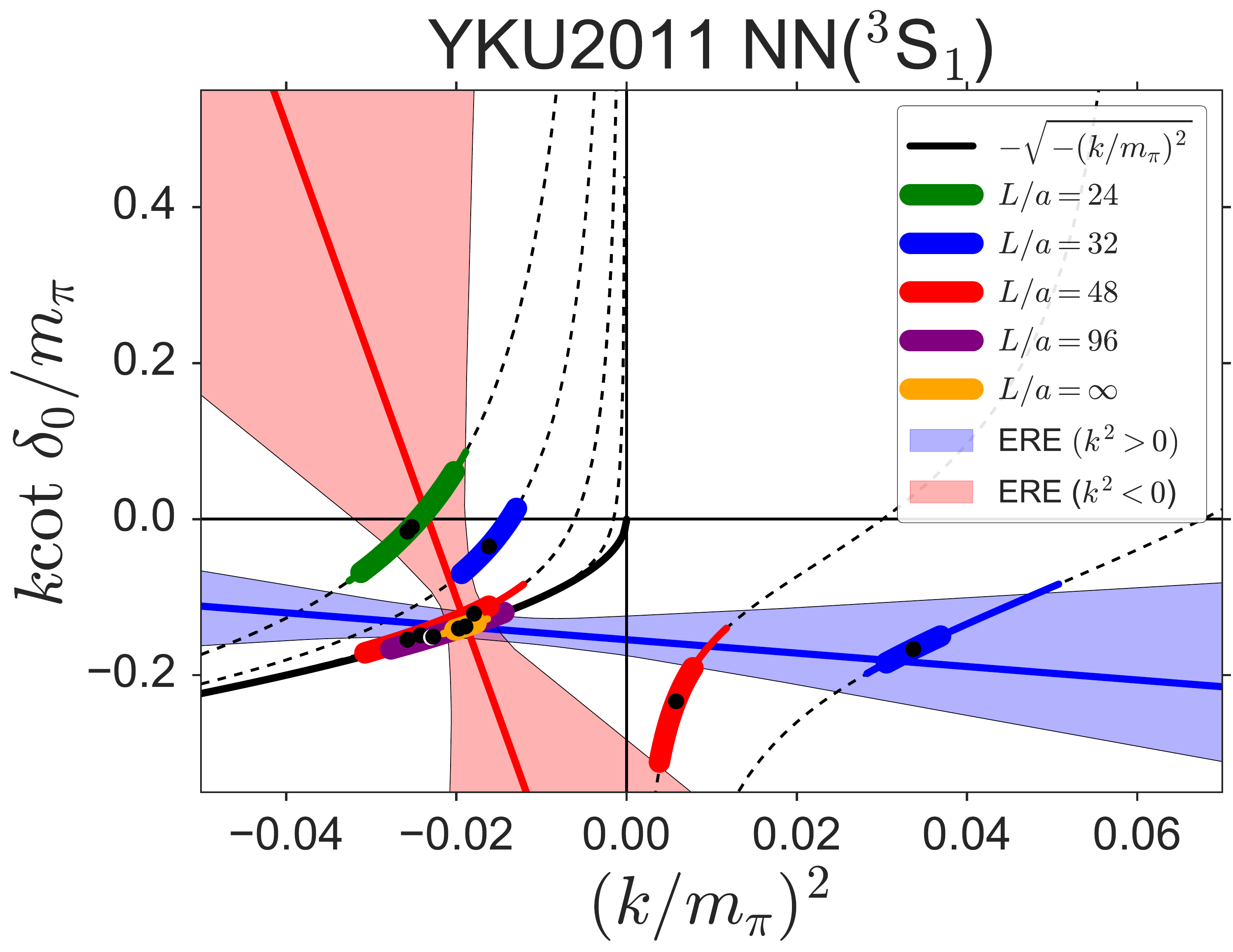}
  \caption{$k\cot\delta_0(k)/m_\pi$ as a function of $(k/m_\pi)^2$ for $NN$($^1$S$_0$) (Left) and
    $NN$($^3$S$_1$) (Right) of  YKU2011.
    Black dashed lines correspond to L\"uscher's formula for each  volume, while the black solid line
    represents $ -\sqrt{-(k/m_\pi)^2}$.
    EREs corresponding to \EREp and \EREn are shown by the light blue band 
  and light red band, respectively, with statistical and systematic errors added in quadrature.
  }
  \label{fig:kcot_YKU2011}
\end{figure}

Fig.~\ref{fig:kcot_YKU2011} shows $k\cot\delta_0(k)/m_\pi$ as a function of
$(k/m_\pi)^2$ for $NN$($^1$S$_0$) (Left) and $NN$($^3$S$_1$) (Right).
The existence of the bound state was claimed for both $NN(^1S_0)$ and $NN(^3S_1)$ 
by the infinite volume extrapolation from a subset of the data at
$k^2 < 0$ fitted with the asymptotic form ~\cite{Beane:2003da,Sasaki:2006jn} of the L\"uscher's formula.

The sanity check on YKU2011 immediately exposes a similar symptom as one observed in NPL2015:
The ERE behaviors are inconsistent between
those at $k^2 > 0$ and $k^2 < 0$ in both $NN(^1S_0)$ and $NN(^3S_1)$ channels.
In fact, $\Delta E$ for the ground states is found to be almost independent of the volume,
and thus data at $k^2 < 0$ align on a nearly  vertical line.
On the other hand, 
data at $k^2 > 0$ align on a nearly horizontal  line in the figure.

In order to quantify the inconsistency of  YKU2011 data,
we perform two different ERE analyses
in the same manner as  those performed for NPL2015 data
\footnote{
Correlations among data points are neglected in these fits.
For \EREn,
we perform an additional fit to a part of data
  ($L/a = 32$ and $48$ from the two-state analysis in Ref.~\cite{Yamazaki:2011nd}),
  which are manifestly uncorrelated.
  We confirm that obtained ERE are consistent with those given in Fig.~\ref{fig:kcot_YKU2011}
  in both $NN(^1S_0)$ and $NN(^3S_1)$ channels.
}.
In Fig.~\ref{fig:kcot_YKU2011},
the ERE lines for \EREp and \EREn are shown with light blue band and light red band, respectively.
Also in Tab.~\ref{tab:ERE-YKU2011},
the scattering parameters are summarized together with scattering lengths evaluated in YKU2011 paper.

\begin{table}[t]
\centering
\begin{ruledtabular}
\begin{tabular}{cc|cc|cc} 
Name & Ref.                                            & \multicolumn{2}{c|}{$NN(^1S_0)$}               & \multicolumn{2}{c}{$NN(^3S_1)$}                \\
      &                                                & $(a_0 m_\pi)^{-1}$             & $r_0 m_\pi$    & $(a_0 m_\pi)^{-1}$              & $r_0 m_\pi$   \\ \hline
      $a_0^{k^2 >0, L=32}$ & \cite{Yamazaki:2011nd}   &   $-0.137(^{+0.020}_{-0.027})(^{+0.118}_{-0.045})$ & 0 & $-0.164(^{+0.019}_{-0.025})(^{+0.077}_{-0.029})$ &    0          \\
        $a_0^{k^2 >0, L=48}$ & \cite{Yamazaki:2011nd}   & $-0.152(^{+0.020}_{-0.026})(^{+0.046}_{-0.001})$ &   0            &  $-0.235(^{+0.044}_{-0.069})(^{+0.082}_{-0.017})$        &    0          \\
        \EREp            & this work                           & $-0.12(^{+0.01}_{-0.01})(^{+0.02}_{-0.01})$  & $-1.69(^{+0.81}_{-0.97})(^{+2.20}_{-0})$  & $-0.15(^{+0.01}_{-0.01})(^{+0.02}_{-0.01})$  & $-1.72(^{+0.67}_{-0.89})(^{+2.00}_{-0.67})$    \\
\EREn            & this work                           & $-0.53(^{+0.25}_{-1.09})(^{+0.12}_{-0.40})$  & $-72.7(^{+39.4}_{-166.7})(^{+15.8}_{-52.6})$ & $-0.71(^{+0.32}_{-1.66})(^{+0.15}_{-1.47})$ & $-60.6(^{+32.8}_{-169.2})(^{+13.1}_{-144.1})$   \\
\end{tabular}
\end{ruledtabular}
\caption{
Same as Tab.~\ref{tab:ERE-NPL2015} but from YKU2011 data~\cite{Yamazaki:2011nd}.
YKU2011~\cite{Yamazaki:2011nd}  evaluated scattering lengths assuming  
$r_0 m_\pi = 0$.
}
\label{tab:ERE-YKU2011}
\end{table}

Inconsistency between \EREp and \EREn is apparent in both channels.
Quantitatively one  notices that the parameters in \EREn are very singular: 
 $r_0 m_\pi$ are one to two orders of magnitude larger (with negative signs) 
than their natural value, $r_0 m_\pi \sim {\cal O}(1)$. The  singular behavior is caused by the fact that
$\Delta E$ are almost independent of the volume, while
 claimed binding energies are shallow compared to the size of lattice volumes.
To the best of our knowledge, 
such singular ERE parameters together with the existence of one shallow bound state
are very difficult to be realized by any reasonable interactions.

As in the case of NPL2015,
the finite volume effect is unlikely to be the origin of the above inconsistency,
 since $m_\pi L\ge 12$ and also
 $\Delta E$ for $k^2 < 0$ is almost independent on $L$.
 The breakdown of the ERE is also unlikely, since $(k/m_\pi)^2$ for YKU2011 data are much smaller than $\vert (k/m_\pi)^2\vert =0.25$.
 Again, the most plausible explanation is that
$\Delta E$ in YKU2011 suffer serious excited state contaminations.

To summarize, the  inconsistency between \EREp and \EREn
  exposed by our sanity check indicates that 
  $\Delta E$ in YKU2011 is not reliable enough to
  claim the existence of $NN$ bound states at $m_{\pi}=0.80$ GeV.

\subsection{YIKU2012 and YIKU2015}

\begin{figure}[t]
  \centering
  \includegraphics[width=0.46\textwidth]{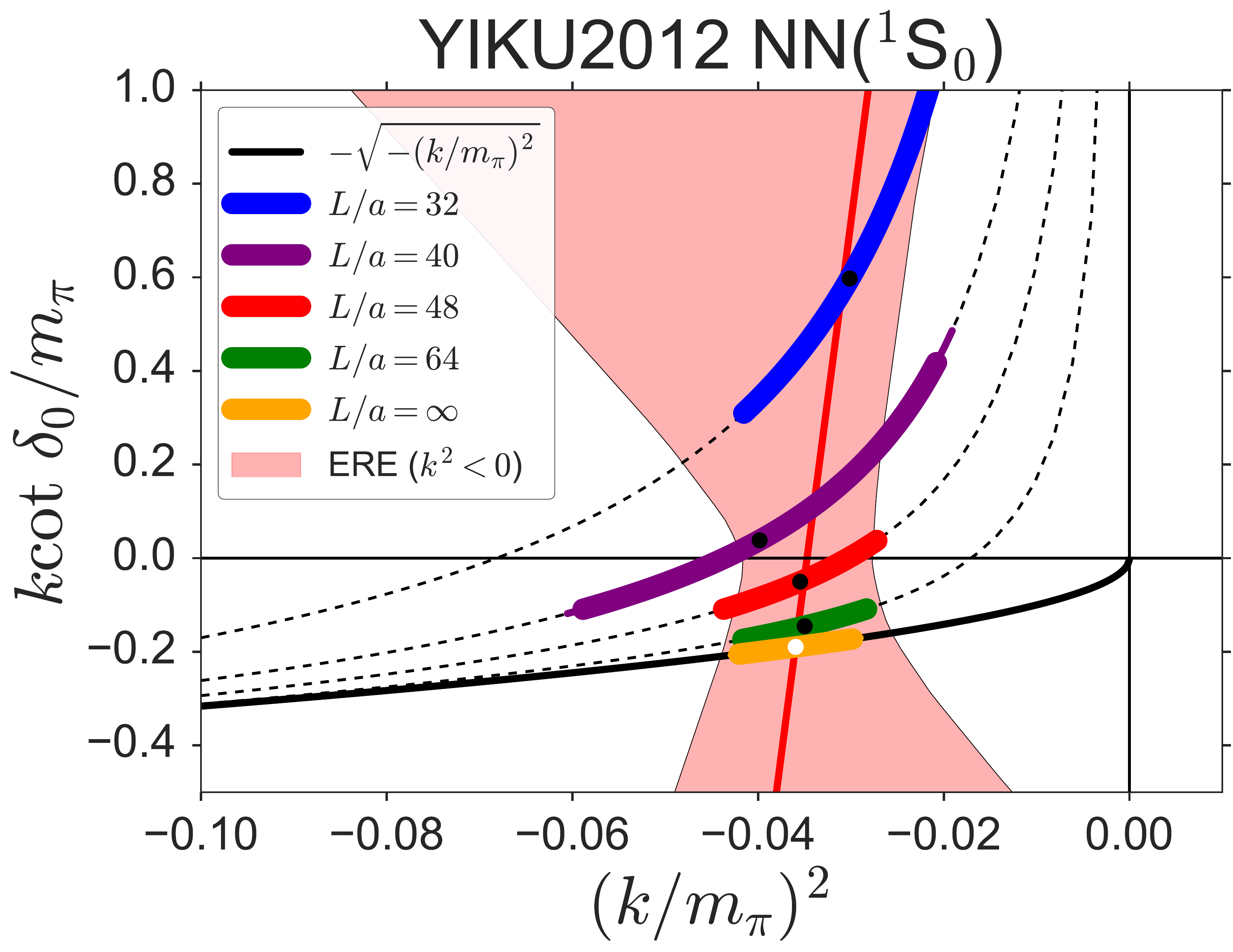}
  \includegraphics[width=0.46\textwidth]{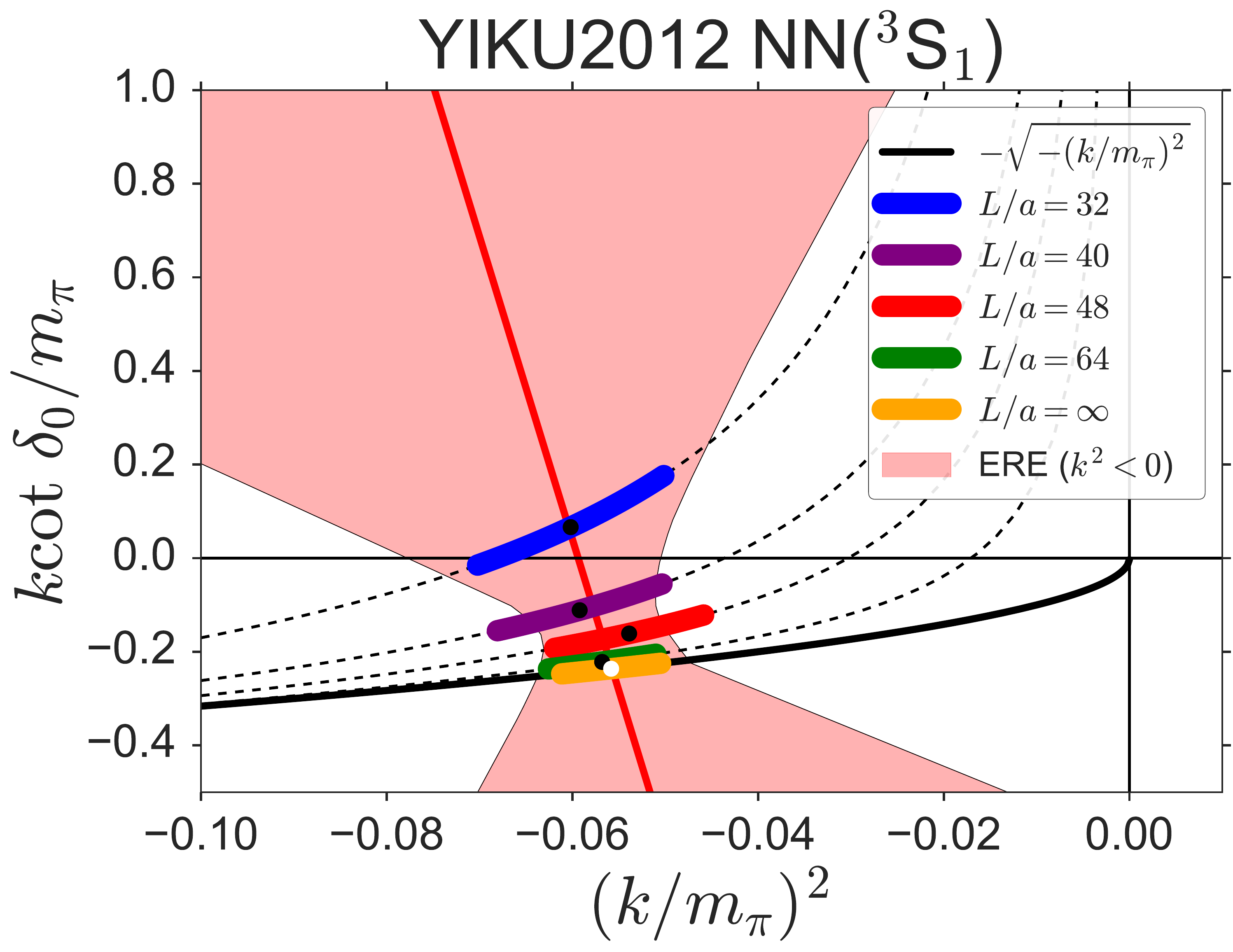}
  \caption{$k\cot\delta_0(k)/m_\pi$ as a function of $(k/m_\pi)^2$ for $NN$($^1$S$_0$) (Left) and
    $NN$($^3$S$_1$) (Right) for data on each volume from YIKU2012, together with
    YIKU2012's infinite volume extrapolation.
    Black dashed lines correspond to the L\"uscher's formula for each finite volume, while the black solid line
    represents $ -\sqrt{-(k/m_\pi)^2}$.
    NLO ERE fits to finite volume data are shown by red lines, together with 
    light red bands corresponding to statistical and systematic errors added in quadrature.
  }
  \label{fig:kcot_YKU2012}
\end{figure}

\begin{figure}[t]
  \centering
  \includegraphics[width=0.46\textwidth]{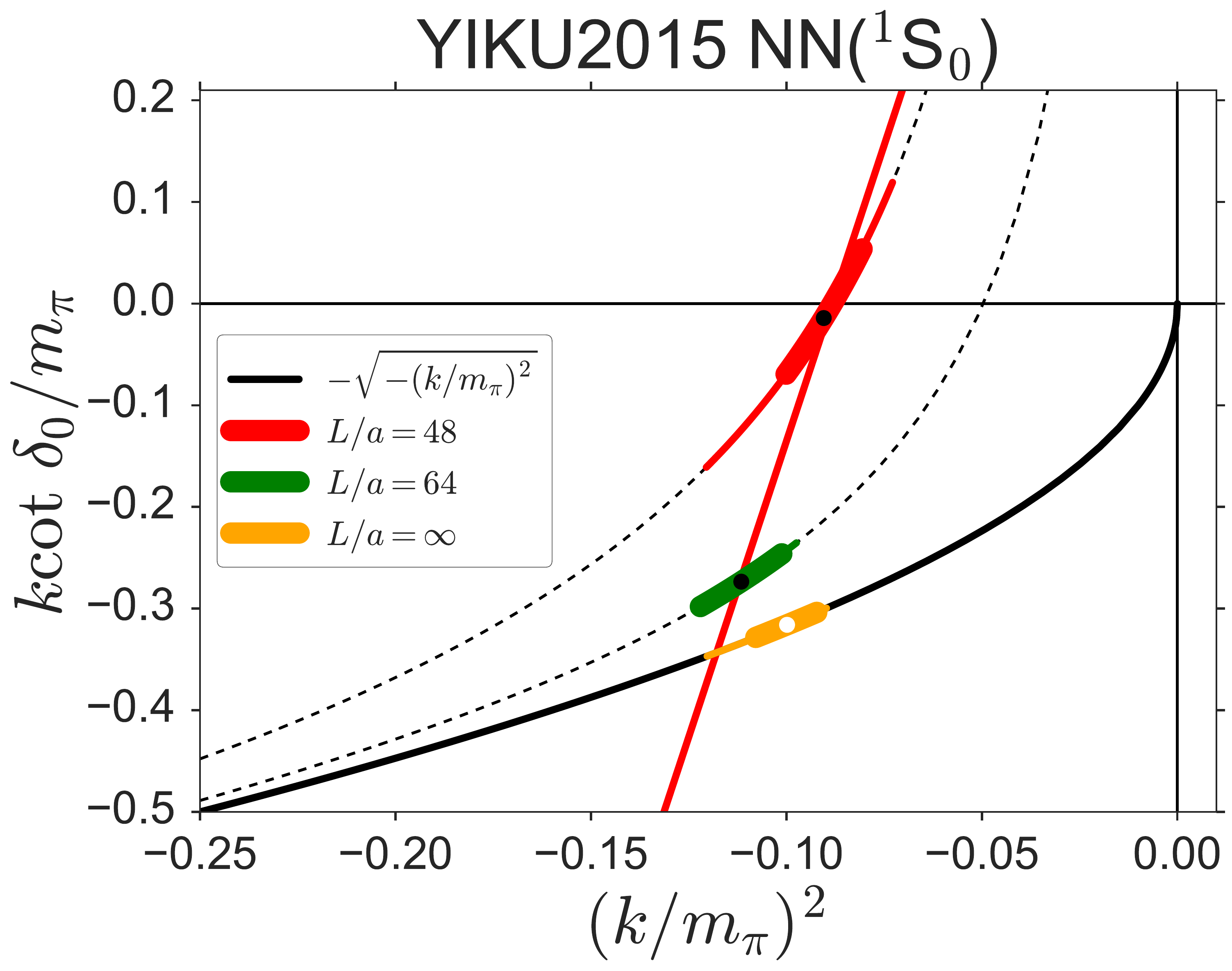}
  \includegraphics[width=0.46\textwidth]{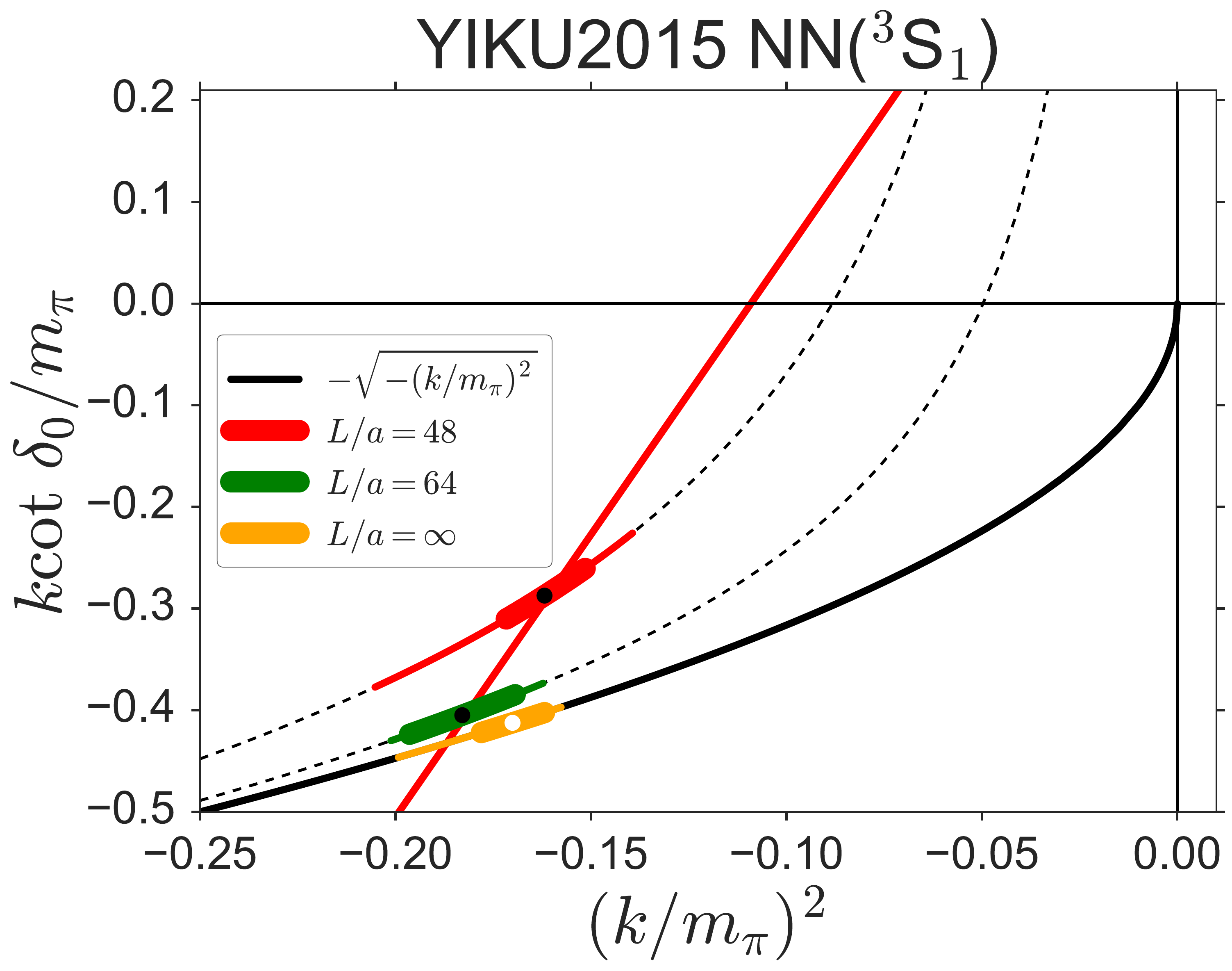}
  \caption{
    Same as Fig.~\ref{fig:kcot_YKU2012}, but from YIKU2015.
    Red lines correspond to NLO ERE fits.
  }
  \label{fig:kcot_YKU2015}
\end{figure}

The ground states for $NN(^1S_0)$ and $NN(^3S_1)$ were studied 
in (2+1)-flavor QCD at $m_\pi=0.51$ GeV (YIKU2012) and $m_\pi=0.30$ GeV (YIKU2015).
Since the excited states were not studied in these works,
we only consider the behavior of $k\cot\delta_0(k)$ for $k^2 < 0$.
Figs.~\ref{fig:kcot_YKU2012} and \ref{fig:kcot_YKU2015}
show $k\cot\delta_0(k)/m_\pi$ as a function of
$(k/m_\pi)^2$ for $NN$($^1$S$_0$) (Left) and $NN$($^3$S$_1$) (Right)
from YIKU2012 and YIKU2015, respectively.
The existence of the bound states in both channels
was claimed by the infinite volume extrapolation 
 with the asymptotic expansion of the L\"uscher's formula
(YIKU2012) or with the constant fit (YIKU2015).

As can be seen from these figures, data show singular behaviors
 in $^1S_0$ and  $^3S_1$ channels for both YIKU2012 and YIKU2015:
 Since $\Delta E$ is almost independent of the volume,
data align almost vertically.
Such behavior leads to very singular ERE parameters, i.e.  
divergent values of $r_0 m_\pi$ and sometimes of $(a_0 m_\pi)^{-1}$. 

We perform the NLO ERE fit to quantify the singular behavior 
 in terms of the scattering parameters.
In the case of YIKU2012, 
the results are plotted in Fig.~\ref{fig:kcot_YKU2012} by the red lines with the light red bands where statistical and systematic errors added in quadrature. 
Although total errors of the ERE fits are rather large, the central values show the singular behaviors:
$((a_0 m_\pi)^{-1}, r_0 m_\pi) = (5.27, 303.6)$  in $NN(^1S_0)$ channel
and 
$((a_0 m_\pi)^{-1}, r_0 m_\pi) = (-3.84, -129.3)$ in $NN(^3S_1)$ channel.
In addition, the red line in the $^1S_0$ channel violates Eq.~(\ref{eq:kcotd:bound}), which must be satisfied for the physical bound state.
 The fake plateaux problem of $\Delta E$ found in Ref.~\cite{Iritani:2016jie} certainly lead to these singular $\kcotd$.

In the case of YIKU2015,
there are only two finite volume data and thus 
degrees of freedom in the NLO ERE fit is zero.
We therefore obtain only the central values for ERE parameters,
$((a_0 m_\pi)^{-1}, r_0 m_\pi) = (1.0, 23.3)$  in $NN(^1S_0)$ channel
and 
$((a_0 m_\pi)^{-1}, r_0 m_\pi) = (0.61, 11.1)$ in $NN(^3S_1)$ channel,
where corresponding ERE lines are plotted in Fig.~\ref{fig:kcot_YKU2015} by red lines.
In both channels, 
the violations of the physical condition Eq.~(\ref{eq:kcotd:bound}) for the intersections
and/or
the singular ERE behaviors
are observed.

Since $\vert(k/m_\pi)^2\vert$ for these data are smaller than 0.25,
singular  ERE behaviors
are very difficult to be realized by any reasonable interactions.
We therefore conclude that the values of $\Delta E$ in YIKU2012 and YIKU2015 are unreliable,
most probably due to the excited state contaminations.

\subsection{NPL2012}
\label{sec:NPL2012}

\begin{figure}[t]
  \centering
  \includegraphics[width=0.49\textwidth]{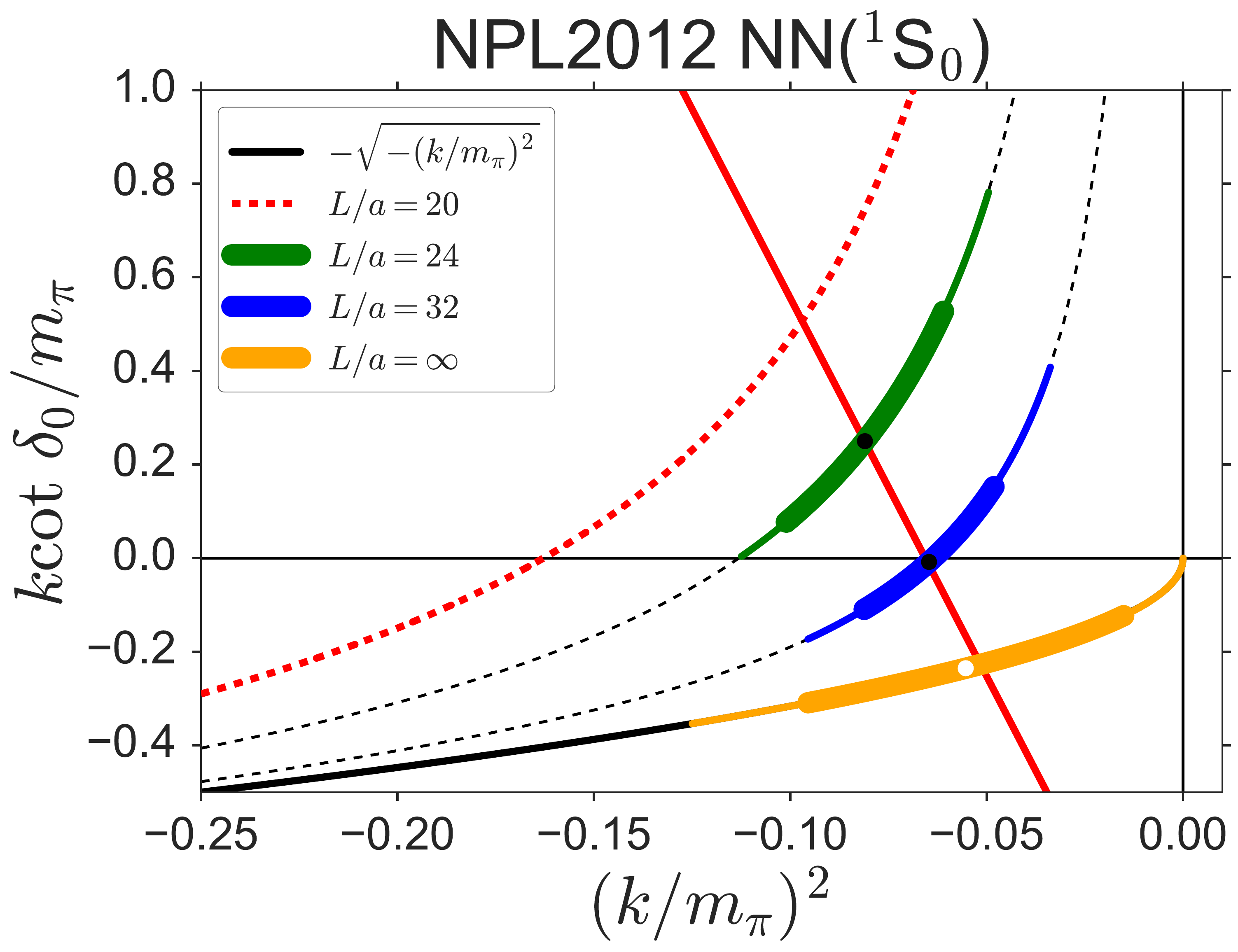}
  \includegraphics[width=0.49\textwidth]{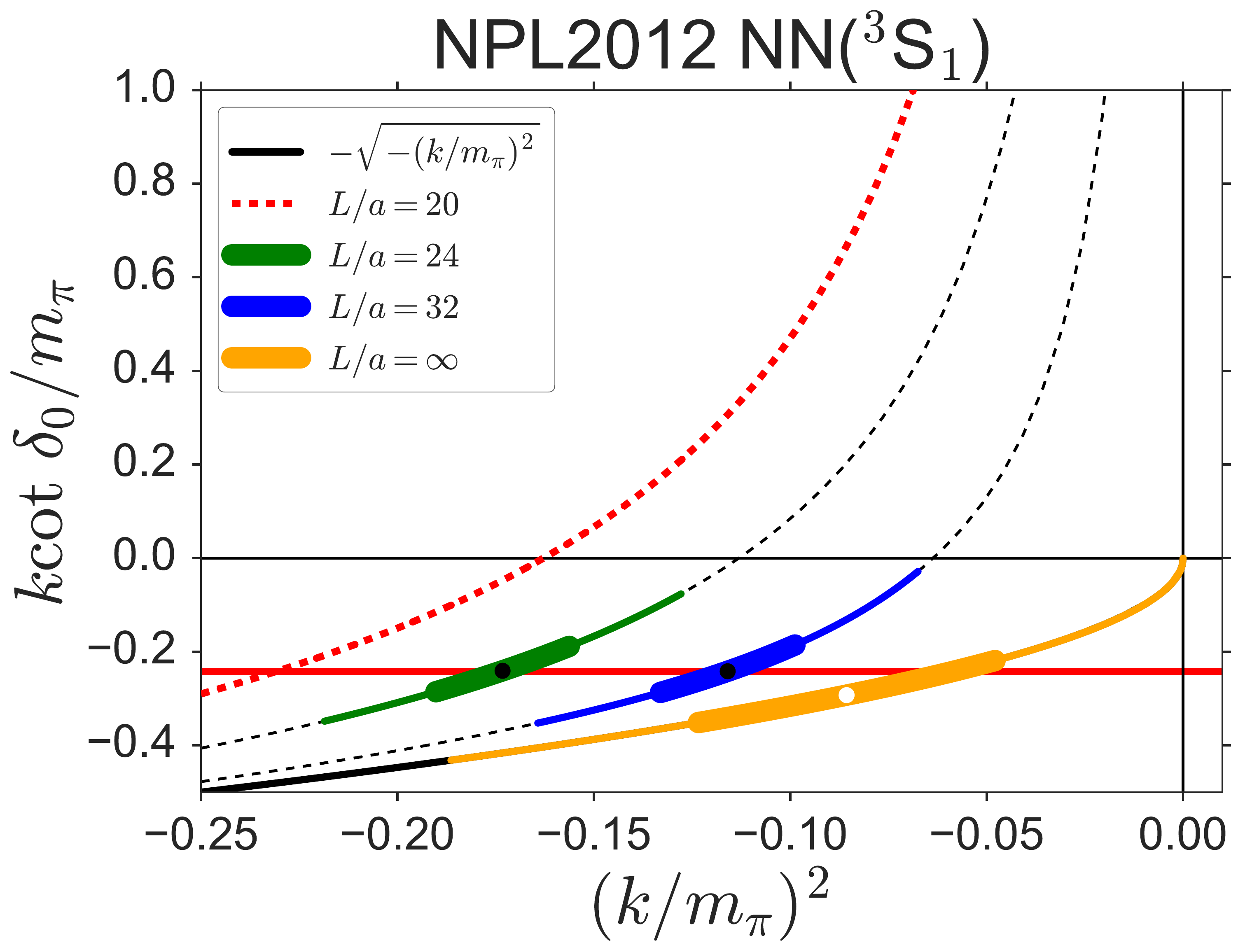}
  \caption{Same as Fig.~\ref{fig:kcot_YKU2012}, but from NPL2012.
    Red lines correspond to the NLO ERE fits. 
    Red dashed line represents the L\"uscher's formula for $L/a = 20$.
    Lattice data around $(k/m_\pi)^2=0$ at $L/a=20$~\cite{Beane:2009py} are located 
    way out of the plot region of the figures.
  }
  \label{fig:kcot_NPL2012}
\end{figure}

We perform the sanity check on  NPL2012 data in (2+1)-flavor QCD at $m_\pi=0.39$ GeV.
Similar to YIKU2012 and YIKU2015, only data for the ground state are available in NPL2012.
Fig.~\ref{fig:kcot_NPL2012} 
shows $k\cot\delta_0(k)/m_\pi$ as a function
of $(k/m_\pi)^2$ 
for $NN$($^1$S$_0$) (Left), $NN$($^3$S$_1$) (Right).
In NPL2012, the binding energies were determined by the infinite volume extrapolation 
with the asymptotic expansion of the L\"uscher's formula.

In $NN$($^1$S$_0$) channel,
we observe
a singular ERE behavior similar to (but somewhat milder than) those observed in YKU2011, YIKU2012 and YIKU2015.
As shown in Fig.~\ref{fig:kcot_NPL2012} (Left),
$k\cot\delta_0(k)/m_\pi$ at $(k/m_\pi)^2 < 0$ decreases vertically as the volume increases.
The NLO ERE fit for data at $L/a=24, 32$ gives
$((a_0 m_\pi)^{-1}, r_0 m_\pi) = (-1.06, -32.3)$ , and the  
corresponding ERE is plotted in Fig.~\ref{fig:kcot_NPL2012} (Left) by the red line.

In $NN$($^3$S$_1$) channel,
 values for ERE parameters are rather reasonable,
$((a_0 m_\pi)^{-1}, r_0 m_\pi) = (-0.24, 0.0)$,
as shown 
by the red line in Fig.~\ref{fig:kcot_NPL2012} (Right).
Even if a reasonable behavior is observed, however,
it does not guarantee that the data are reliable. 
Indeed,  as seen in appendix~\ref{app:NPL2013-CalLat2017}, NPL2013 and CalLat2017
give non-singular but manifestly source-dependent  $\kcotd$ behaviors.

NPLQCD Collaboration reported~\cite{Beane:2009py} 
the small positive  values for $(k/m_\pi)^2$
with the same lattice setup but on a smaller volume ($L/a=20$),
\begin{eqnarray}
(k/m_\pi)^2 = 0.030(13)(20) \ {\rm for} \  NN(^1S_0) , \  \ (k/m_\pi)^2 = 0.012(20)(33) \ {\rm for} \  NN(^3S_1) . 
\end{eqnarray}
Such results clearly  conflict with the ERE behaviors obtained from $L/a=24, 32$:
\footnote{
  The L\"uscher's formula for $L/a = 20$
intersects with the NLO ERE at $(k/m_\pi)^2 = -0.097$ for $NN$($^1$S$_0$)
and $(k/m_\pi)^2 = -0.231$ for $NN$($^3$S$_1$), respectively.}
In Fig.~\ref{fig:kcot_NPL2012}, we only show the lines corresponding to the L\"uscher's formula for $L/a=20$,
as the lattice data around $(k/m_\pi)^2 =0$ are located way beyond the plot range of the figure.

Again the sanity check reveals that at least some of the data 
in NPL2012 (and  their earlier result~\cite{Beane:2009py}) are unreliable.
Provided that all $\Delta E$ in NPL2012 were obtained by 
the plateau identification at early times slices,
further investigations which do not rely on the plateau identification
are necessary before claiming the existence of $NN$ bound states.

\section{Conclusion and discussion}
\label{sec:conclusion}

In this paper, we have introduced a simple test (sanity check) to 
inspect the reliability of the energy shift of two-hadron systems in lattice QCD
on the basis of the L\"uscher's finite volume formula.
We have argued that useful information on the hadron-hadron interactions
can be extracted from the lattice data in the region of
not only positive squared momentum $k^2 > 0$
but also negative squared momentum $k^2 < 0$.
Consistency with the effective range expansion (ERE) around $k^2=0$ 
exposes the problem of the lattice data
which otherwise hidden in the energy shift $\Delta E$.

We have applied the sanity check to  lattice results from which 
 the existence of the $NN$ bound state(s) for heavy quark masses are concluded in the literature.
All of them employ the direct method, in which $\Delta E$ is obtained 
by the plateau identification at early time slices.
In Tab.~\ref{tab:Summary}, we summarize our sanity checks, together with source independences of the plateaux (the mirage problem) 
discussed in~\cite{Iritani:2016jie} and reviewed in Sec.~\ref{sec:introduction}.
In the table,  ``Source independence'' means that whether $\Delta E$ is physical in the sense that 
 it is independent of the nucleon source operators,  ``Sanity check (i)'' means that 
whether  \EREp and \EREn are consistent with each other,  ``Sanity check (ii)''  means that whether the 
scattering parameters obtained by ERE is non-singular, 
 and  ``Sanity check (iii)'' means that whether the bound state pole has a physical residue 
 in Eq.~(\ref{eq:kcotd:bound}).
  As can be seen from the table, none of these results is free from either the plateau problem 
or the ERE problem, or both.

\begin{table}[tbh]
\centering
\begin{ruledtabular}
\begin{tabular}{|c||c|c|c|c|c|c|c|c|}
& \multicolumn{4}{c|}{$NN(^1S_0)$} & \multicolumn{4}{c|}{$NN(^3S_1)$} \\ 
\hline
 Data & Source & \multicolumn{3}{c|}{Sanity check} & Source & \multicolumn{3}{c|}{Sanity check} \\
     & independence & (i) & (ii) & (iii) & independence  & (i)  & (ii) & (iii) \\
\hline\hline
YKU2011 \cite{Yamazaki:2011nd}  & $\dagger$ &   No        &   No       &  $\ast$   &$\dagger$&  No          &    No     &  $\ast$ \\  
YIKU2012 \cite{Yamazaki:2012hi}  & No            &$\dagger$&  No       & $\ast$ &  No         & $\dagger$ &   No     & $\ast$    \\
YIKU2015 \cite{Yamazaki:2015asa}  & $\dagger$ & $\dagger$&  No      &  $\ast$      &$\dagger$&  $\dagger$ &   No     & No  \\
NPL2012 \cite{Beane:2011iw}  & $\dagger$ & $\dagger$ &   No     &   $\ast$     &   $\dagger$& $\dagger$ &  $\ast$   &  $\ast$   \\
NPL2013 \cite{Beane:2013br,Beane:2012vq}   & No      & $\ast$    & $\ast$   &No    &   No      &  $\ast$      &  $\ast$    &  ?         \\
NPL2015 \cite{Orginos:2015aya}   & $\dagger$ &   No         &  $\ast$         & No   &   $\dagger$& No   &  $\ast$ & No\\
CalLat2017 \cite{Berkowitz:2015eaa} & No             &  ? & $\ast$  & No     &  No             & ?   &   $\ast$    &  No \\
\end{tabular}
\end{ruledtabular}
\caption{A summary of sanity checks (i) consistency between \EREp and \EREn,
  (ii) non-singular ERE parameters and (iii) physical residue for the bound
  state pole, together with the  source independence of $\Delta E$. Here ``No''
  means that the source independency/sanity check has failed,
  while the symbol $\dagger$ implies there is none or only insufficient
  study on the corresponding item.  
  The symbol $\ast$ means that obvious contradiction is not found within the error bars,  
  while it does not necessarily guarantee that the data are reliable.
  See appendix~\ref{app:NPL2013-CalLat2017}
for the meaning of the symbol ? on the Sanity check 
for NPL2013 and CalLat2017.  }  
\label{tab:Summary}
\end{table}

Results in this paper, together with those in our previous paper~\cite{Iritani:2016jie},
strongly indicate that $\Delta E$ in the direct method,  determined by plateaux at  earlier time slices,
suffer uncontrolled systematic errors from excited state contaminations.
This conclusion brings a serious doubt on the existence of the $NN$ bound states 
for pion masses heavier than 300 MeV,
contrary to the claims of YKU2011, YIKU2012, YIKU2015, NPL2012, NPL2013, NPL2015 and CalLat2017.
In order to determine correct spectra of two nucleon systems at heavier pion masses by the direct method,  
much more sophisticated method than the plateau fitting such as the variational method~\cite{Luscher:1990ck} 
must be employed.

An alternative method to determine spectra of multi hadrons is the HAL QCD method,
which does not suffer from the problem of excited state contaminations in multi hadron systems by the use of 
the space-time correlations instead of the temporal correlations~\cite{HALQCD:2012aa}.
In forthcoming papers~\cite{Iritani2017}, we will investigate the source dependence 
of the potential in the HAL QCD method, which will be also used to analyze the fundamental
 origin of the mirage problem in the direct method.

 After the submission of the present paper, 
 two related articles were posted,
 by Beane et al. \cite{Beane:2017edf}
 and 
 by Wagman et al \cite{Wagman:2017tmp}.
 We confirmed that none of the conclusions of 
 the present paper summarized in Table~\ref{tab:Summary}
 are not affected by these papers.
 Critical comments on  these articles
 can be seen in a recent summary \cite{Aoki:2017byw}.

\acknowledgments
This work is supported in part by the Japanese Grant-in-Aid for Scientific
Research (No. JP24740146, JP25287046, JP15K17667, JP16H03978, JP16K05340, 
(C)26400281), by MEXT as ``Priority
Issue on Post-K computer'' (Elucidation of the Fundamental Laws and Evolution of
the Universe) and by Joint Institute for Computational Fundamental Science
(JICFuS). 
We thank Dr.~A.~Walker-Loud for the detailed information of the smeared quark source.
S.A., T.D., T.~Iritani and H.N. thank the Institute for Nuclear Theory at the University of Washington for its hospitality during the INT 16-1 program and the Department of Energy for partial support during the initiation of this work.
T.Iritani also thanks Dr.~L.~Contessi for his suggestion,
which triggered this study.
T.D. and T.H. were partially supported by RIKEN iTHES Project and iTHEMS Program.

\clearpage
\appendix

\section{The square well potential and $\kcotd$}
\label{app:demo-well}

\begin{figure}[b]
  \centering
  \includegraphics[width=0.46\textwidth,clip]{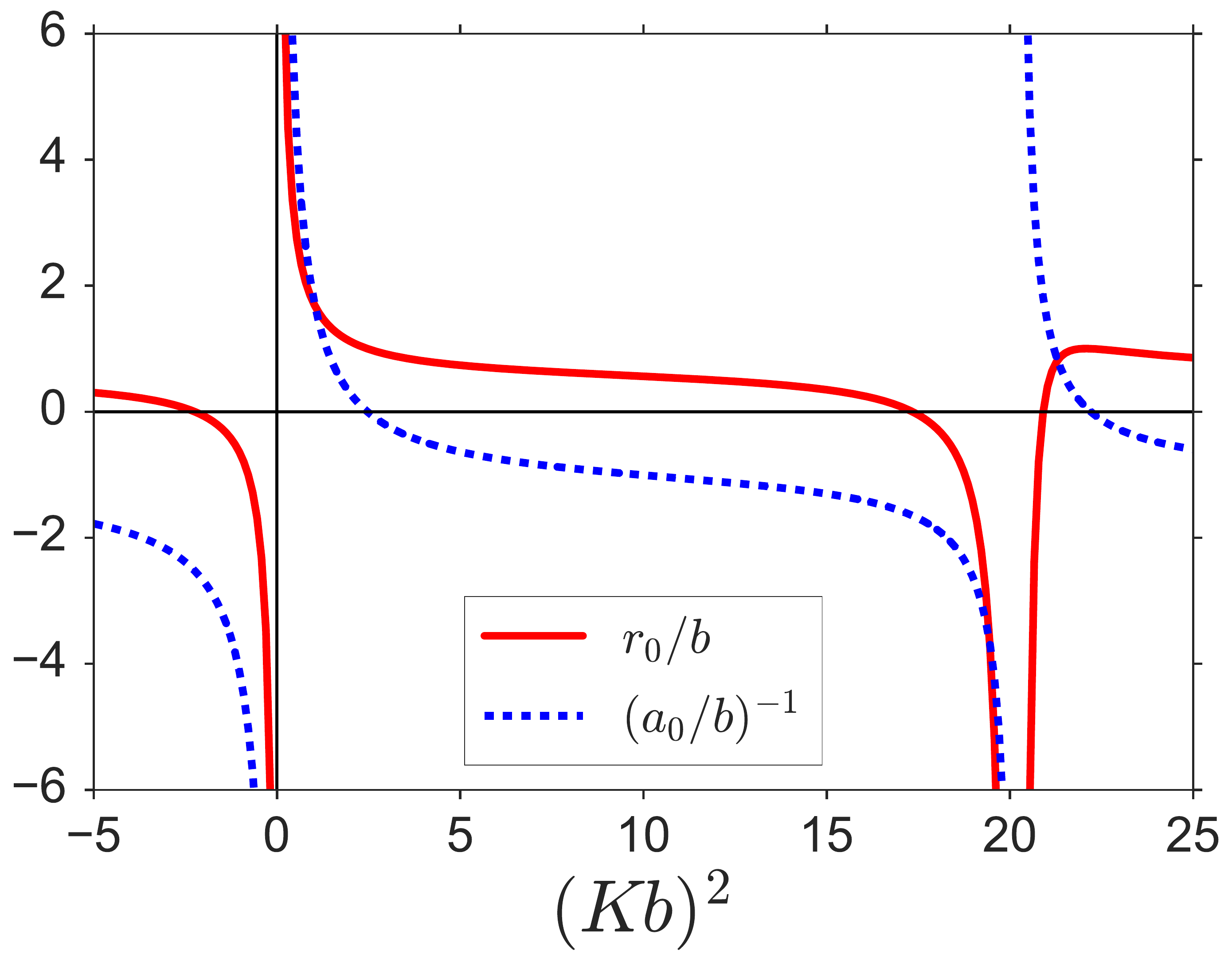}
  \caption{
    \label{fig:square_a0inv_r0}
    The inverse scattering length $a_0^{-1}$ (blue dashed line) and the effective range $r_0$ (red solid line) as a function of $(Kb)^2$.
    The first bound state appears at $(Kb)^2 = (\pi/2)^2$, and the second one at $(Kb)^2 = (3\pi/2)^2$.}
\end{figure}

In this appendix,
we consider two non-relativistic particles with each mass $M$ interacting through the   three-dimensional square well potential, 
  $V(\vec{r}) = -v \cdot \theta(b - |\vec{r}|)$, 
which leads to
\begin{equation}
  k\cot\delta(k) = 
  \frac{k^2 + \sqrt{K^2 + k^2} \cot(\sqrt{K^2 + k^2}b) k\cot(kb)}{k\cot(kb) - \sqrt{K^2 + k^2}\cot(\sqrt{K^2 + k^2}b)},
  \label{}
\end{equation}
where $k^2 = M E $ and $ K^2=M v$.
From the effective range expansion, the scattering length $a_0$ and the effective range $r_0$ are obtained as
\begin{equation}
  a_0/b =  \frac{\tan(Kb)}{Kb} - 1, \qquad
  r_0/b = 1 - \frac{(Kb)^2}{3(\tan(Kb) - Kb)^2}
  + \frac{1}{Kb(\tan(Kb) - Kb)},
  \label{}
\end{equation}
which are plotted as a function of $(Kb)^2$ in
Fig.~\ref{fig:square_a0inv_r0}.
A number of bound states increases  as $(Kb)^2$ does, and  scattering length diverges at $(Kb)^2 = (\pi/2)^2, (3\pi/2)^2, \cdots$.

\begin{figure}[t]
  \centering
  \includegraphics[width=0.47\textwidth,clip]{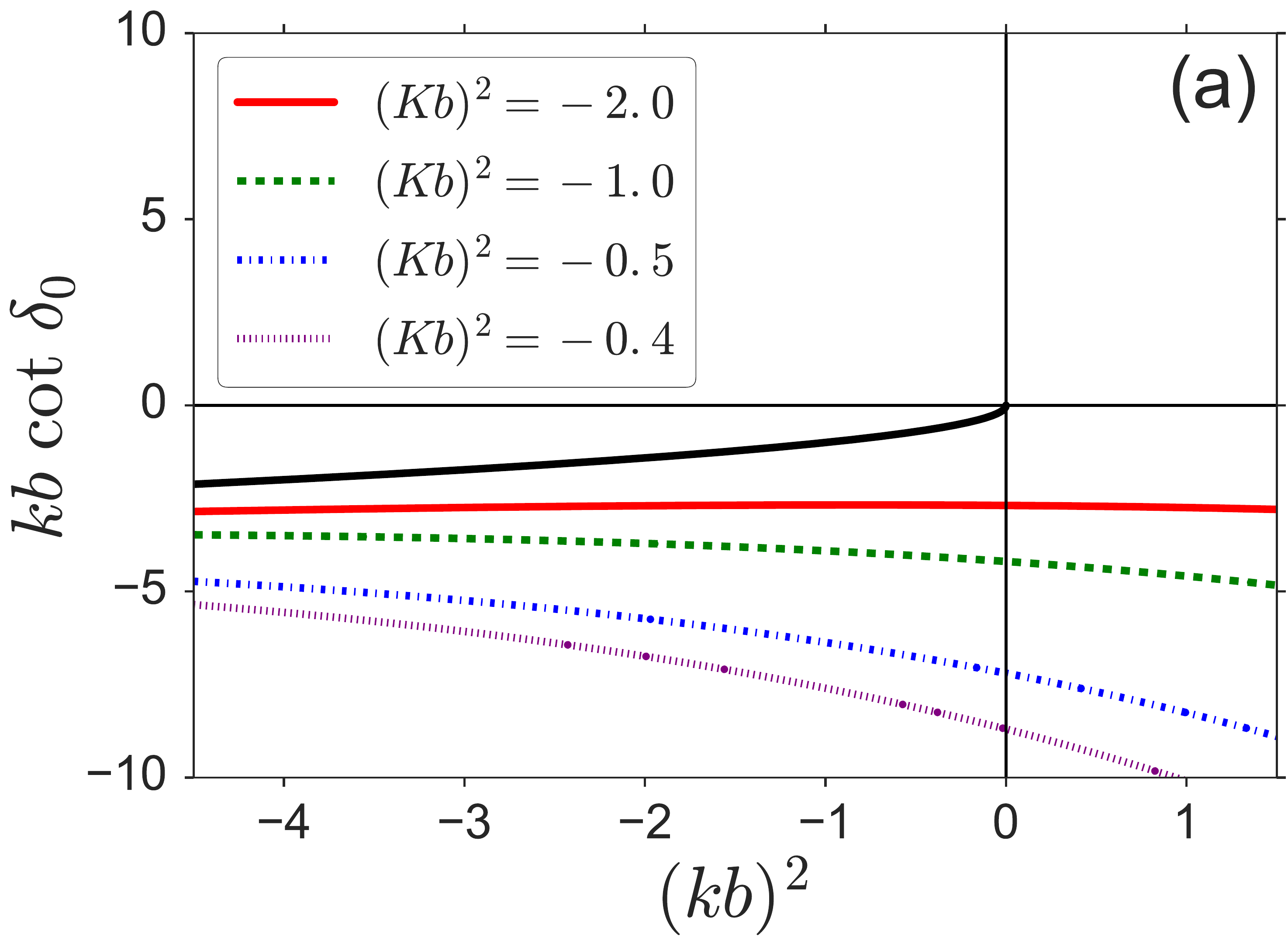}

  \includegraphics[width=0.47\textwidth,clip]{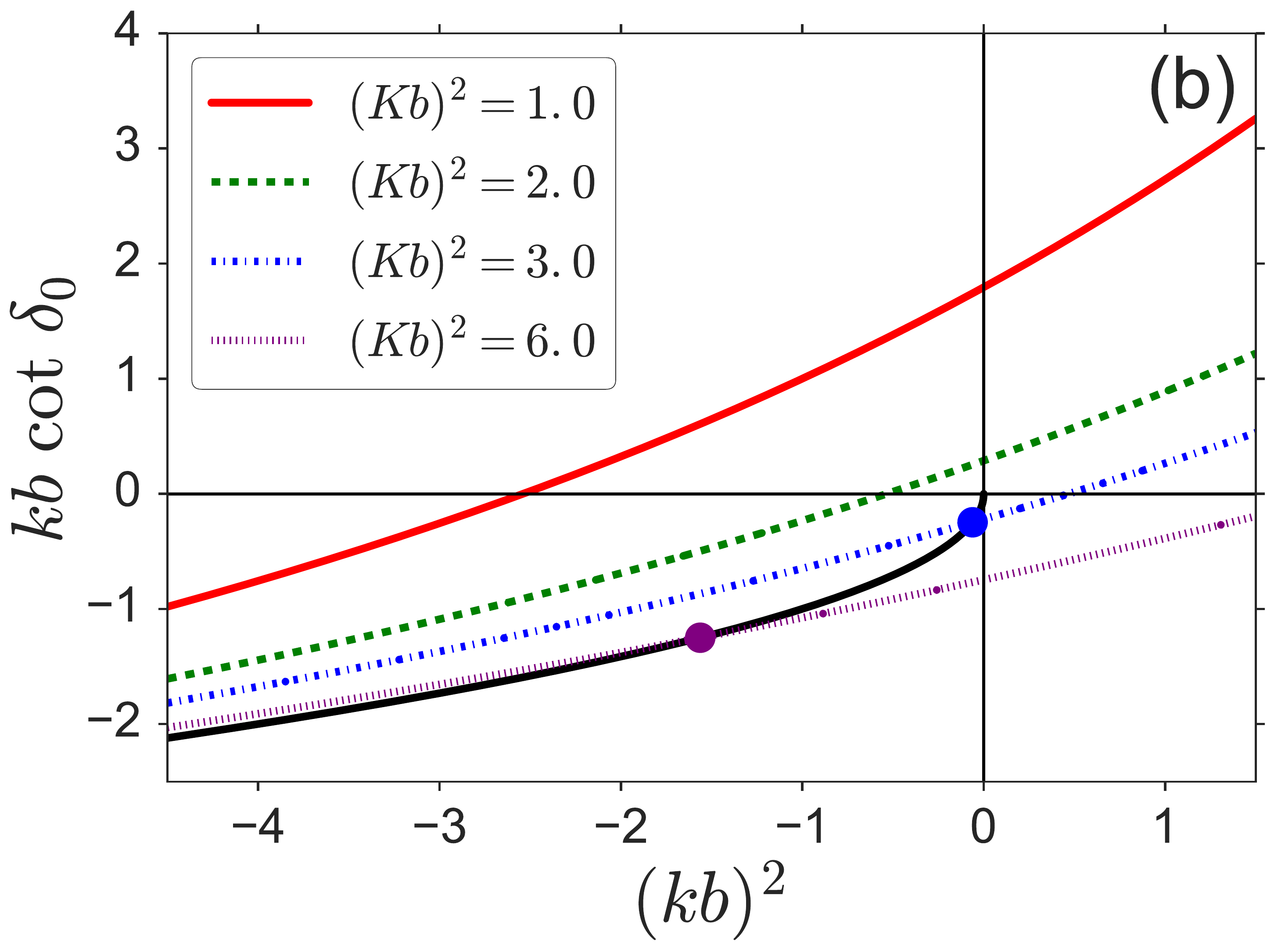}
  \includegraphics[width=0.47\textwidth,clip]{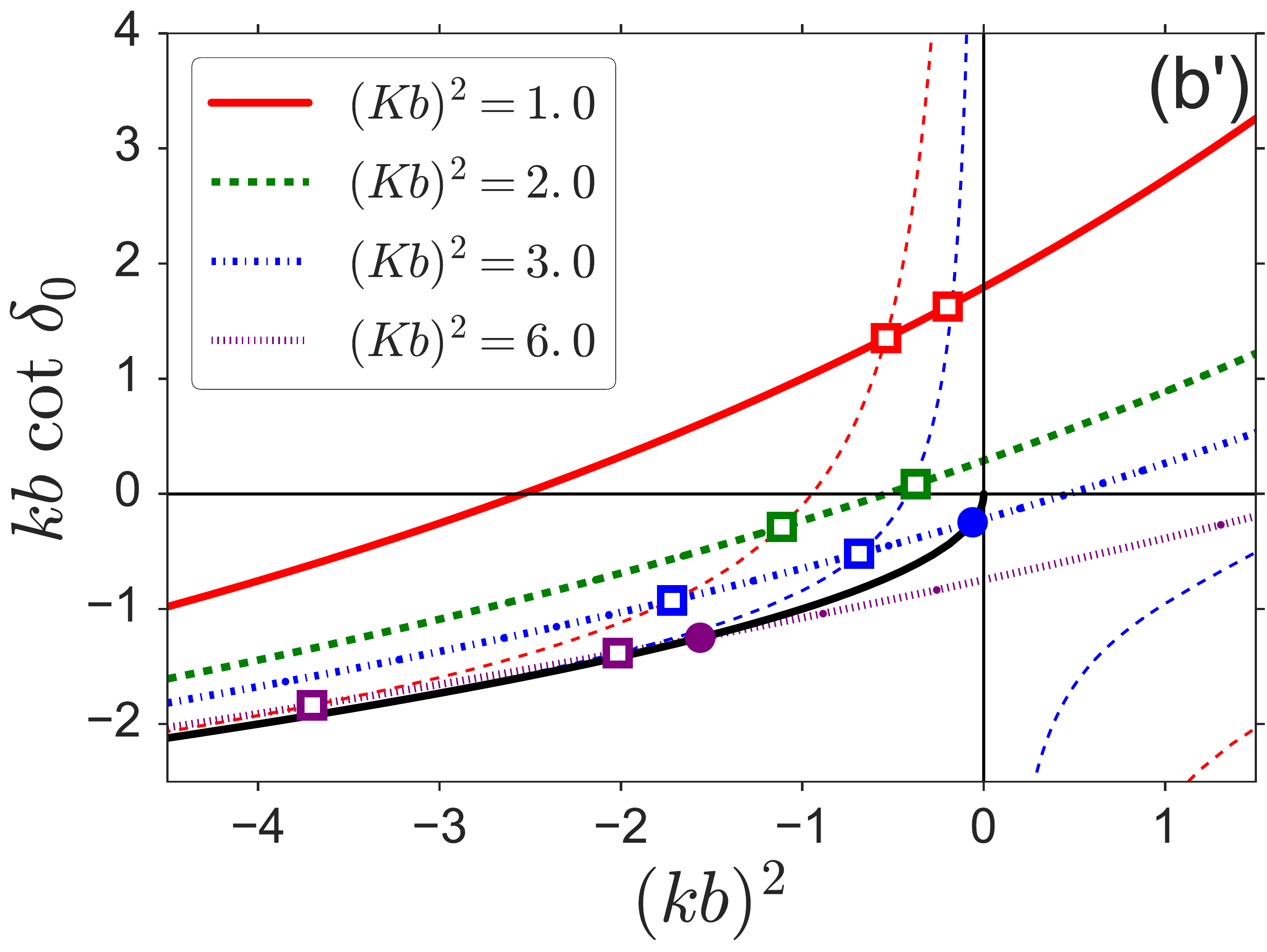}

  \includegraphics[width=0.47\textwidth,clip]{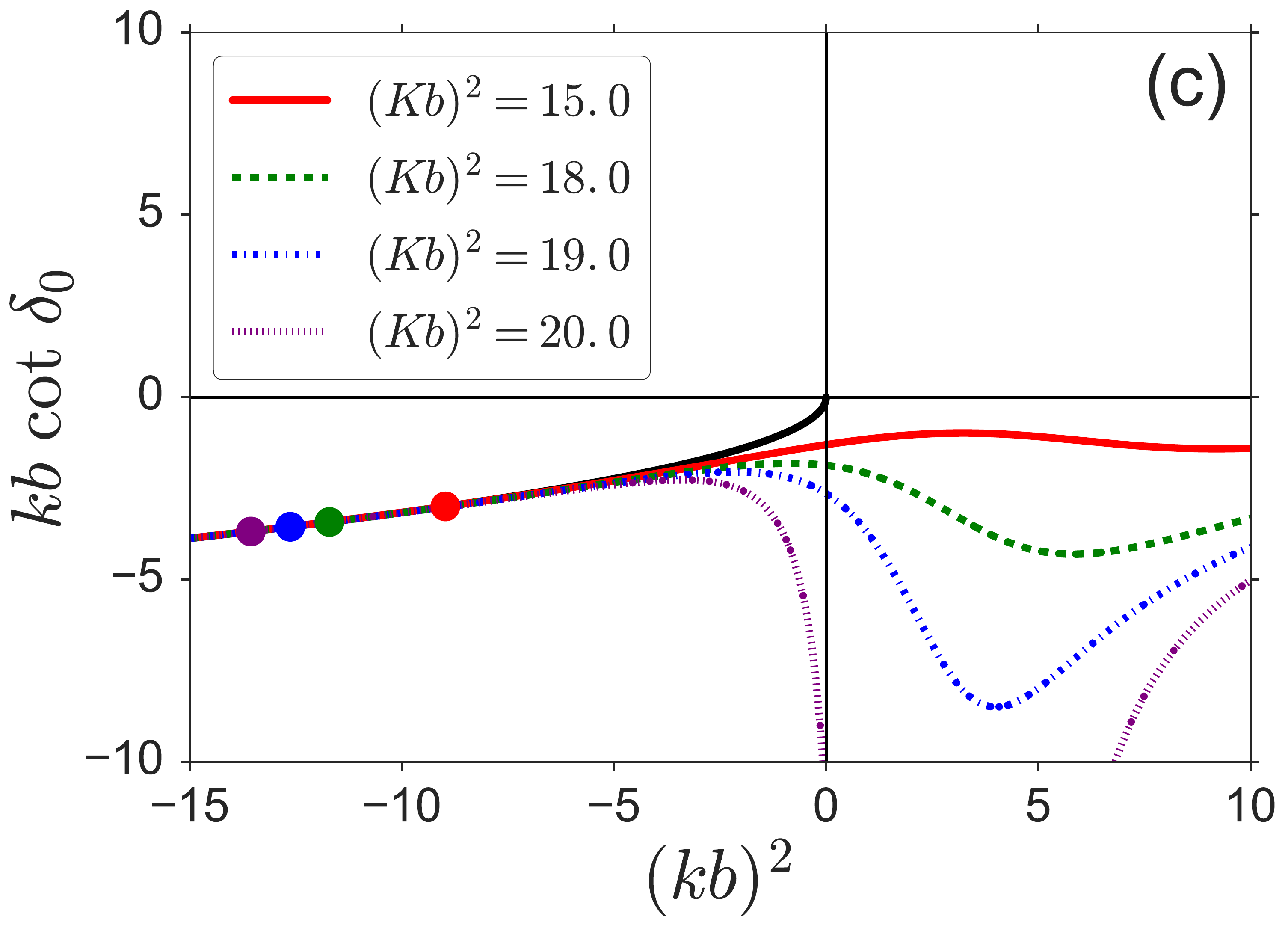}
  \includegraphics[width=0.47\textwidth,clip]{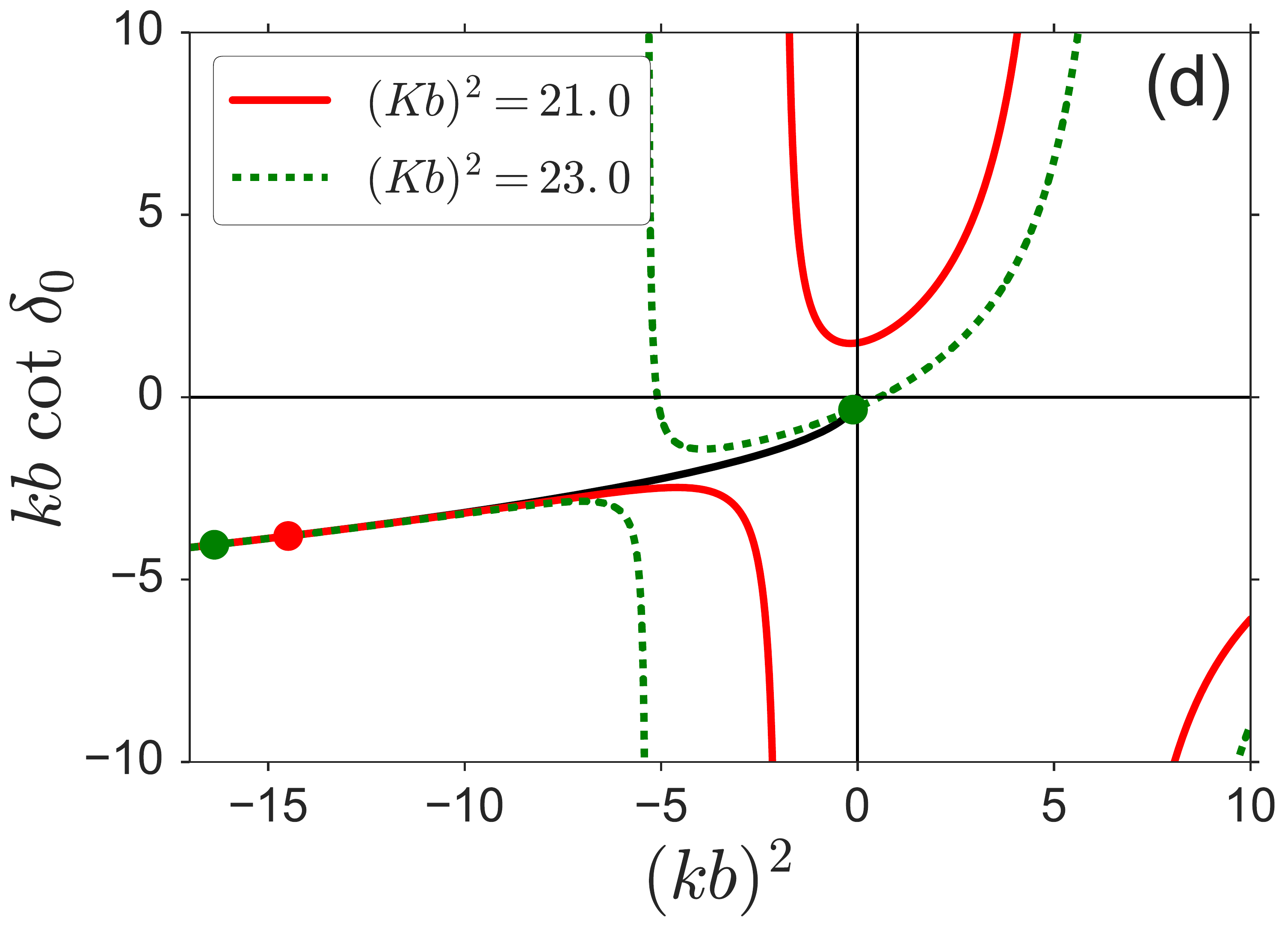}
  \caption{
      \label{fig:square_kcot}
    The  $kb \cot \delta_0(k)$ as a function of $(kb)^2$ are shown by colored lines.
    The black solid lines denote the condition for the bound states,
    and the solid circles correspond to the poles.
    (a) Weak repulsion. (b) Weak attraction. 
    (b') Weak attraction  together with the L\"uscher's formula at $L/b = 2$ $(3)$ by the red (blue) thin dashed line,
    where open squares are finite volume spectra.
    (c)  Moderate attraction. (d) Strong attraction 
    having the 2nd pole at $(Kb)^2 = 23$.
  }
\end{figure}

The $\kcotd$ for several interaction strength are given in Fig.~\ref{fig:square_kcot}:
(a)  weak repulsion with $-2.0 \le (Kb)^2 \le -0.4$, 
(b), (b') weak attraction with $1.0 \le (Kb)^2 \le 6.0$,
(c) moderate attraction with $15.0 \le (Kb)^2 \le 20.0$,
and (d) strong attraction with $21.0 \le (Kb)^2 \le 23.0$.
Solid circles correspond to the  bound state poles. The thin dashed  lines in Fig.~\ref{fig:square_kcot} (b') represent
the L\"uscher's formula, together with finite volume spectra denoted by open squares. 

\clearpage
\section{Sanity check for NPL2013 and CalLat2017}
\label{app:NPL2013-CalLat2017}

\begin{figure}[tbh]
\centering
 \includegraphics[width=0.48\textwidth]{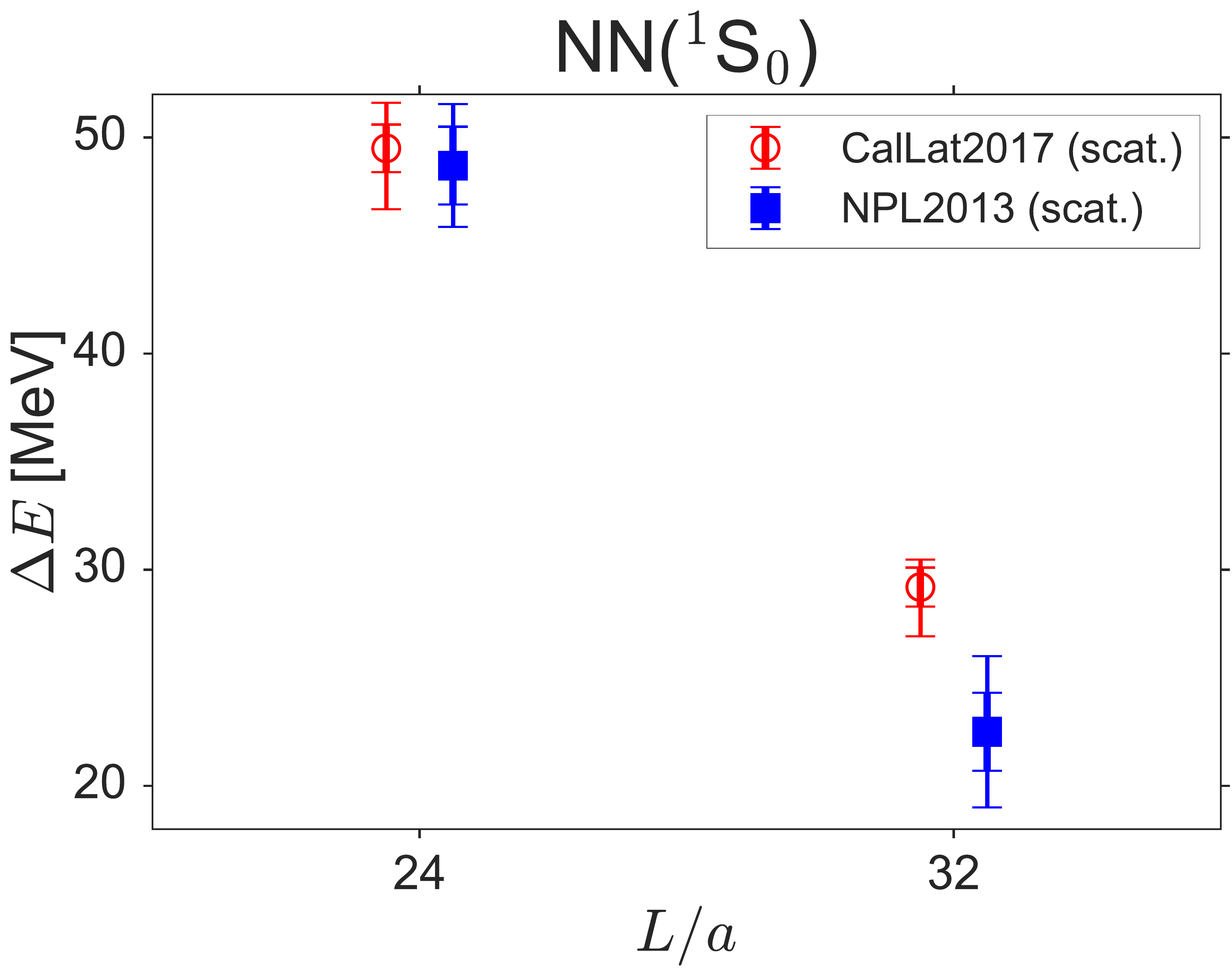}
 \includegraphics[width=0.48\textwidth]{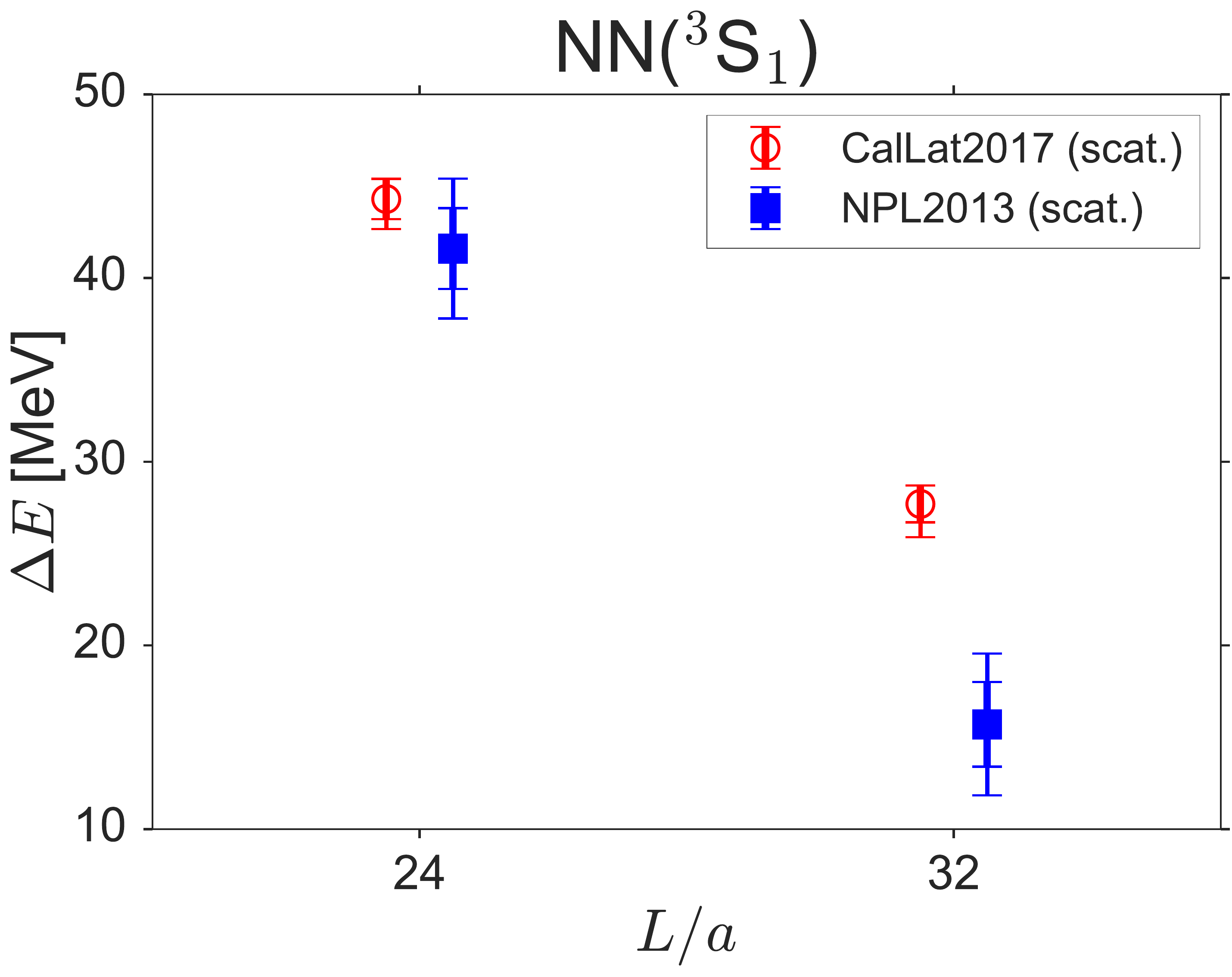}
 \caption{Same as Fig.~\ref{fig:Delta_E:CalLat_vs_NPL2013} but for the excited state ($\Delta E > 0$) from CalLat2017 (red circles) and NPL2013 (blue triangles) in the center of mass system.
}
 \label{fig:Delta_E2:CalLat_vs_NPL2013}
\end{figure}
In NPL2013 and CalLat2017, although the same gauge configurations are employed 
for $L/a=24$ and 32,  mutual and/or self inconsistencies are found for $\Delta E$ at $k^2 < 0$,
as discussed in Sec.~\ref{sec:introduction}
 (See Fig.~\ref{fig:Delta_E:CalLat_vs_NPL2013}).
 As shown in Fig.~\ref{fig:Delta_E2:CalLat_vs_NPL2013}, a similar mutual
inconsistency is also observed for $\Delta E$ at $k^2 > 0$, which are obtained in the center of mass system with a non-zero relative momentum injected between two nucleons at the sink.  Here NPL2013 employed the zero displaced two nucleon source, while the CalLat2017 used the non-zero displaced one. The inconsistency at $L/a=32$, in particular in the $NN(^3S_1)$ channel
indicates that scattering state also fails to satisfy the source independence.\footnote{
The details  of sink operators may also differ between the two.}

\begin{figure}[hbt]
\centering
   \includegraphics[width=0.49\textwidth]{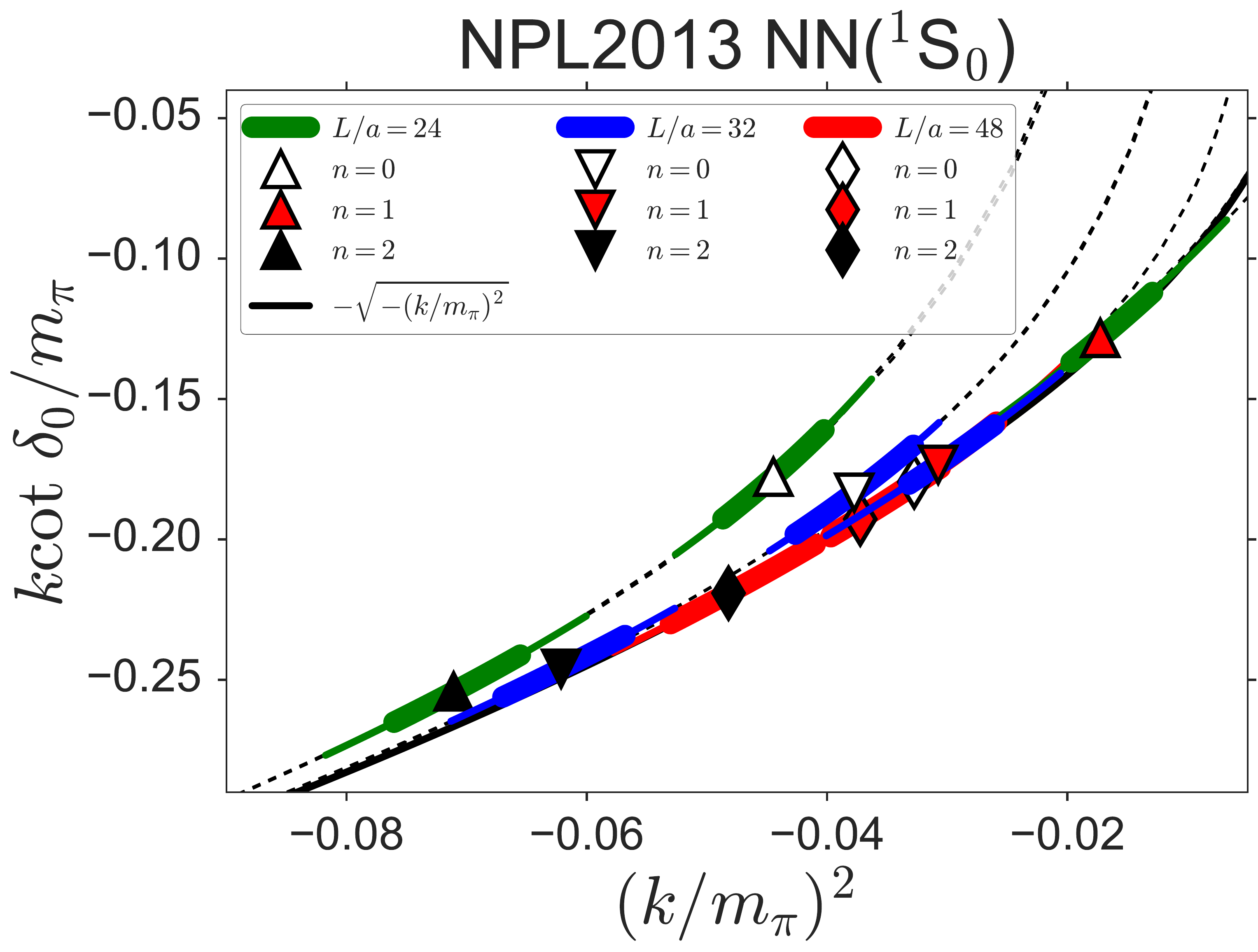}
   \includegraphics[width=0.49\textwidth]{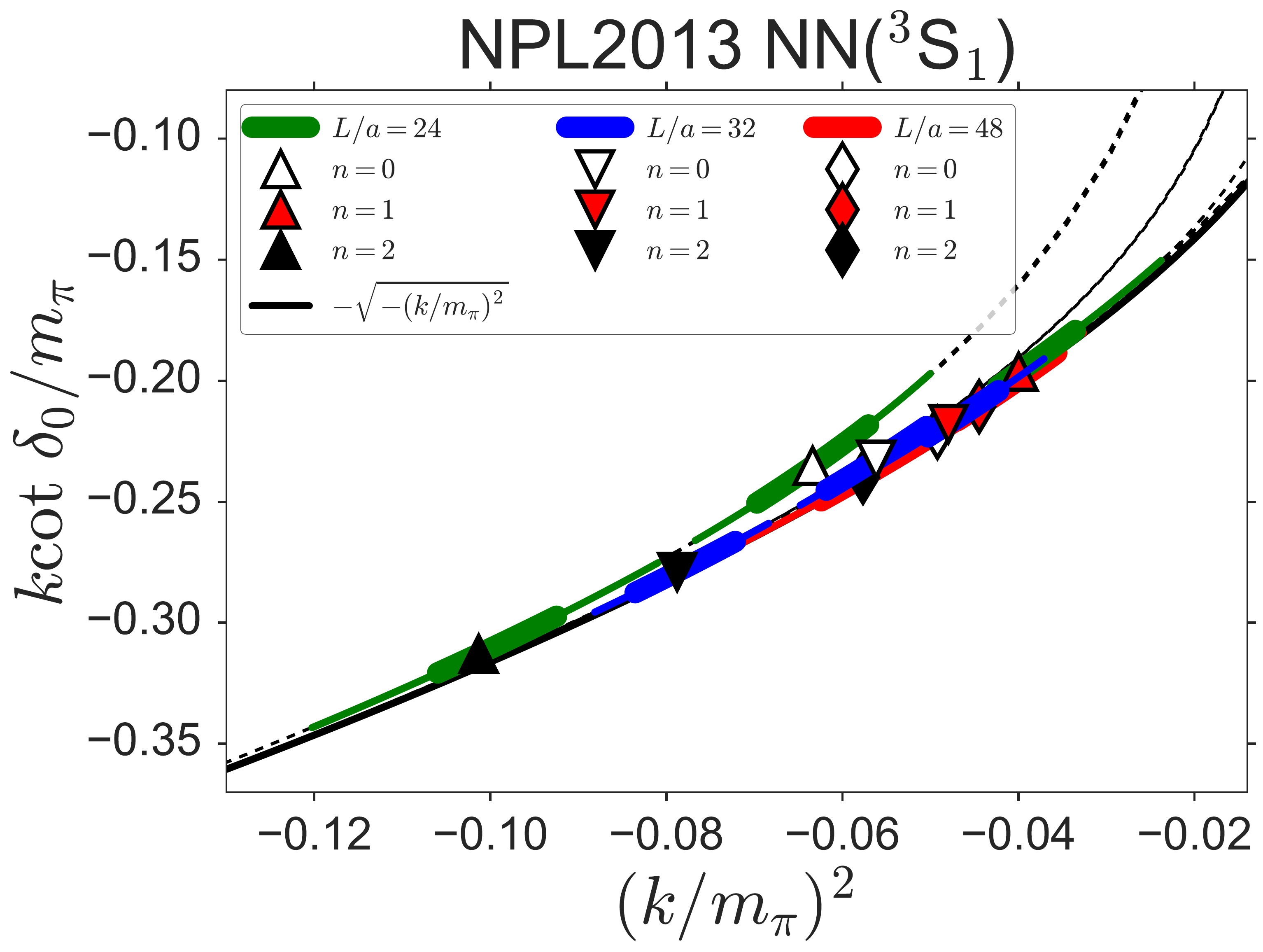}
   \includegraphics[width=0.49\textwidth]{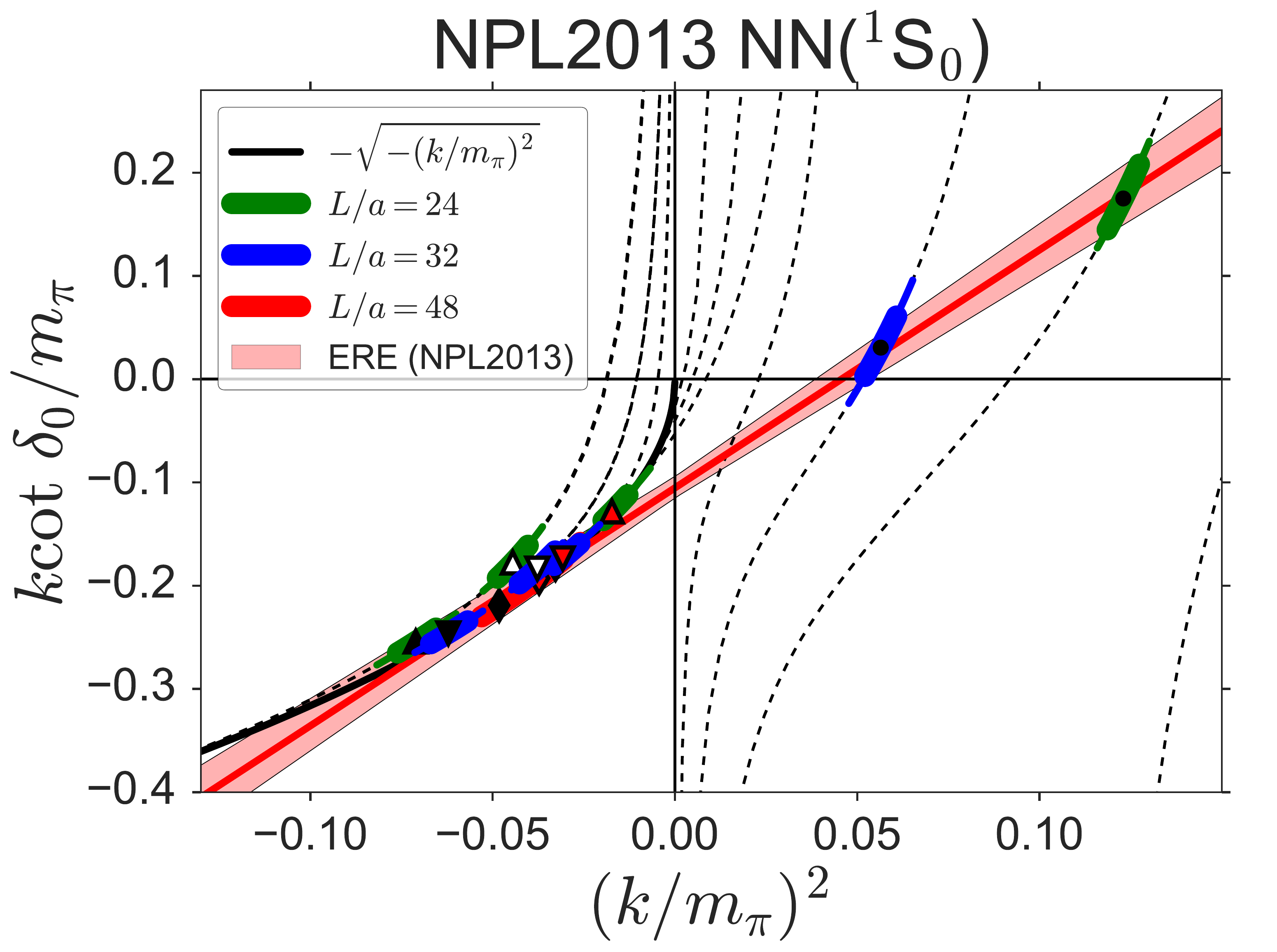}
   \includegraphics[width=0.49\textwidth]{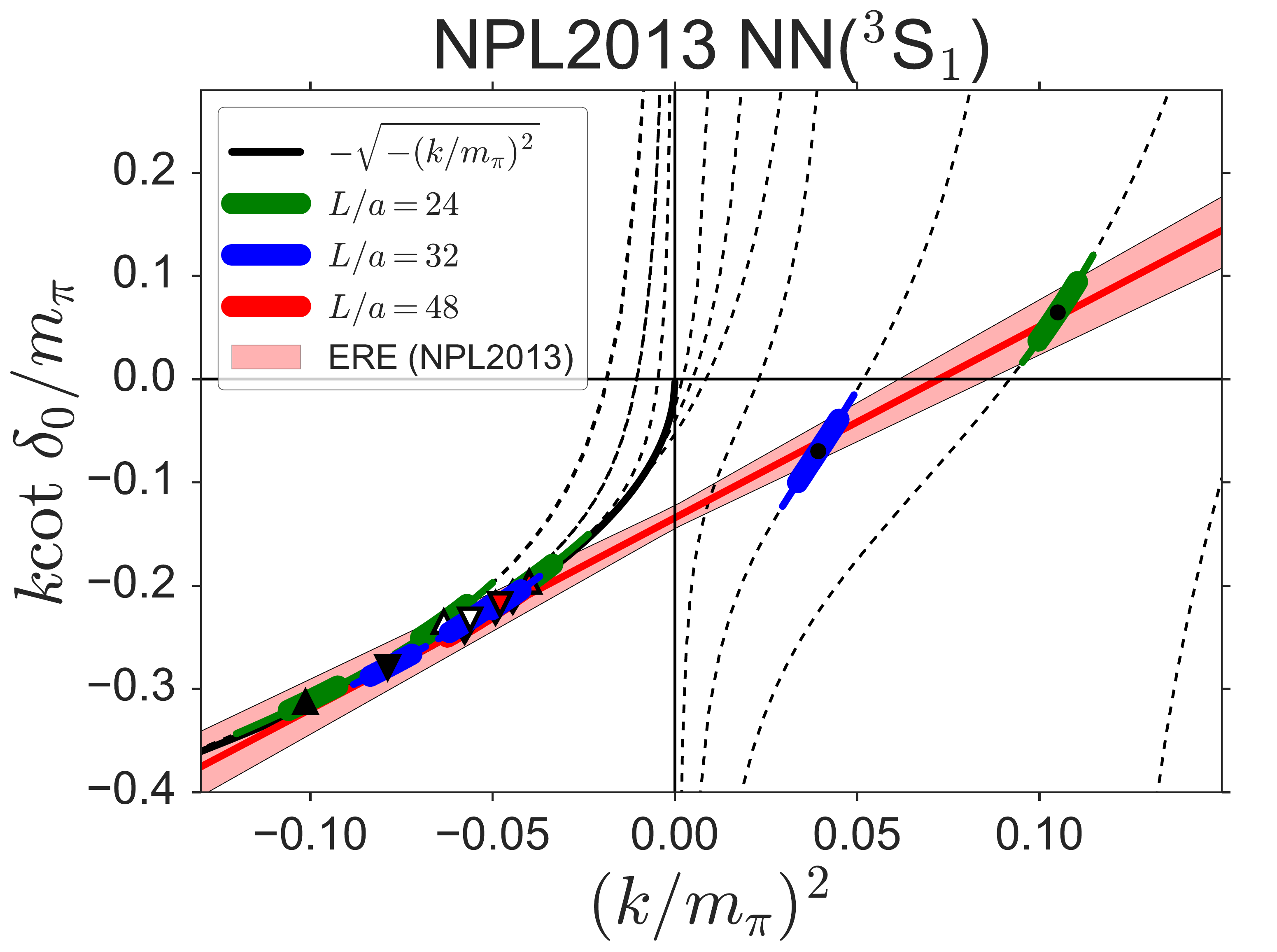}
   \caption{(Upper) Same as Fig.~\ref{fig:kcot_YKU2012}, but from NPL2013.
 (Lower) Same as upper figures but with data from excited states.
 Red bands correspond to \EREp given in NPL2013 with statistical and systematic errors added in quadrature.
}
 \label{fig:kcot_NPL2013}
\end{figure}

\begin{figure}[htb]
  \centering
  \includegraphics[width=0.49\textwidth]{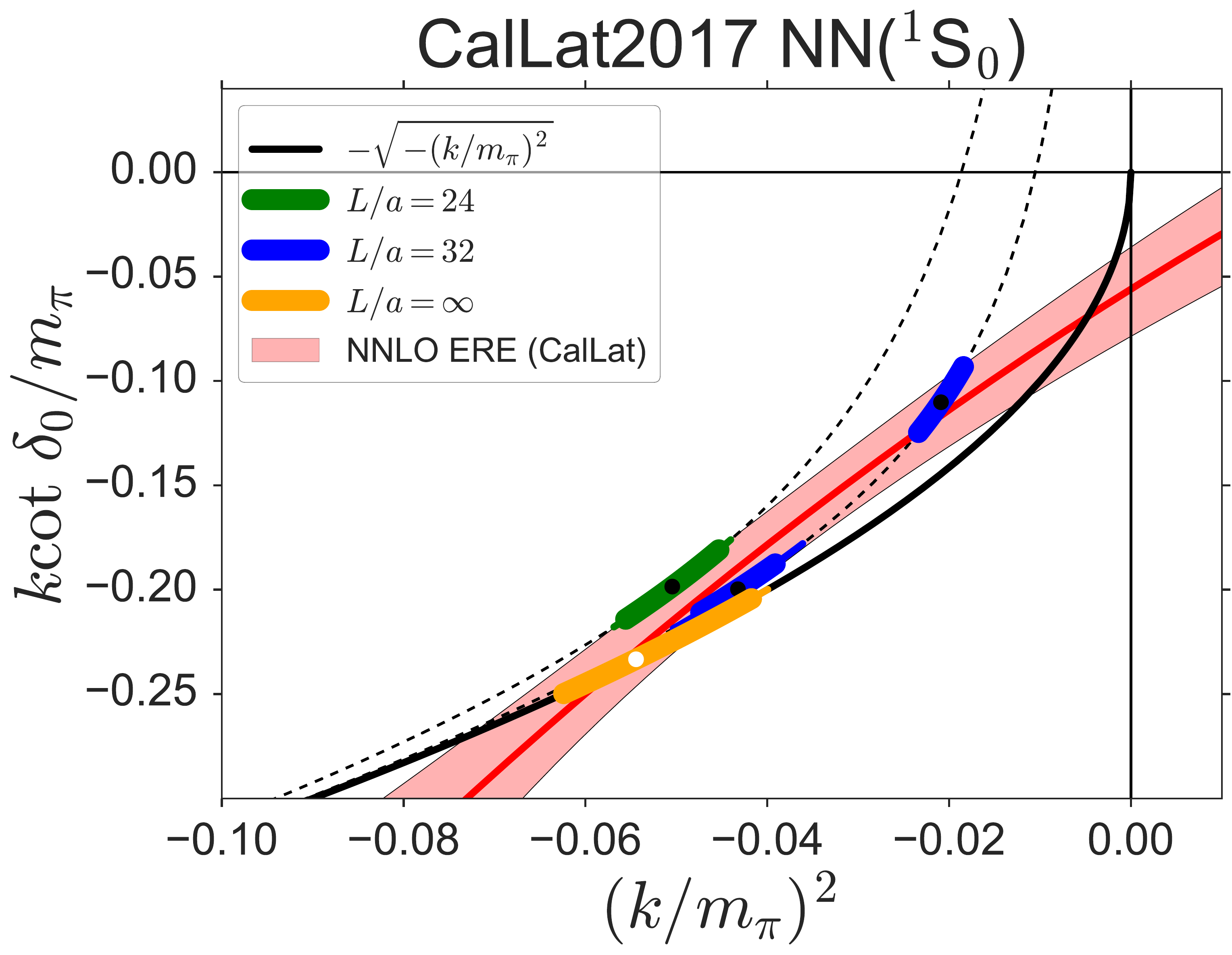}
  \includegraphics[width=0.49\textwidth]{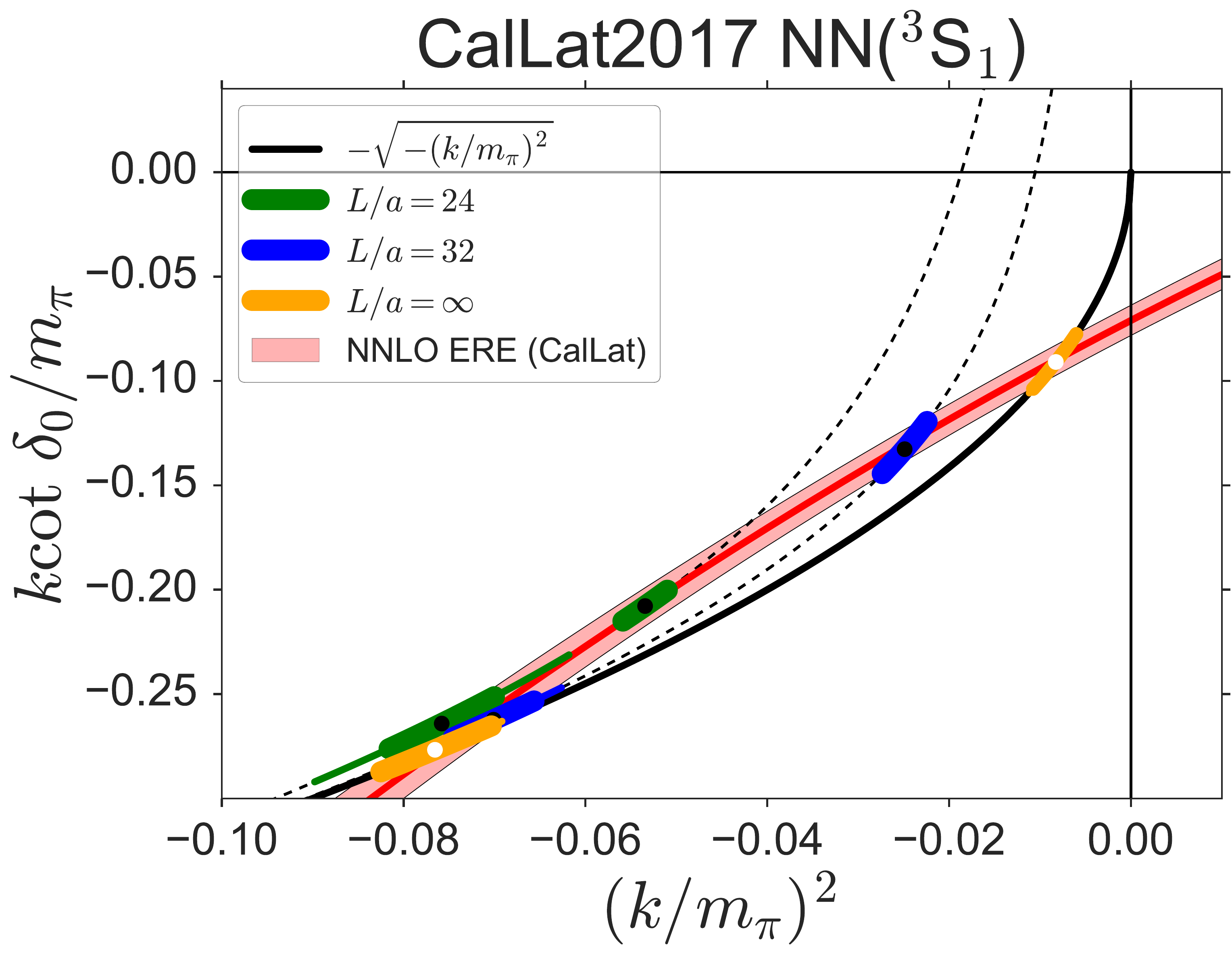}
  \includegraphics[width=0.49\textwidth]{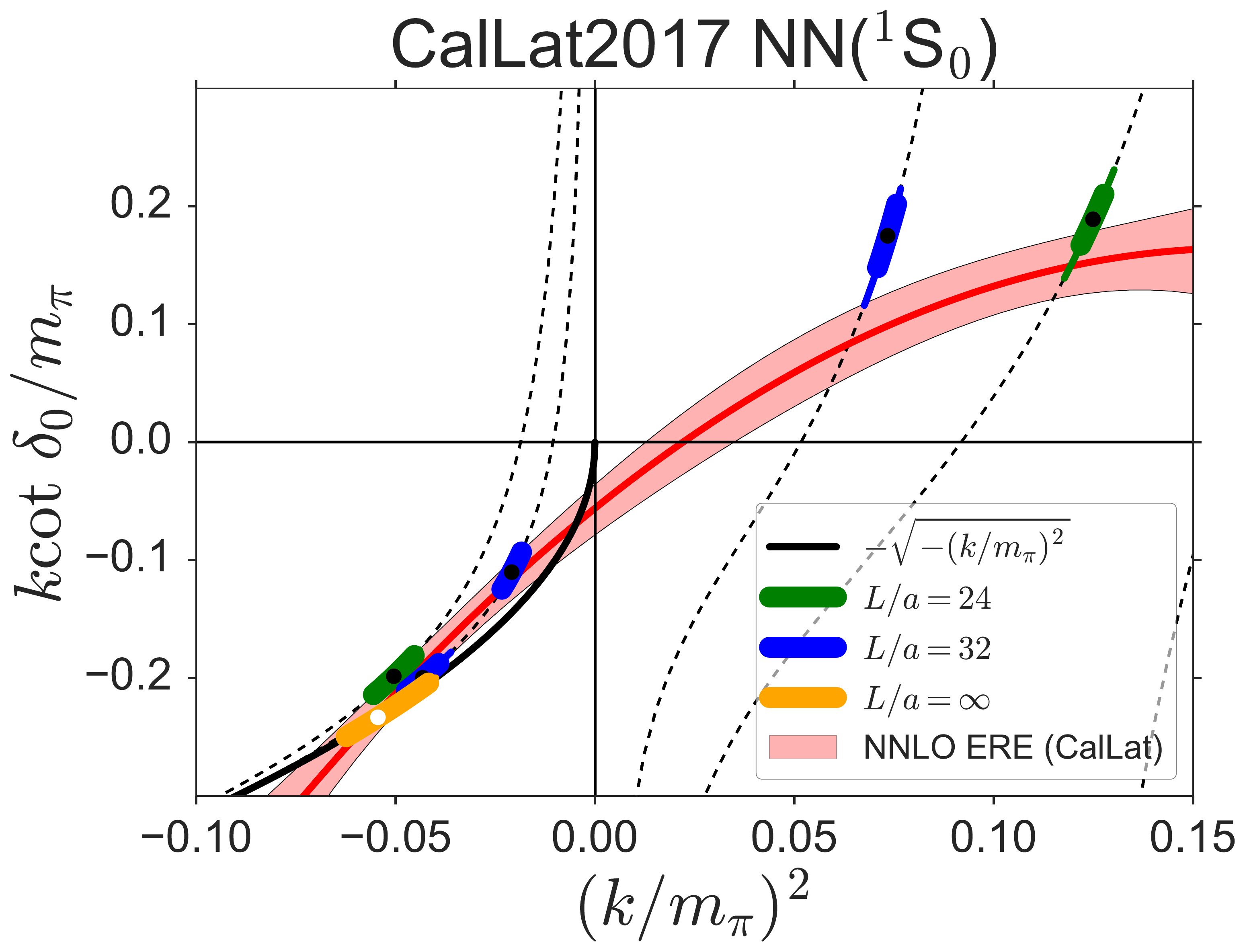}
  \includegraphics[width=0.49\textwidth]{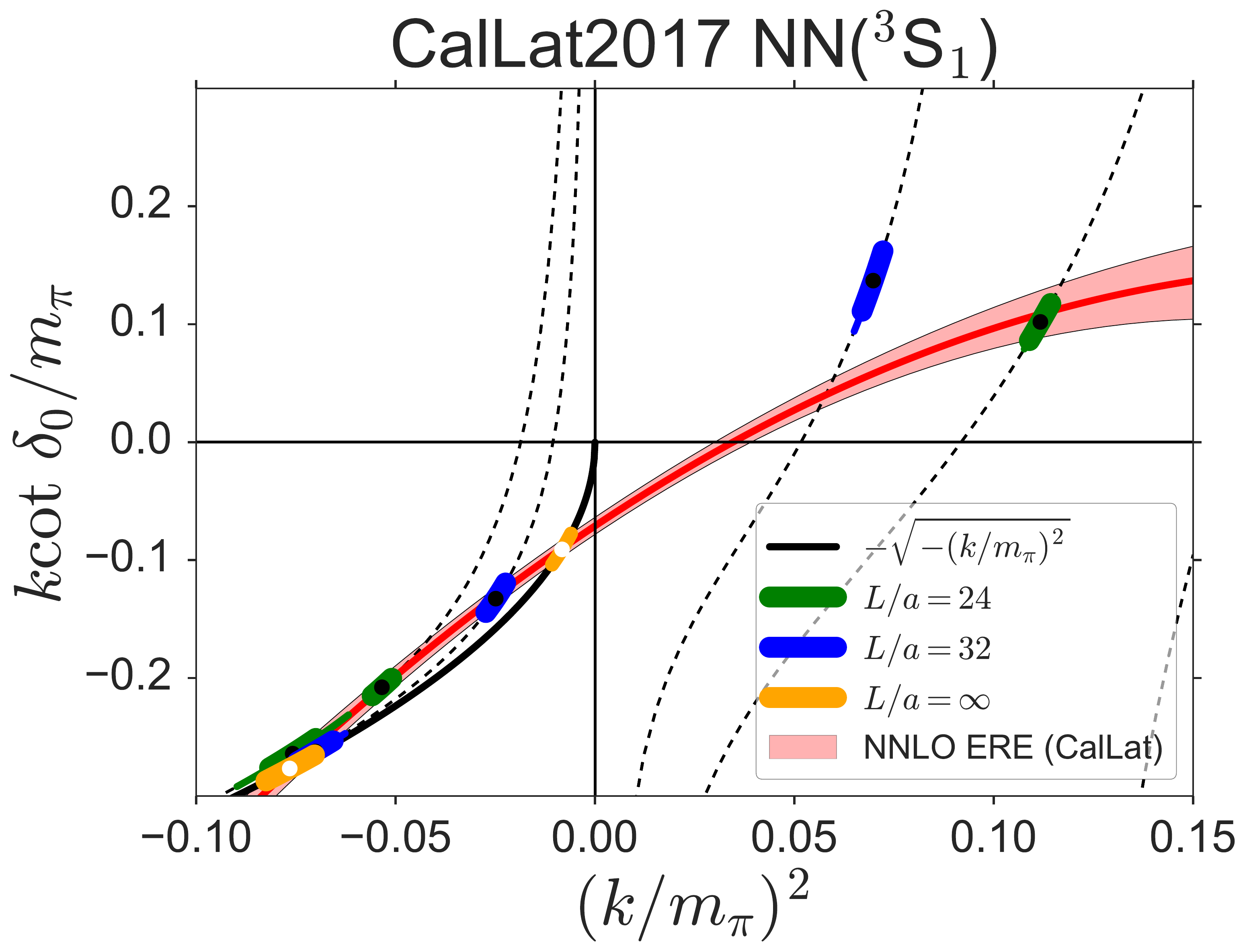}
  \caption{(Upper) Same as Fig.~\ref{fig:kcot_YKU2012}, but from CalLat2017.
    Red bands correspond to NNLO ERE  given in CalLat2017.
    (Lower) Same as upper figures but with data from excited states.
  }
  \label{fig:kcot_CalLat2017}
\end{figure}

In the rest of  this appendix, we analyze these data in terms of $\kcotd$.

Upper panels of Fig.~\ref{fig:kcot_NPL2013} show $\kcotd/m_\pi$ at $(k/m_\pi)^2 < 0 $ 
for $^1S_0$ (Left) and $^3S_1$ (Right) in the case of NPL2013. 
Given $L$, apparent inconsistency between $n=0$ (open symbols) and 
$n=2$ (black solid symbols) data\footnote{$n \equiv |\vec{n}|$ corresponds to
the boost momentum as $\vec{P} = (2\pi/L)\cdot\vec{n}$.}
in both channels can be seen clearly, which confirms the
 discussion in  Sec.~\ref{sec:introduction}.
 Lower panels of Fig.~\ref{fig:kcot_NPL2013} include data at $(k/m_\pi)^2 > 0 $ together with \EREp from NPL2013.  
 While \EREp and data at $(k/m_\pi)^2 < 0$ with $n=0$ show no apparent inconsistency, 
 \EREp themselves violate the physical condition for the residue of the bound state pole in both channels. 
 Considering the uncertainties,
   we put ``No'' and ``?'' for $^1S_0$ and $^3S_1$, respectively,
 about the Sanity check (iii) in Tab.~\ref{tab:Summary}.

Upper panels of Fig.~\ref{fig:kcot_CalLat2017} represent $\kcotd/m_\pi$ at $(k/m_\pi)^2 < 0 $  for CalLat2017,
 while lower panels of Fig.~\ref{fig:kcot_CalLat2017} include data at $(k/m_\pi)^2 > 0 $.
As already discussed in Sec.~\ref{sec:FVF}, the ``naive'' ERE fits by CalLat2017 contradict
 physical pole condition (see the right panel of Fig.~\ref{fig:kcot_demo:2pole}).
 If the two bound-state poles are physical,  $\kcotd$  should diverge
at a very narrow interval of $(k/m_\pi)^2$, between $-0.043$ (left blue point) and $-0.021$ (right blue point) for $^1S_0$ and
 between $-0.070$ (left blue point) and $-0.053$ (right green point)  for $^3S_1$.
 This is unlikely if not impossible, which supports 
 our interpretation that  two data at $k^2<0$ on each volume are the artifact due to
  the source operator dependence. 

Data at $(k/m_\pi)^2 > 0 $ behave rather differently from those at $(k/m_\pi)^2 < 0 $ (Lower panels).  As a consequence, their NNLO ERE fit misses the point at $(k/m_\pi)^2 > 0$ on $L/a=32$ in both channels.
We thus put ``?'' on the sanity check (i) in Tab.~\ref{tab:Summary}.

\clearpage
\section{Sanity check for  lattice data with hyperon(s)}
\label{app:other-BB}

Here we present  two examples  of the sanity check using the data given in appendix~\ref{app:data-table}.

\begin{figure}[bth]
\centering
  \includegraphics[width=0.49\textwidth]{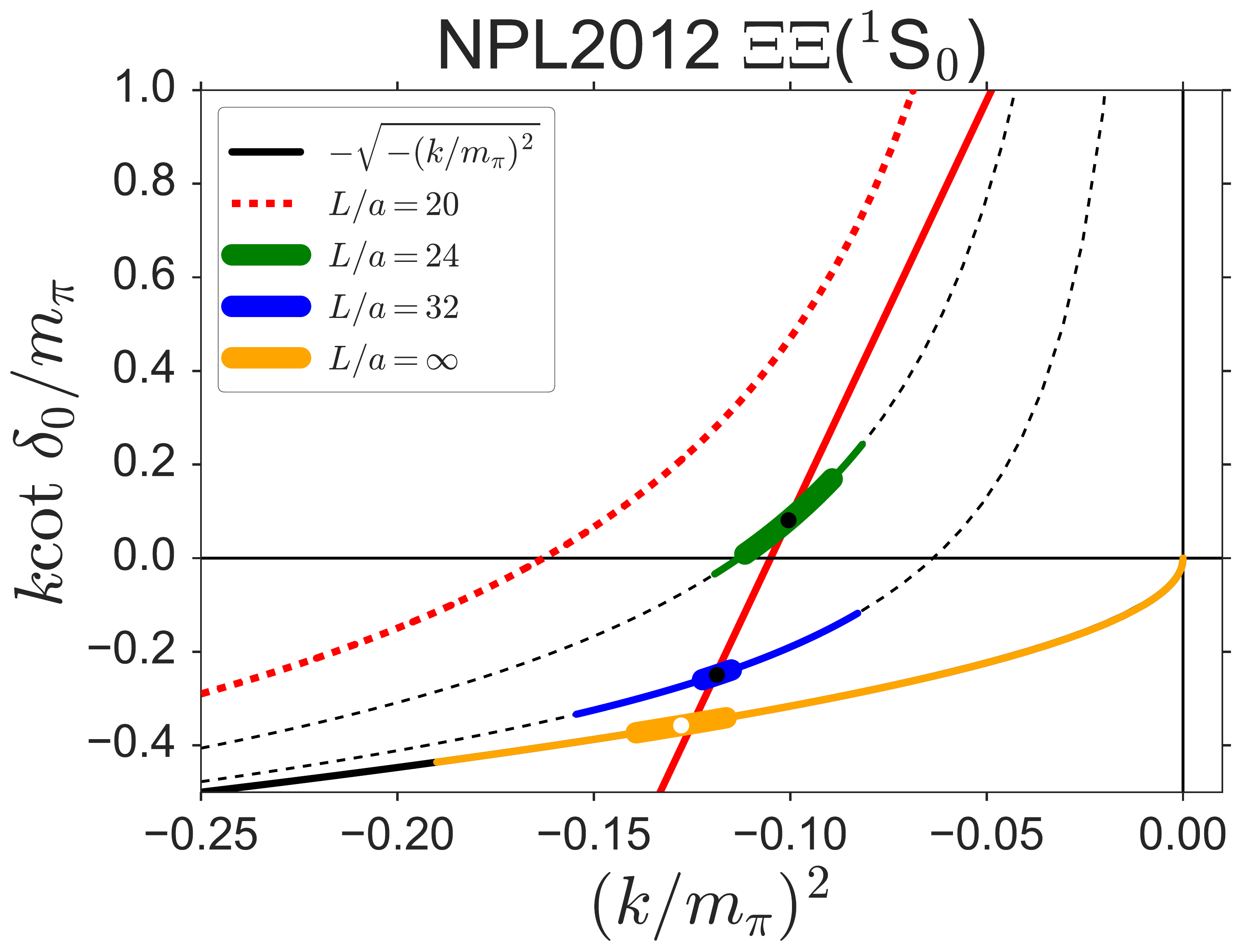}
  \includegraphics[width=0.49\textwidth]{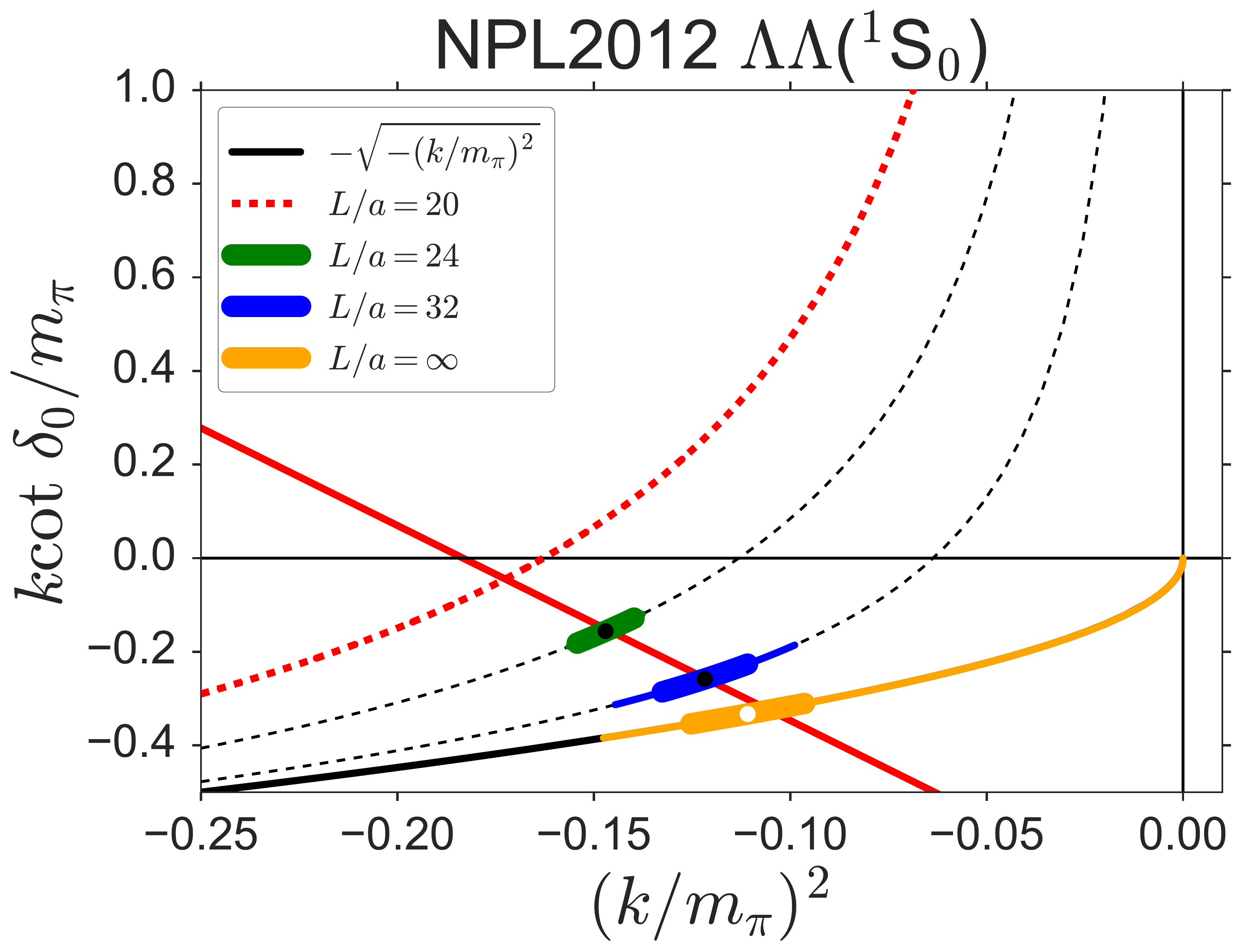}
 \caption{$\kcotd/m_\pi$ as a function of $(k/m_\pi)^2$ in NPL2012
   for $\Xi\Xi$($^1$S$_0$) (Left) and 
   $\Lambda\Lambda$($^1$S$_0$) (Right).
   Red lines correspond to the NLO ERE fit using two volumes.
    Red dashed line represents the L\"uscher's formula for $L/a = 20$,
    while the corresponding lattice data around $(k/m_\pi)^2=0$~\cite{Beane:2009py} are located 
    way out of the plot region of the figures.
}
 \label{fig:kcot_NPL2012:YY}
\end{figure}

Fig.~\ref{fig:kcot_NPL2012:YY} shows $k\cot\delta_0(k)/m_\pi$ as a function
of $(k/m_\pi)^2$ for
$\Xi\Xi$($^1$S$_0$) (Left) and
$\Lambda\Lambda$($^1$S$_0$) (Right) in the case of NPL2012.
The best NLO fit for data at $L/a=24, 32$ for $\Xi\Xi$($^1$S$_0$),
$( (a m_\pi)^{-1}, r m_\pi ) = (1.87, 35.6)$,
violates the physical pole condition Eq.~(\ref{eq:kcotd:bound}),
while that for $\Lambda\Lambda$($^1$S$_0$) does not violate the condition and gives 
$( (a m_\pi)^{-1}, r m_\pi ) = (-0.76, -8.33)$.

We also note that the earlier paper by NPLQCD Collaboration~\cite{Beane:2009py} reported 
the results with the same lattice setup but on a smaller volume ($L/a=20$),
\begin{eqnarray}
  (k/m_\pi)^2 = 0.0247(94)(77) \ {\rm for} \  \Xi\Xi(^1\mathrm{S}_0) , \  \ 
  (k/m_\pi)^2 = -0.033(09)(11) \ {\rm for} \  \Lambda\Lambda(^1\mathrm{S}_0) .
\end{eqnarray}
Such results clearly conflict with the ERE behaviors obtained from $L/a=24, 32$,
which intersects with the L\"uscher's formula for $L/a = 20$ (red dashed line)
at $(k/m_\pi)^2 = -0.173$ for $\Lambda\Lambda$($^1$S$_0$),
while it has no intersection for $\Xi\Xi$($^1$S$_0$) at $(k/m_\pi)^2 < 0$
\footnote{
  $(k/m_\pi)^2 = 0.0247(94)(77)$ for $\Xi\Xi$($^1$S$_0$) corresponds to
  $k\cot\delta_0/m_\pi = -5.11({}^{+1.26}_{-2.79})({}^{+0.83}_{-3.40})$,
which is also incompatible with the ERE from $L/a = 24, 32$.}.

\clearpage
\section{Data of $\Delta E$, $(k/m_\pi)^2$ and $k\cot\delta_0(k)/m_\pi$}
\label{app:data-table}

\begin{table}[hb]
\centering
\begin{tabular}{|c|c|c|c|c|c|}
\hline
Label   & state            & $L/a$ & $\Delta E$ [MeV]            & $(k/m_\pi)^2$           & $k\cot\delta_0(k)/m_\pi$            \\
\hline\hline
YKU2011 & $NN$ ($^1$S$_0$) & 32        & -3.0(1.7)($^{+0.3}_{-0.7}$) & -0.008(4)($^{+1}_{-2}$) & 0.17($^{+45}_{-14}$)($^{+21}_{-5}$) \\
        & two-states       & 48        & -4.5(0.9)($^{+2.1}_{-0.1}$) & -0.011(2)($^{+5}_{-1}$) & -0.08($^{+2}_{-2}$)($^{+9}_{-1}$)   \\
        &                  & $\infty$  & -4.4(0.6)(1.0)              & -0.011(1)($^{+3}_{-3}$) & -0.11($^{+1}_{-1}$)($^{+1}_{-1}$)   \\
                \cline{3-6}
        &                  & 32        & 15.8(1.6)($^{+9.6}_{-0.3}$) & 0.040(4)($^{+25}_{-2}$) & -0.13($^{+2}_{-2}$)($^{+12}_{-1}$)  \\
        &                  & 48        & 4.2(0.8)($^{+2.1}_{-0.0}$)  & 0.011(2)($^{+5}_{-1}$)  & -0.15($^{+2}_{-2}$)($^{+5}_{-1}$)   \\
                \cline{2-6}
        & $\mathcal{O}_1$  & 24        & -6.1(2.3)(2.2)              & $-0.02(1)(^{+1}_{-1})$  & $0.17(^{+26}_{-12})(^{+39}_{-10})$  \\
        &                  & 48        & -5.2(2.6)(0.8)              & $-0.01(1)(^{+0}_{-0})$  & $-0.09(^{+8}_{-4})(^{+4}_{-1})$     \\
        &                  & 96        & -4.6(2.0)(1.1)              & $-0.012(5)(^{+3}_{-3})$ & $-0.11(^{+3}_{-2})(^{+2}_{-1})$     \\
                \cline{2-6}
        & $\mathcal{O}_2$  & 24        & -8.4(1.5)(0.5)              & $-0.021(4)(^{+1}_{-1})$ & $0.05(^{+7}_{-5})(^{+3}_{-2})$      \\
        &                  & 48        & -6.4(2.0)(0.8)              & $-0.016(5)(^{+2}_{-2})$ & $-0.11(^{+4}_{-2})(^{+2}_{-1})$     \\
        &                  & 96        & -6.0(1.9)(0.5)              & $-0.015(5)(^{+1}_{-1})$ & $-0.12(2)(^{+1}_{-1})$              \\
                \cline{2-6}
        & $NN$ ($^3$S$_1$) & 32        & -6.4(1.3)($^{+0.1}_{-0.7}$) & -0.016(3)($^{+1}_{-2}$) & -0.03($^{+5}_{-3}$)($^{+2}_{-2}$)   \\
        & two-states       & 48        & -7.1(0.7)($^{+2.2}_{-0.1}$) & -0.018(2)($^{+6}_{-1}$) & -0.12($^{+1}_{-1}$)($^{+4}_{-0}$)   \\
        &                  & $\infty$  & -7.5(0.5)(0.9)              & -0.019(1)($^{+2}_{-2}$) & -0.14($^{+1}_{-0}$)($^{+1}_{-1}$)   \\
                \cline{3-6}
        &                  & 32        & 13.3(1.3)($^{+6.6}_{-1.7}$) & 0.034(3)($^{+17}_{-4}$) & -0.17($^{+2}_{-2}$)($^{+8}_{-3}$)   \\
        &                  & 48        & 2.3(0.8)($^{+2.2}_{-0.1}$)  & 0.006(2)($^{+6}_{-1}$)  & -0.23($^{+4}_{-7}$)($^{+8}_{-4}$)   \\
                \cline{2-6}
        & $\mathcal{O}_1$  & 24        & -10.2(2.2)(1.6)             & $-0.03(1)(^{+0}_{-0})$  & $-0.02(^{+8}_{-5})(^{+7}_{-4})$     \\
        &                  & 48        & -9.6(2.6)(0.9)              & $-0.02(1)(^{+0}_{-0})$  & $-0.15(^{+3}_{-2})(^{+1}_{-1})$     \\
        &                  & 96        & -7.8(2.1)(0.4)              & $-0.02(1)(^{+0}_{-0})$  & $-0.14(2)(^{+0}_{-0})$              \\
                \cline{2-6}
        & $\mathcal{O}_2$  & 24        & -10.0(1.5)(0.5)             & $-0.025(4)(^{+1}_{-1})$ & $-0.01(^{+5}_{-4})(^{+2}_{-1})$     \\
        &                  & 48        & -10.2(2.0)(0.8)             & $-0.026(5)(^{+2}_{-2})$ & $-0.15(2)(^{+1}_{-1})$              \\
        &                  & 96        & -9.0(2.0)(0.5)              & $-0.023(5)(^{+1}_{-1})$ & $-0.15(2)(^{+1}_{-0})$              \\

\hline
\end{tabular}
\caption{Summary of the data from YKU2011 \cite{Yamazaki:2011nd}. 
 Corresponding $(k/m_\pi)^2$ and $k\cot \delta_0(k) /m_\pi$ are calculated by using Eq.~(\ref{eq:kcot_delta}). 
}
\label{tab:data1}
\end{table}

\begin{table}
\centering
\begin{tabular}{|c|c|c|c|c|c|}
\hline
Label    & state            & $L/a$ & $\Delta E$ [MeV]             & $(k/m_\pi)^2$           & $k\cot\delta_0(k)/m_\pi$           \\
\hline\hline
YIKU2012 & $NN$ ($^1$S$_0$) & 32       & -6.2(2.4)(0.5)               & -0.03(1)($^{+0}_{-0}$)  & 0.60($^{+63}_{-29}$)($^{+24}_{-7}$) \\
         &                  & 40       & -8.2(4.0)(1.5)               & -0.04(2)($^{+1}_{-1}$)  & 0.04($^{+38}_{-15}$)($^{+24}_{-5}$) \\
         &                  & 48       & -7.3(1.7)(0.5)               & -0.04(1)($^{+0}_{-0}$)  & -0.05($^{+9}_{-6}$)($^{+3}_{-2}$)  \\
         &                  & 64       & -7.2(1.4)(0.3)               & -0.03(1)($^{+0}_{-0}$)  & -0.15($^{+4}_{-3}$)($^{+1}_{-1}$)  \\
         &                  & $\infty$ & -7.4(1.3)(0.6)               & -0.04(1)($^{+0}_{-0}$)  & -0.19($^{+2}_{-2}$)($^{+1}_{-1}$)  \\
                \cline{2-6}
         & $NN$ ($^3$S$_1$) & 32       & -12.4(2.1)(0.5)              & -0.06(1)($^{+0}_{-0}$)  & 0.07($^{+11}_{-8}$)($^{+4}_{-2}$)  \\
         &                  & 40       & -12.2(1.9)(0.6)              & -0.06(1)($^{+0}_{-0}$)  & -0.11($^{+6}_{-4}$)($^{+3}_{-2}$)  \\
         &                  & 48       & -11.1(1.7)(0.3)              & -0.05(1)($^{+0}_{-0}$)  & -0.16($^{+4}_{-3}$)($^{+1}_{-1}$)  \\
         &                  & 64       & -11.7(1.2)(0.5)              & -0.06(1)($^{+0}_{-0}$)  & -0.22($^{+2}_{-1}$)($^{+1}_{-1}$)  \\
         &                  & $\infty$ & -11.5(1.1)(0.6)              & -0.06(1)($^{+0}_{-0}$)  & -0.24($^{+1}_{-1}$)($^{+1}_{-1}$)  \\
\hline  
YIKU2015 & $NN$ ($^1$S$_0$) & 48       & -7.7(0.9)($^{+1.2}_{-2.4}$)  & -0.09(1)($^{+1}_{-3}$)  & -0.01($^{+7}_{-6}$)($^{+11}_{-14}$)  \\
         &                  & 64       & -9.5(0.9)($^{+0.8}_{-0.5}$)  & -0.11(1)($^{+1}_{-1}$)  & -0.27($^{+3}_{-2}$)($^{+3}_{-1}$)  \\
         &                  & $\infty$ & -8.5(0.7)($^{+0.5}_{-1.6}$)  & -0.10(1)($^{+1}_{-2}$)  & -0.32($^{+1}_{-1}$)($^{+1}_{-3}$)  \\
                \cline{2-6}
         & $NN$ ($^3$S$_1$) & 48       & -13.8(0.9)($^{+1.7}_{-3.6}$) & -0.16(1)($^{+2}_{-4}$)  & -0.29($^{+3}_{-2}$)($^{+6}_{-9}$)  \\
         &                  & 64       & -15.6(1.2)($^{+1.3}_{-1.0}$) & -0.18(1)($^{+2}_{-1}$)  & -0.40($^{+2}_{-2}$)($^{+2}_{-2}$)  \\
         &                  & $\infty$ & -14.5(0.7)($^{+0.8}_{-2.4}$) & -0.17(1)($^{+1}_{-3}$)  & -0.41($^{+1}_{-1}$)($^{+1}_{-3}$)  \\
\hline
\end{tabular}
\caption{Summary of the data from YIKU2012 \cite{Yamazaki:2012hi}  and YIKU2015 \cite{Yamazaki:2015asa}. 
 Corresponding $(k/m_\pi)^2$ and $k\cot \delta_0(k) /m_\pi$ are calculated by using Eq.~(\ref{eq:kcot_delta}). 
}
\label{tab:data1b}
\end{table}

\begin{table}
\centering
\begin{tabular}{|c|c|c|c|c|c|}
\hline
Label   & state                     & $L/a$   & $\Delta E$ [MeV] & $(k/m_\pi)^2$             & $k\cot\delta_0(k)/m_\pi$              \\
\hline\hline
NPL2012 & $NN$ ($^1$S$_0$)          & 24       & -10.4(2.6)(3.1)  & -0.08(2)($^{+2}_{-2}$)    & 0.25($^{+28}_{-17}$)($^{+45}_{-18}$)  \\
        &                           & 32       & -8.3(2.2)(3.3)   & -0.06(2)($^{+3}_{-3}$)    & -0.01($^{+17}_{-10}$)($^{+38}_{-13}$) \\
        &                           & $\infty$ & -7.1(5.2)(7.3)   & -0.06(4)($^{+6}_{-6}$)    & -0.24($^{+11}_{-7}$)($^{+21}_{-9}$)$\ast$   \\
                \cline{2-6}
        & $NN$ ($^3$S$_1$)          & 24       & -22.3(2.3)(5.4)  & -0.17(2)($^{+4}_{-4}$)    & -0.24($^{+5}_{-5}$)($^{+15}_{-10}$)   \\
        &                           & 32       & -14.9(2.3)(5.8)  & -0.12(2)($^{+5}_{-5}$)    & -0.24($^{+5}_{-4}$)($^{+21}_{-10}$)   \\
        &                           & $\infty$ & -11.0(5.0)(12.0) & -0.09(4)($^{+9}_{-9}$)   & -0.29($^{+7}_{-6}$)($^{+28}_{-13}$)$\ast$   \\
                \cline{2-6}
        & $\Lambda\Lambda(^1$S$_0$) & 24       & -17.5(0.9)(0.7)  & -0.15(1)($^{+1}_{-1}$)    & -0.16($^{+3}_{-3}$)($^{+2}_{-2}$)     \\
        &                           & 32       & -14.5(1.3)(2.4)  & -0.12(1)($^{+2}_{-2}$)    & -0.26($^{+3}_{-3}$)($^{+6}_{-5}$)     \\
        &                           & $\infty$ & -13.2(1.8)(4.0)  & -0.11(1)($^{+3}_{-3}$)    & -0.33($^{+2}_{-2}$)($^{+6}_{-5}$)     \\
                \cline{2-6}
        & $\Xi\Xi (^1$S$_0)$        & 24       & -11.0(1.3)(1.6)  & -0.10(1)($^{+2}_{-2}$)    & 0.08($^{+9}_{-7}$)($^{+14}_{-9}$)   \\
        &                           & 32       & -13.0(0.5)(3.9)  & -0.119(4)($^{+36}_{-36}$) & -0.25($^{+1}_{-1}$)($^{+13}_{-8}$)    \\
        &                           & $\infty$ & -14.0(1.4)(6.7)  & -0.13(1)($^{+6}_{-6}$)    & -0.36($^{+2}_{-2}$)($^{+10}_{-8}$)     \\
\hline
\end{tabular}
\caption{Same as Table~\ref{tab:data1}, but from NPL2012 \cite{Beane:2011iw}.
  To evaluate the systematic errors  for $k\cot\delta_0(k)/m_\pi$ with the * symbol,
  we impose a constraint that the corresponding $(k/m_\pi)^2$ is negative,
  since the pole condition $k\cot\delta_0(k) = - \sqrt{-k^2}$ is meaningful only for negative $k^2$.
}
\label{tab:data3}
\end{table}

\begin{table}
\centering
\begin{tabular}{|c|c|c|c|c|c|c|}
\hline
Label   &                           state & $n$ & $L/a$ & $\Delta E$ [MeV] &           $(k/m_\pi)^2$ & $k\cot\delta_0(k)/m_\pi$           \\
\hline\hline
NPL2013 &           {\bf 27}  ($^1$S$_0$) & 0   &    24 & -17.8(1.7)(2.8)  & -0.044(4)($^{+7}_{-7}$) & -0.18($^{+2}_{-1}$)($^{+3}_{-2}$)  \\
        &                                 &     &    32 & -15.1(2.0)(2.0)  & -0.038(5)($^{+5}_{-5}$) & -0.18($^{+2}_{-1}$)($^{+2}_{-2}$)  \\
        &                                 &     &    48 & -13.1(2.8)(4.3)  &  -0.03(1)($^{+1}_{-1}$) & -0.18($^{+2}_{-2}$)($^{+4}_{-3}$)  \\
                  \cline{4-7}
        &                                 &     &    24 & 48.7(1.8)(2.2)   &  0.123(4)($^{+6}_{-6}$) & 0.18($^{+3}_{-3}$)($^{+5}_{-4}$)   \\
        &                                 &     &    32 & 22.5(1.8)(3.0)   &  0.056(4)($^{+8}_{-8}$) & 0.03($^{+3}_{-3}$)($^{+6}_{-5}$)   \\
                  \cline{3-7}
        &                                 & 1   &    24 & -6.9(1.8)(3.8)   & -0.017(4)($^{+10}_{-10}$) & -0.13($^{+2}_{-1}$)($^{+4}_{-3}$)  \\
        &                                 &     &    32 & -12.3(1.9)(3.6)  & -0.031(5)($^{+9}_{-9}$) & -0.17($^{+1}_{-1}$)($^{+3}_{-2}$)  \\
        &                                 &     &    48 & -14.9(2.7)(2.7)  &  -0.04(1)($^{+1}_{-1}$) & -0.19($^{+2}_{-1}$)($^{+2}_{-2}$)  \\
                  \cline{3-7}
        &                                 & 2   &    24 & -28.5(2.3)(3.8)  &  -0.07(1)($^{+1}_{-1}$) & -0.25($^{+1}_{-1}$)($^{+2}_{-2}$)  \\
        &                                 &     &    32 & -24.9(2.2)(3.1)  &  -0.06(1)($^{+1}_{-1}$) & -0.25($^{+1}_{-1}$)($^{+2}_{-2}$)  \\
        &                                 &     &    48 & -19.3(2.9)(3.3)  &  -0.05(1)($^{+0}_{-1}$) & -0.22($^{+2}_{-1}$)($^{+2}_{-2}$)  \\
                  \cline{2-7}
        & $\overline{\bf10}$  ($^3$S$_1$) & 0   &    24 & -25.4(2.6)(4.7)  &  -0.06(1)($^{+1}_{-1}$) & -0.24($^{+2}_{-2}$)($^{+3}_{-3}$)  \\
        &                                 &     &    32 & -22.5(2.3)(2.6)  &  -0.06(1)($^{+1}_{-1}$) & -0.23($^{+1}_{-1}$)($^{+2}_{-1}$)  \\
        &                                 &     &    48 & -19.7(3.1)(4.1)  &  -0.05(1)($^{+1}_{-1}$) & -0.22($^{+2}_{-2}$)($^{+3}_{-2}$)  \\
                  \cline{4-7}
        &                                 &     &    24 & 41.6(2.2)(3.1)   &   0.10(1)($^{+1}_{-1}$) & 0.06($^{+3}_{-3}$)($^{+5}_{-4}$)   \\
        &                                 &     &    32 & 15.7(2.3)(3.1)   &   0.04(1)($^{+1}_{-1}$) & -0.07($^{+3}_{-3}$)($^{+4}_{-4}$)  \\
                  \cline{3-7}
        &                                 & 1   &    24 & -16.0(2.7)(5.9)  &  -0.04(1)($^{+1}_{-1}$) & -0.20($^{+2}_{-1}$)($^{+4}_{-3}$)  \\
        &                                 &     &    32 & -19.2(2.3)(3.7)  &  -0.05(1)($^{+1}_{-1}$) & -0.22($^{+1}_{-1}$)($^{+2}_{-2}$)  \\
        &                                 &     &    48 & -17.8(3.6)(3.1)  &  -0.04(1)($^{+1}_{-1}$) & -0.21($^{+2}_{-1}$)($^{+2}_{-2}$)  \\
                  \cline{3-7}
        &                                 & 2   &    24 & -40.7(3.6)(7.4)  &  -0.10(1)($^{+2}_{-2}$) & -0.31($^{+1}_{-1}$)($^{+3}_{-3}$)  \\
        &                                 &     &    32 & -31.6(2.7)(3.2)  &  -0.08(1)($^{+1}_{-1}$) & -0.28($^{+1}_{-1}$)($^{+2}_{-1}$)  \\
        &                                 &     &    48 & -23.1(3.9)(5.5)  &  -0.06(1)($^{+1}_{-1}$) & -0.24($^{+2}_{-1}$)($^{+3}_{-3}$)  \\
\hline
\end{tabular}
\caption{Same as Table~\ref{tab:data1}, but from NPL2013 \cite{Beane:2012vq,Beane:2013br}.  
  $\mathbf{27}$($^1$S$_0$) and $\overline{\mathbf{10}}$($^3$S$_1$) irreducible representations of flavor SU(3)
  correspond to $NN$($^1$S$_0$) and $NN$($^3$S$_1$), respectively.
  $n \equiv |\vec n|$ in the Table is related to the
 boost momentum as $\vec P = (2\pi/L) \vec n$.
}
\label{tab:data3a}
\end{table}

\begin{table}
\centering
\begin{tabular}{|c|c|c|c|c|c|c|}
\hline
Label & state & $n$ & $L/a$ & $\Delta E$ [MeV] & $(k/m_\pi)^2$ & $k\cot\delta_0(k)/m_\pi$ \\
\hline\hline
NPL2013 (continued) & {\bf 1}     & 0 & 24 & -77.7(1.8)(3.2) & -0.192(4)($^{+8}_{-8}$) & -0.438($^{+5}_{-5}$)($^{+9}_{-9}$) \\
                   &             &   & 32 & -76.0(2.3)(2.8) & -0.19(1)($^{+1}_{-1}$)  & -0.43($^{+1}_{-1}$)($^{+1}_{-1}$)  \\
                   &             &   & 48 & -73.7(3.3)(5.1) & -0.18(1)($^{+1}_{-1}$)  & -0.43($^{+1}_{-1}$)($^{+2}_{-1}$)  \\
                  \cline{3-7}
                   &             & 1 & 24 & -67.2(2.5)(2.5) & -0.17(1)($^{+1}_{-1}$)  & -0.41($^{+1}_{-1}$)($^{+1}_{-1}$)  \\
                   &             &   & 32 & -70.3(2.3)(3.1) & -0.17(1)($^{+1}_{-1}$)  & -0.42($^{+1}_{-1}$)($^{+1}_{-1}$)  \\
                   &             &   & 48 & -73.7(4.4)(7.6) & -0.18(1)($^{+2}_{-2}$)  & -0.43($^{+1}_{-1}$)($^{+2}_{-2}$)  \\
                  \cline{3-7}
                   &             & 2 & 24 & -85.0(3.1)(4.0) & -0.21(1)($^{+1}_{-1}$)  & -0.46($^{+1}_{-1}$)($^{+1}_{-1}$)  \\
                   &             &   & 32 & -79.6(2.6)(3.9) & -0.20(1)($^{+1}_{-1}$)  & -0.44($^{+1}_{-1}$)($^{+1}_{-1}$)  \\
                   &             &   & 48 & -75.4(3.3)(3.3) & -0.19(1)($^{+1}_{-1}$)  & -0.43($^{+1}_{-1}$)($^{+1}_{-1}$)  \\
                  \cline{2-7}
                   & {\bf 8$_A$} & 0 & 24 & -40.1(1.7)(2.9) & -0.100(4)($^{+7}_{-7}$) & -0.31($^{+1}_{-1}$)($^{+1}_{-1}$)  \\
                   &             &   & 32 & -38.5(2.3)(4.4) & -0.10(1)($^{+1}_{-1}$)  & -0.31($^{+1}_{-1}$)($^{+2}_{-2}$)  \\
                   &             &   & 48 & -38.7(2.9)(2.9) & -0.10(1)($^{+1}_{-1}$)  & -0.31($^{+1}_{-1}$)($^{+1}_{-1}$)  \\
                  \cline{3-7}
                   &             & 1 & 24 & -26.5(1.8)(3.6) & -0.066(4)($^{+9}_{-9}$) & -0.25($^{+1}_{-1}$)($^{+2}_{-2}$)  \\
                   &             &   & 32 & -34.0(2.6)(3.4) & -0.08(1)($^{+1}_{-1}$)  & -0.29($^{+1}_{-1}$)($^{+2}_{-1}$)  \\
                   &             &   & 48 & -34.6(2.8)(3.1) & -0.09(1)($^{+1}_{-1}$)  & -0.29($^{+1}_{-1}$)($^{+1}_{-1}$)  \\
                  \cline{3-7}
                   &             & 2 & 24 & -46.7(2.0)(3.2) & -0.116(5)($^{+8}_{-8}$) & -0.34($^{+1}_{-1}$)($^{+1}_{-1}$)  \\
                   &             &   & 32 & -45.2(3.0)(3.1) & -0.11(1)($^{+1}_{-1}$)  & -0.33($^{+1}_{-1}$)($^{+1}_{-1}$)  \\
                   &             &   & 48 & -39.7(3.0)(2.7) & -0.10(1)($^{+1}_{-1}$)  & -0.31($^{+1}_{-1}$)($^{+1}_{-1}$)  \\
                  \cline{2-7}
                   & {\bf 10}    & 0 & 24 & -11.4(1.8)(4.0) & -0.029(4)($^{+10}_{-10}$) & -0.10($^{+3}_{-3}$)($^{+11}_{-5}$)  \\
                   &             &   & 32 & -10.5(2.5)(4.1) & -0.03(1)($^{+1}_{-1}$)  & -0.14($^{+3}_{-3}$)($^{+8}_{-4}$)  \\
                   &             &   & 48 & -6.6(3.4)(4.1)  & -0.02(1)($^{+1}_{-1}$)  & -0.12($^{+6}_{-3}$)($^{+17}_{-4}$) \\
                  \cline{3-7}
                   &             & 1 & 24 & - 6.3(1.9)(4.4) & -0.016(5)($^{+11}_{-11}$) & -0.12($^{+2}_{-2}$)($^{+5}_{-4}$)  \\
                   &             &   & 32 & - 1.1(2.4)(4.2) & -0.003(6)($^{+11}_{-11}$) & -0.06($^{+4}_{-3}$)($^{+10}_{-5}$)  \\
                   &             &   & 48 & - 2.8(3.1)(4.1) & -0.01(1)($^{+1}_{-1}$)  & -0.08($^{+6}_{-4}$)($^{+15}_{-4}$) \\
                  \cline{3-7}
                   &             & 2 & 24 & -15.3(2.2)(4.5) & -0.04(1)($^{+1}_{-1}$)  & -0.15($^{+3}_{-2}$)($^{+7}_{-4}$)  \\
                   &             &   & 32 & -12.9(2.6)(4.5) & -0.03(1)($^{+1}_{-1}$)  & -0.16($^{+3}_{-2}$)($^{+6}_{-4}$)  \\
                   &             &   & 48 & - 7.0(3.4)(3.7) & -0.02(1)($^{+1}_{-1}$)  & -0.13($^{+5}_{-3}$)($^{+10}_{-3}$)  \\
\hline  
\end{tabular}
\caption{Same as Table~\ref{tab:data1}, but from NPL2013 \cite{Beane:2012vq,Beane:2013br} (continued). 
}
\label{tab:data4}
\end{table}

\begin{table}
\centering
\begin{tabular}{|c|c|c|c|c|c|}
\hline
Label & state & $L/a$ & $\Delta E$ [MeV] & $(k/m_\pi)^2$ & $k\cot\delta_0(k)/m_\pi$ \\
\hline\hline
NPL2015 & $NN$ ($^1$S$_0$) & 24         & -24.1(1.5)(4.5)              & -0.15(1)($^{+3}_{-3}$) & -0.23($^{+3}_{-3}$)($^{+10}_{-7}$)     \\
        &                  & 32         & -18.4(1.5)(3.3)              & -0.11(1)($^{+2}_{-2}$) & -0.27($^{+2}_{-2}$)($^{+6}_{-5}$)     \\
        &                  & 48         & -11.8(1.9)(3.1)              & -0.07(1)($^{+2}_{-2}$) & -0.25($^{+3}_{-3}$)($^{+6}_{-4}$)     \\
        &                  & $\infty$   & -12.5($^{+1.9}_{-1.7}$)($^{+4.5}_{-2.5}$) & -0.08(1)($^{+3}_{-2}$) & -0.28($^{+2}_{-2}$)($^{+6}_{-3}$)     \\
                 \cline{3-6}
        &                  & 32$^{(\ast)}$ & 7.9(2.1)($^{+3.3}_{-3.3}$)   & 0.05(1)($^{+2}_{-2}$)  & 0.13($^{+10}_{-8}$)($^{+14}_{-8}$)     \\
        &                  & 48         & 33.2(1.8)($^{+4.7}_{-4.4}$)  & 0.21(1)($^{+3}_{-3}$)  & 0.87($^{+36}_{-23}$)($^{+379}_{-41}$) \\
                 \cline{2-6}
        & $NN$ ($^3$S$_1$) & 24         & -19.6(1.2)(1.6)              & -0.12(1)($^{+1}_{-1}$) & -0.14($^{+3}_{-3}$)($^{+5}_{-4}$)     \\
        &                  & 32         & -17.5(1.5)(1.6)              & -0.11(1)($^{+1}_{-1}$) & -0.25($^{+3}_{-2}$)($^{+3}_{-2}$)     \\
        &                  & 48         & -13.3(2.0)(3.2)              & -0.08(1)($^{+2}_{-2}$) & -0.27($^{+3}_{-2}$)($^{+5}_{-4}$)     \\
        &                  & $\infty$   & -14.4($^{+1.8}_{-1.6}$)($^{+1.8}_{-2.7}$) & -0.09(1)($^{+1}_{-2}$) & -0.30($^{+2}_{-2}$)($^{+2}_{-3}$)     \\
                 \cline{3-6}
        &                  & 32$^{(\ast)\dagger}$ & 11.9(2.4)($^{+3.7}_{-5.0}$)  & 0.07(1)($^{+2}_{-2}$)  & 0.35($^{+21}_{-18}$)($^{+46}_{-18}$)  \\
        &                  & 48$^{\dagger}$      & 29.4(5.0)($^{+0.2}_{-0.2}$)  & 0.18(3)($^{+1}_{-1}$)    & 0.44($^{+66}_{-25}$)($^{+42}_{-9}$)    \\
\hline  
\end{tabular}
\caption{Same as Table~\ref{tab:data1}, but from NPL2015 \cite{Orginos:2015aya}.  For the data with $^{(\ast)}$, the 
 boost momentum $n = 1$ is taken.   
 $^\dagger$ Errors for $((k/m_\pi)^2, k\cot\delta_0(k)/m_\pi)$ given in Ref.~\cite{Orginos:2015aya}
  seem to be inconsistent between their Table VII and their Fig.~19.  
  In the above Table, 
  we assume their Fig.~19  is correct, and reevaluated the errors for $(k/m_\pi)^2$. Central values are unchanged.
}
\label{tab:data5}
\end{table}

\begin{table}[h]
  \centering
  \begin{tabular}{|c|c|c|c|c|c|}
    \hline
    Label & state & $L/a$ & $\Delta E$ [MeV] & $(k/m_\pi)^2$ & $k\cot\delta_0(k)/m_\pi$ \\
    \hline\hline
    CalLat2017 & $NN$($^1$S$_0$) & 24 & $-20.2(2.1)(1.5)$ & $-0.05(1)(^{+0}_{-0})$ & $-0.20(2)(^{+1}_{-1})$ \\
               &           & 32 & $-17.3(1.7)(2.3)$ & $-0.043(4)(^{+6}_{-6})$ & $-0.20(1)(^{+2}_{-2})$ \\
               &           & $\infty$ & $-21.8(^{+5.1}_{-3.2})(^{+2.8}_{-0.8})$ & $-0.054(^{+13}_{-8})(^{+7}_{-2})$ & $-0.233(^{+29}_{-16})(^{+17}_{-4})$ \\
               \cline{3-6}
               &           & 32 & $-8.3(1.0)(0.5)$ & $-0.021(2)(^{+1}_{-1})$ & $-0.11(^{+2}_{-1})(^{+1}_{-1})$ \\
               \cline{3-6}
& & 24$^\dagger$ & $49.5(1.1)(^{+1.8}_{-2.6})$ & $0.125(3)(^{+5}_{-7})$ & $0.19(2)(^{+4}_{-4})$ \\
& & 32$^\dagger$ & $29.2(0.9)(^{+0.9}_{-2.1})$ & $0.073(2)(^{+2}_{-5})$ & $0.18(3)(^{+3}_{-5})$ \\
\cline{2-6}
& $NN$($^3$S$_1$)  & 24 & $-30.4(2.4)(5.1)$ & $-0.08(1)(^{+1}_{-1})$ & $-0.26(1)(^{+3}_{-3})$ \\
& & 32 & $-28.1(1.8)(2.4)$ & $-0.070(4)(^{+6}_{-6})$ & $-0.26(1)(^{+1}_{-1})$ \\
& & $\infty$ & $-30.7(^{+2.5}_{-2.4})(^{+1.6}_{-0.5})$ & $-0.077(6)(^{+4}_{-1})$ & $-0.277(11)(^{+7}_{-2})$ \\
 \cline{3-6}
& & 24 & $-21.4(1.0)(0.5)$ & $-0.053(2)(^{+1}_{-1})$ & $-0.21(1)(^{+0}_{-0})$ \\
& & 32 & $-10.0(1.0)(0.4)$ & $-0.025(2)(^{+1}_{-1})$ & $-0.13(1)(^{+1}_{-0})$ \\
 & & $\infty$ & $-3.3(^{+0.9}_{-1.0})(^{+0.2}_{-0.6})$ & $-0.008(^{+2}_{-3})(^{+1}_{-2})$ & $-0.091(13)(^{+3}_{-7})$ \\
  \cline{3-6}
& & 24$^\dagger$ & $44.3(1.1)(^{+0}_{-1.2})$ & $0.112(3)(^{+1}_{-3})$ & $0.10(2)(^{+0}_{-2})$ \\
& & 32$^\dagger$ & $27.7(1.0)(^{+0}_{-1.5})$ & $0.070(2)(^{+1}_{-4})$ & $0.14(3)(^{+1}_{-3})$ \\
 \hline
\hline
  \end{tabular}
  \caption{Same as Table~\ref{tab:data1}, but from CalLat2017 \cite{Berkowitz:2015eaa}.
    $^\dagger$ The values for the scattering states are read from the figures in Ref \cite{Berkowitz:2015eaa}.
  }
  \label{tab:data6}
\end{table}

\end{document}